\documentclass{aa}

\usepackage{float}
\usepackage{graphicx}
\usepackage{natbib}
\usepackage{txfonts}
\usepackage{color}
\usepackage{enumerate}
\usepackage{listings}
\usepackage{bm}
\usepackage{hyperref}
\hypersetup{
  colorlinks=true,
  pdfborder=2 2 0.,
  linkcolor=blue,
  anchorcolor=black,
  citecolor=blue,
  urlcolor=black,
}

\begin{document} 

\title{Where do they come from? \\Identification of escaped globular cluster  stars}

\titlerunning{GC escapees}
\authorrunning{Xu  et~al.}

\author{Cheng Xu\inst{1,2}, Baitian Tang\inst{1,2}, Chengyuan Li\inst{1,2}, Jos\'e G. Fern\'andez-Trincado\inst{3}, Jing Zhong\inst{4},\\ 
Long Wang\inst{1,2}, Hao Tian\inst{5,6} and Yang Huang\inst{5,7}}
   
\institute{Department of Astronomy, School of Physics and Astronomy, Sun Yat-sen University, Zhuhai, Guangdong Province, China \\
            \email{tangbt@sysu.edu.cn}
\and{CSST Science Center for the Guangdong-Hong Kong-Macau Greater Bay Area, Zhuhai 519082, China}
\and{Instituto de Astronom\'ia, Universidad Cat\'olica del Norte, Av. Angamos 0610, Antofagasta, Chile}
\and{Key Laboratory for Research in Galaxies and Cosmology, Shanghai Astronomical Observatory, Chinese Academy of Sciences, 80 Nandan Road, Shanghai 200030, China}
\and{School of Astronomy and Space Science, University of Chinese Academy of Sciences, Beijing 100049, China}
\and{Institute for Frontiers in Astronomy and Astrophysics, Beijing Normal University, Beijing, 102206, China}
\and{Key Laboratory of Space Astronomy and Technology, National Astronomical Observatories, Chinese Academy of Sciences, Beijing 100101, China}}

\date{Received --, --; accepted --, --}

 

\abstract{Globular clusters (GCs), as old as our Galaxy, constantly lose their members to the field as they cross through the Milky Way (MW). These escaped GC stars (or escapees) are thought to contribute significantly to the MW halo. If a star left the host GC a long time ago, chemical finger prints (e.g., N enrichment) may reveal its origin. In this work we aim to establish dynamical connections between N-rich field stars recently identified by LAMOST and the existing MW GCs. By constructing the full action distribution in combination  with metallicity, we found 29 potential GC progenitors for 15 N-rich field stars. In particular, some of them may be related to MW accretion events. On the other hand, if a star has recently left its host GC via tidal evaporation, it still maintains the kinematic properties of the cluster. Here we identify extra-tidal candidates based on their spatial locations, proper motions (PMs), and their positions on color-magnitude diagrams (CMDs). We successfully identified more than 1600 extra-tidal candidates in the vicinity of six Gaia-Enceladus (GE)-related GCs: NGC 1851, NGC 1904, NGC 6205, NGC 6341, NGC 6779, NGC 7089. The density map of the extra-tidal candidates is confirmed to be an efficient way to find extra-tidal structures. The two possible  density peaks at opposite sides of the inner boundary is a good indicator for a long stellar stream. Among 95 extra-tidal candidates with spectroscopic radial velocities and metallicity, 54 of them are confirmed to be GC escaped stars as they share similar properties to host GCs. These extra-tidal candidates are ideal targets for follow-up spectroscopic observation as it greatly improves the scientific outcome. Once a statistically significant number of spectroscopic radial velocities and metallicities are available, the GC dynamical evolution (e.g., mass loss, rotation) can be carefully investigated.}


\maketitle
%

\section{Introduction}

In recent years, large spectroscopic surveys, such as the Apache Point Observatory Galactic Evolution Experiment (APOGEE; \citealt{2022ApJS..259...35A}), the Gaia-ESO survey (GES; \citealt{2022A&A...666A.121R}), the Large Sky Area Multi-Object Fibre Spectroscopic Telescope (LAMOST; \citealt{2012RAA....12.1197C,2012RAA....12..735D,2012RAA....12..723Z}),  Galactic Archeology with HERME (GALAH; \citealt{2021MNRAS.506..150B}), and the Radial Velocity Experiment (RAVE; \citealt{2020yCat.3283....0S}), have provided radial velocities and chemistry of a large number of stars in the MW. At the same time, the Gaia space mission \citep{2016A&A...595A...1G} has also delivered astrometry, radial velocity, and chemical information for billions or millions of stars. The combining dataset yields six-dimensional space-velocity information and multiple dimensions in chemical abundances, which not only helps us evaluate the dynamical evolution of different scale of stellar systems (e.g., star clusters, galaxies), but also promotes the development of Galactic archaeology.

With such a rich dataset and with such unprecedented detail, the MW is finally showing its true nature to us.  It  devoured nearby dwarf galaxies during its growth, such as the recent victim, the Sagittarius dwarf elliptical galaxy \citep{1994Natur.370..194I,2006ApJ...642L.137B}. There are other digested casualties, for example the Gaia-Enceladus \citep[GE,][]{2018Natur.563...85H, 2018MNRAS.478..611B} and Sequoia Galaxies \citep{2019MNRAS.488.1235M}, which are now part of the inner MW halo and thick disk. During this process, some GCs, which originally formed in dwarf galaxies, were later brought into the MW as the dwarf galaxies were accreted by the MW.

Due to the tidal evaporation and binary ejection, GC stars are continuously lost to the field \citep{2015MNRAS.451.1229A,2016MNRAS.457.2078A,2019MNRAS.489.4565K,2021A&A...645A.116K,2021A&A...647A..64F,2020MNRAS.495.2222S}. The main escape mechanism is tidal evaporation \citep[$\sim80$\%,][]{2023ApJ...946..104W}, where stars   preferentially escape  through Lagrangian 1 and Lagrangian 2 points formed by the combined potentials of  GCs and the MW. As these escaped stars (or escapees) rotate differentially around the MW, stellar streams are created \citep{2001ApJ...548L.165O,2006ApJ...643L..17G,2010A&A...522A..71J,2018MNRAS.473.2881K,2018MNRAS.481.3442M,2019NatAs...3..667I,2020Natur.583..768W}. Stellar streams do not persist forever as they eventually disperse to become part of the MW. These GC escapees are the key to answering questions related to MW-GC coevolution, for example determining the percentage of GC escapees in the MW halo and the initial mass distribution for GCs, understanding how galaxy accretion affects GC evolution, and whether these escapees also show signatures of dense star clusters, for example multiple populations (MPs; see below).

Tracing these escapees in the enormous number of MW stars is not an easy task, the astronomical community started the identification only recently using kinematic and chemical properties, thanks to the rich survey data. As tidal evaporation is the main escape mechanism, we should expect that the escapees around GCs share similar kinematic properties with their host GCs. For example, using the data from Gaia DR2 \citep{2018A&A...616A...1G} and SDSS DR14 \citep{2018ApJS..235...42A}, \citet{2020A&A...637A..98H} established associations between 151 extra-tidal stars in the neighborhood of eight GCs by estimating the dynamical likelihood between modeled stellar streams and extra-tidal stars. Following this, by utilizing the Gaia EDR3 \citep{2021A&A...649A...1G}, \citet{2021A&A...645A.116K,2022A&A...665A...8K} obtained proper motions (PMs) and constructed color-magnitude diagrams
(CMDs) to search within 1-5$r_t$ of multiple GCs, leading to the identification of numerous extra-tidal candidates. While some of these candidates exhibited tidal tails near specific GC (e.g., NGC 2808), the majority of extra-tidal candidates showed no significant external structures in the  GC's vicinity. 

On the other hand, the chemical signatures of GCs can also help the identification of GC escapees, as chemistry is more stable than kinematics. The distinct chemical signature of GCs is known as multiple populations \citep{2010A&A...516A..55C,2015AJ....149..153M,2015ApJ...808...51M,2015AJ....149..153M,2018ARA&A..56...83B,2017MNRAS.465...19T,2018ApJ...855...38T,2021ApJ...908..220T,2021ApJ...906..133L}. This phenomenon refers to the variations in light element abundances in GC members. Traditionally, GC stars with enhanced N and  Na (and sometimes He and  Al) and reduced C and  O (and sometimes Mg) are called the second population (2P), while the primordial population is called 1P. It is generally accepted that such chemical anomalies are formed mostly in dense environments, such as GCs \citep{2012A&ARv..20...50G,2018ARA&A..56...83B}. So searching for 2P-like stars is a feasible way to find the association between GC escapees and the ejecting and/or dissolving GCs. Nitrogen is an obvious targeted element as it is connected to many molecular lines in the optical and near-infrared (NIR), such as NH or CN molecules, making its abundance easier to obtain. 
A large number of N-rich field stars have been discovered through high-resolution spectral surveys (see, e.g., \citealt{2015A&A...575L..12L,2016ApJ...825..146M,2017AJ....154...94M,2017MNRAS.465..501S,2016ApJ...833..132F,2017ApJ...846L...2F,2019MNRAS.488.2864F,2019ApJ...886L...8F,2020A&A...644A..83F,2020ApJ...903L..17F,2020A&A...643L...4F,2020MNRAS.495.4113F,2021ApJ...918L..37F,2022A&A...663A.126F}). Alternately, low-resolution spectra observations can expand the search to fainter and larger sample of stars \citep{2010A&A...519A..14M,2011A&A...534A.136M,2019A&A...625A..75K}. In this quest, LAMOST DR5 offered $\sim$100 newly identified N-rich metal-poor field stars \citet{2019ApJ...871...58T,2020ApJ...891...28T}. A larger sample size of N-rich field stars increases the credibility of the drawn statistical results.

There are still debated questions related to GC escapees, for example, whether it is possible to  find the host GC for a given N-rich field star so that we can trace each escaped incident throughout the galaxy evolution or if we can find escapees in the vicinity of GCs with similar kinematic properties so that we can estimate the star loss.  We try to address these two questions in this work. This is also part of our recently proposed project ``Scrutinizing {\bf GA}laxy-{\bf ST}a{\bf R} cluster coevoluti{\bf ON} with cheom{\bf O}dyna{\bf MI}cs ({\bf GASTRONOMI}),'' which   investigates the interplay between the  MW, nearby satellite dwarf galaxies, and star clusters using photometric and spectroscopic data. This project bridges star evolution and galaxy evolution with solid foundations, which may also help us understand exotic objects in the early universe.

This paper is structured as follows. We first describe the methodology (Section \ref{sec:method}),  and then we present and discuss the possible associations between escaped N-rich field stars and their host GCs (Section \ref{sec:result-method1}). We found extra-tidal candidates of six GE GCs using PM and CMDs, and we carefully discuss them case by case, for example checking metallicities ([Fe/H]) and RVs of the extra-tidal star candidates to confirm their origin (Section \ref{sec:result-method2}). Finally, we summarize our results and give suggestions for further studies (Section \ref{sec:conclusion}).

\section{Methods}
\label{sec:method}

\subsection{Method \uppercase \expandafter{\romannumeral1}: Integrals of motion}
\label{sec:method1}

To identify the host GC for a given N-rich field star, we calculate their integral of motion (IoM), and find if they match in this parameter space. There are two assumptions here. First, the MW potential is axisymmetric and steady-state because orbits in such a potential have a conservative IoM \citep{2008gady.book.....B}. The second assumption is that N-rich field stars escaped GCs with evaporation, so that they share  similar kinematic properties. The latter assumption is  true for $\sim80$\% of escapees, according to \cite{2023ApJ...946..104W}.

\subsubsection{Data}
In this work we try to establish the connection between host GC and N-rich field stars from \citet{2019ApJ...871...58T,2020ApJ...891...28T}. In order to simulate their orbits and find their IoMs, we need to first compile a catalog of kinematic information of existing GCs and N-rich field star sample. Metallicity is also used to constrain their consistency.
The N-rich field stars are metal-poor red giants with metallicity ([Fe/H]) in the range [-1.8, -1.0] and  effective temperature (T$_{eff}$) in the range [4000K, 5500K]. We retrieved their stellar parameters and radial velocity (RV) values from \citet{2020ApJ...891...28T}, while their sky positions and PMs come from Gaia EDR3 \citep{2021A&A...649A...1G}. We used the Bayesian spectrophotometric distances with no assumptions about the underlying populations \citep{2015AJ....150....4C} presented in \citet{2020ApJ...891...28T}.

Next, we retrieved PMs, distances, radial velocities, and the associated uncertainties of 158 GCs from \citet{2021MNRAS.505.5978V} and \citet{2021MNRAS.505.5957B}.  We obtained their metallicity from \citet{1996AJ....112.1487H} (2010 version), but there are still a few GCs (13) without literature metallicity values. As a result, a total of 145 GCs possess complete 6D kinematic parameters along with metallicity information.

\subsubsection{Orbital model}
We integrated orbits and computed IoMs for  the stars and for the clusters using a Python package for galactic dynamics: \textsc{galpy} \citep{2015ApJS..216...29B}. We adopted a steady-state and axisymmetric MW potential: MWPotential2014. The Galactic coordinates of the Sun are (8.27, 0, 0) kpc. The circular velocity at the solar radius is $V_C$(R$_0$) = 220 km$\cdot$s$^{-1}$ and the
peculiar velocity of the Sun is [11.1, 12.24, 7.25] km$\cdot$s$^{-1}$ \citep{2010MNRAS.403.1829S}. Alternative choices of circular velocity and solar peculiar velocities are also tested (e.g., \citealt{2015MNRAS.449..162H,2016MNRAS.463.2623H,2023ApJ...946...73Z}), and  the  results hold very well. To calculate the orbital energy E, the vertical component of the angular momentum $L_{z}$, the radial actions $J_{r}$, and the vertical actions $J_{z}$, we used the St\"ackel fudge approximation in \textsc{galpy} \citep{2012MNRAS.426.1324B}. We set the backward integration time to 5 Gyr, with a time step of 5 Myr.

To estimate the uncertainties of the actions for each object (N-rich field star and GC), we ran \textsc{galpy} with random kinematic parameters drawn from Gaussian distributions (expected values equal to values presented above, and standard deviations equal to their uncertainties). 
Given that PMRA and PMDEC measured by Gaia are correlated, their Gaussian distributions are also correlated.
Then we acquired the Gaussian-liked distributions of actions for each object, which were used for the following comparison.

To establish the connection between N-rich field stars and GCs, we assumed they are consistent in actions. In addition, stars escaped from GCs should also share the same metallicity ([Fe/H]) with the GC progenitor. As done in \citet{2019A&A...624L...9S}, we set the metallicity uncertainties to 0.3 dex for individual stars, which is reasonable for measurements obtained from low-resolution spectra. Finally, we calculated the matching probability of each pair of N-rich field stars and GCs, with  4D Gaussian-like distributions $(J_r, J_z, L_z, [Fe/H])^T$  \citep[also see][]{2020A&A...637A..98H}. During this process, the correlations between actions are also taken into account.

\begin{table*}
\caption{Kinematic parameters of six GCs studied in this work.}              
\label{Tabel:GC parameters}      
\centering                                      
\begin{tabular}{cccccccccc}         
\hline\hline
\\[-5pt]

Name & RA & DEC & $r_t^a$ &  $r_J^b$& $\mu_\alpha cos\delta$ & $\mu_\delta$ & $\sigma_{\mu_\alpha}$ & $\sigma_{\mu_\delta}$ & $\rho$\\

& [deg] & [deg] & [$'$] & [$'$] & [mas yr$^{-1}$] & [mas yr$^{-1}$] & [mas yr$^{-1}$] & [mas yr$^{-1}$]\\

\hline\\[-8pt] 
NGC 1851 & 78.528 & $-40.047$ & 6.52 & 36.56 & 2.13 & $-0.62$ & 0.38 & 0.46 & $-0.04$\\ 
NGC 1904 & 81.046 & $-24.524$ & 8.02 & 17.26 & 2.47 & $-1.59$ & 0.37 & 0.50 & 0.02\\ 
NGC 6205 & 250.422 & 36.460 & 21.01     & 62.2 & $-3.12$        & $-2.54$ & 0.32 & 0.4 & 0.17\\ 
NGC 6341 & 259.281 & 43.136 & 12.44 & 48.88 & $-4.93$ & $-0.66$ & 0.4 & 0.42 & 0.18\\ 
NGC 6779 & 289.148 & 30.183     & 10.55 & 31.93 & $-2$   & 1.63 & 0.38 & 0.39 & $-0.01$\\ 
NGC 7089 & 323.363 & $-0.823$ & 12.45   & 32.78 & 3.58 & $-2.19$ & 0.6 & 0.38 & $-0.11$\\ 

\hline                                             
\end{tabular}

\tablefoot{$a$:\citet{1996AJ....112.1487H}, 2010 edition;\ $b$: \cite{2021MNRAS.505.5978V, 2021MNRAS.505.5957B};\ $\rho$: Correlation between two components of PM.}
\end{table*}
 %

\begin{table*}
\caption{Parameters of the GC used to generate the isochrones and ZAHBs.}              
\label{Tabel:GC iso}      
\centering                                      
\begin{tabular}{ccccccc}         
\hline\hline
\\[-5pt]

Name & Age$^1$ & [Fe/H]$^2$ & Distance$^3$ & [$\alpha$/Fe] & E(B-V)$^2$ & Initial Mass on HB (IMHB)\\

& [Gyr] & [dex] & [Kpc] & & [mag] & [M$_\odot$]\\

\hline\\[-8pt] 
NGC 1851 & 9.98 $\pm$ 0.5       & $-1.18$       & 11.95 & 0.1 & 0.02 & 0.8\\ 
NGC 1904 & 11.14 $\pm$ 0.5 & $-1.6$ & 13.08 & 0.2 & 0.01 & 0.7\\ 
NGC 6205 & 11.65 $\pm$ 0.5 & $-1.53$ & 7.42 & 0.3 & 0.02 & 0.7\\ 
NGC 6341 & 13.18 $\pm$ 0.5 & $-2.31$ & 8.5 & 0.3 & 0.02 & 0.9\\ 
NGC 6779 & 13.7 $\pm$ 0.5       & $-1.98$       & 10.43 & 0.2 & 0.26 & 0.75\\ 
NGC 7089 & 11.78 $\pm$ 0.5 & $-1.65$ & 11.69 & 0.2 & 0.06 & 0.7\\ 

\hline                                             
\end{tabular}

\tablefoot{$1$: \citet{2010MNRAS.404.1203F};\ $2$: \citet{1996AJ....112.1487H}, 2010 edition;\ $3$: \cite{2021MNRAS.505.5957B}. The [$\alpha$/Fe] and IMHB are obtained by fitting the model. The model requires [$\alpha$/Fe] in the range 0-0.3 and IMHB in the range 0.7-0.9.}
\end{table*}

\subsection{Method \uppercase \expandafter{\romannumeral2}: Stars in globular cluster's vicinity}
\label{sec:method2}

If a star is evaporated out of a GC, and does not leave its vicinity, the kinematic properties of this star should be consistent with the host GC. The stable stellar chemical evolution also provides another constraint in its magnitude and color. Here we try to identify escaped stars in the vicinity of six GE-related GCs with PM and CMD, which should be similar to their host GCs. We pick GCs related to GE dwarf galaxies first because some of the N-rich field stars may originate from GE \citep{2021ApJ...913...23Y} and second because these GCs are brought inside our MW during the accretion, which may experience stronger tidal evaporation, and thus more escaped stars. The six GCs are NGC 1851, NGC 1904, NGC 6205, NGC 6341, NGC 6779, NGC 7089, which are thought to be related to GE, according to \citet{2019A&A...630L...4M}, \citet{2022ApJ...935..109L} and \citet{2022ApJ...926..107M}. The basic parameters of these six GCs are shown in Table \ref{Tabel:GC parameters}.

\begin{figure*}[htbp]
    \centering
    \includegraphics[width=0.85\textwidth]{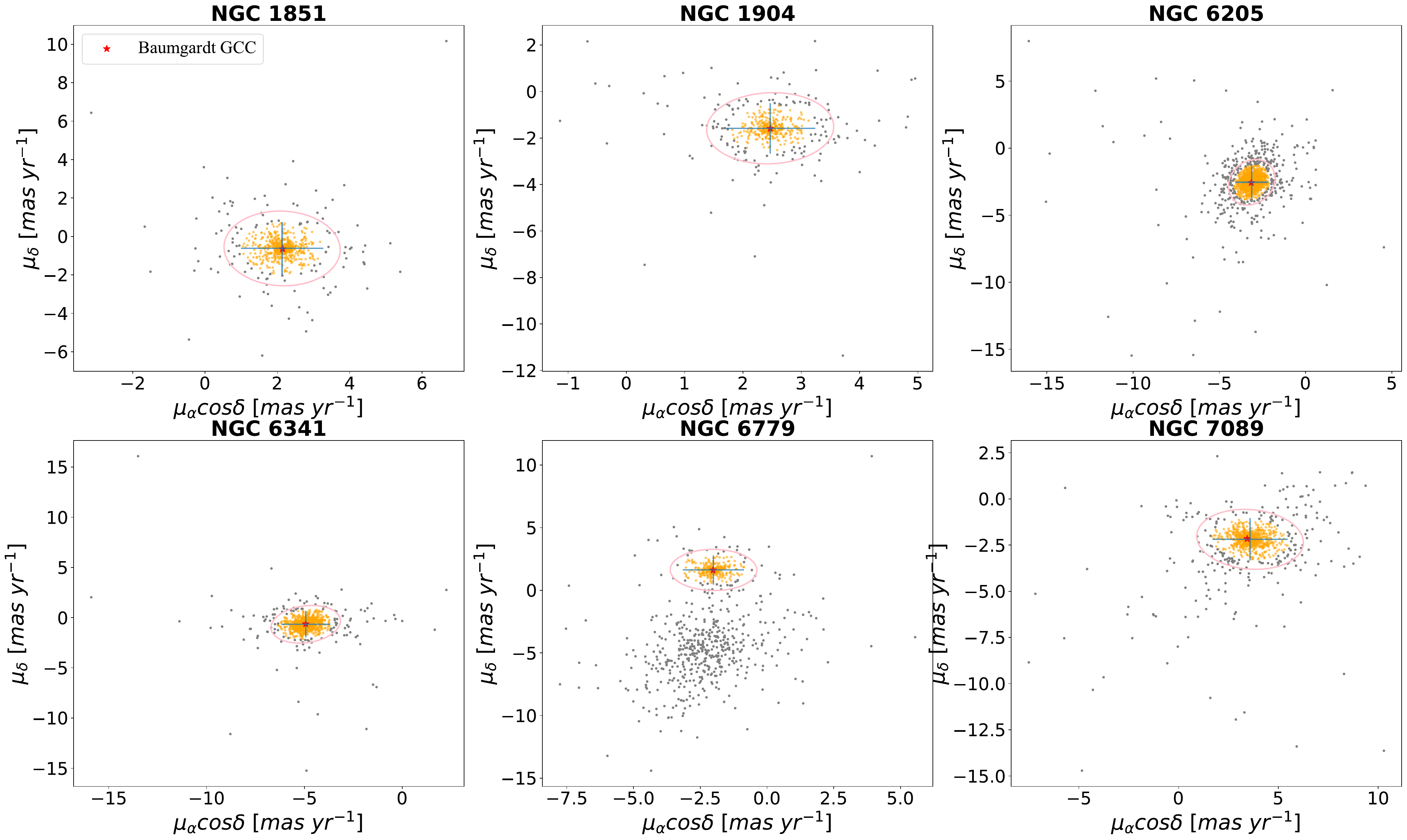}
    \caption{PM distribution with its ±3$\sigma$ regions (blue lines, including 99.7\% of the sample stars) of each cluster, determined as the GMM fit of the cluster stars within 1 $r_t$ (orange dot), as well as the field population (gray dots). The red star in each panel represents the mean PM of clusters obtained by \citet{2021MNRAS.505.5957B}.}
    \label{fig:GMM PM}%
\end{figure*}

\subsubsection{Data quality selection}

To identify escaped stars by PM and CMD with high confidence, we retrieved high-quality astrometric and photometric data from Gaia EDR3. We adopted the following selection criteria recommended by Gaia \citep{2016A&A...595A...1G, 2021A&A...649A...1G}:
\begin{enumerate}[(1)] 
\item $\texttt{RUWE \textless \ 1.4}$ 

RUWE is the re-normalized unit weight error. A small value means a good  fit by a single star model.

\item $\texttt{ASTROMETRIC\_EXCESS\_NOISE\_SIG \textless \ 2}$

Excess noise refers to the extra noise in each observation that causes the residual scatter in the astrometric solution. If $\texttt{ASTROMETRIC\_EXCESS\_NOISE\_SIG}$ is greater than two, then this excess noise cannot be ignored.

\item $\texttt{ASTROMETRIC\_GOF\_AL \textless \ 3}$

This parameter represents the goodness of fit between the astrometric model and the observed data. A high  value indicates a poor fit.

\item $\texttt{VISIBILITY\_PERIODS\_USED \textgreater \ 10}$

This parameter represents the number of visibility periods used in the astrometric solution. A high number of visibility periods is a better indicator of an astrometrically well-observed source.\footnote{https://gea.esac.esa.int/archive/documentation/GEDR3/}

\item $\texttt{0.001+0.039(BP-RP) \textless \ log10 excess\_flux\ \textless \ }$ $\texttt{0.12+0.039(BP-RP)}$

This parameter primarily constrains whether the sources have consistent fluxes. The sources that are outside of this range may be  extended sources.

\item $\texttt{Plx \textless \ 0.5}$

Due to the large distance of the selected GCs, it becomes very difficult to estimate the distance via the Gaia parallax. However, we can exclude foreground stars with large parallax (Plx > 0.5).

\end{enumerate}
For each GC, we applied these criteria to stars within five times the  King tidal radius ($r_t$). 
The escaped stars are defined to satisfy these criteria: 1) in the range  1-5$r_t$; 2) having similar PM to stars inside 1 $r_t$; 3) on the same CMD outlined by stars inside 1 $r_t$. Such stars are traditionally called ``extra-tidal candidates'' in the literature. We show the $r_t$ and the Jacobi radius ($r_J$)  of six GCs in Table \ref{Tabel:GC parameters}. The stars beyond the $r_J$ are completely out of the gravitational potential of the cluster. However, the stars that are between the $r_t$ and the $r_J$ may be loosely bound to the cluster (see, e.g., \citealt{2000MNRAS.318..753F, 2010MNRAS.401.1832B, 2010MNRAS.407.2241K}).

\subsubsection{PM selection}

Before identifying extra-tidal candidates via PM, we need to estimate the PM distribution of stars inside $r_t$. Here we utilized the Python package \textbf{Scikit-Learn} \citep{2011JMLR...12.2825P} to build a Gaussian mixture model (GMM) with two different distributions, one for the cluster stars with a smaller dispersion in PM and another for field stars with a larger dispersion. We set up a full covariance to account for asymmetric distribution. We obtained the mean, standard deviation, and correlation coefficient of the two best-fitting   Gaussian distributions, representing the average PM and corresponding dispersion of cluster stars and field stars, respectively.

The best-fitting  results are shown in Fig. \ref{fig:GMM PM}, with mean PM and dispersion listed in Table \ref{Tabel:GC parameters}. Among our investigated GCs, the mean PMs of the cluster stars are consistent with the observed cluster PMs from \citet{2021MNRAS.505.5957B} (red stars). We also tested multiple-Gaussian models, for example  with  three or four components \citep{2022A&A...665A...8K}, but we did not find a significant difference compared to the case with two components.

We found that the PM distribution of cluster stars was not a simple symmetric distribution, indicating that PMs in RA and DEC are correlated. Therefore, we could not simply adopt the observed cluster PM and its dispersion. Instead, we needed to estimate the PM distribution of the cluster stars, and extrapolate to the surrounding area of the cluster (within 1-5$r_t$). To this end, we estimated their probabilities with the method and formulas presented in Section 4 of \citet{2015A&A...584A..59S}. In their formulas the mean and dispersion of PMs for both cluster and field stars were obtained from the averages and standard deviations calculated by GMM. The correlation between the two PM components was also determined by the cluster stars. This method may identify high-probability extra-tidal candidates, although with a potential drawback of underestimating their true sample size \citep{2022A&A...665A...8K}.

We applied this method to six clusters and obtained a batch of extra-tidal candidates (based on PMs) for each one. Finally,  for further analysis we only kept stars whose probability of having PMs compatible with the cluster PM  is greater than 80\%.

\begin{figure*}[htbp]
    \centering
    \includegraphics[width=1.\textwidth, height=520pt]{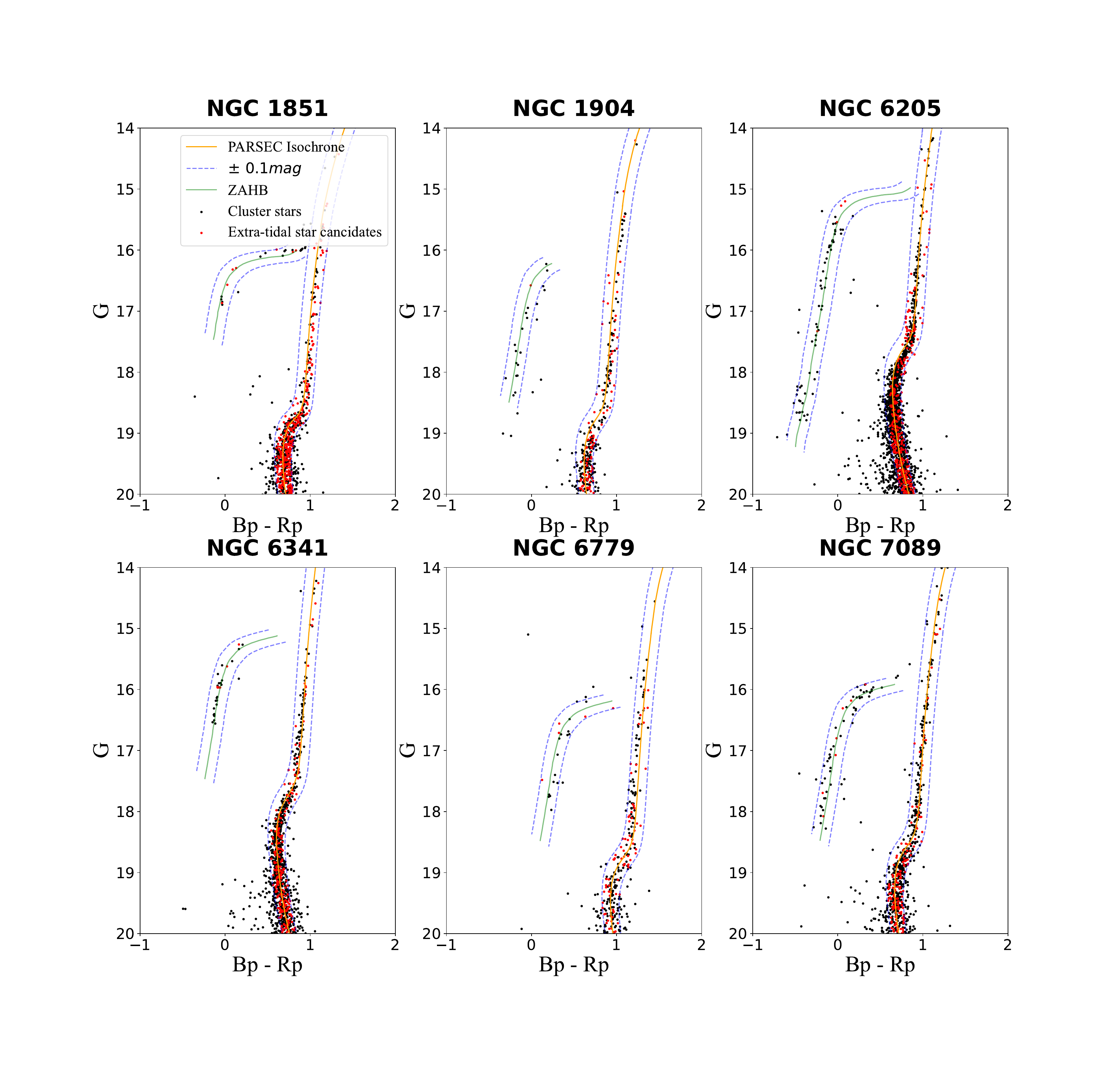}
    \caption{Gaia EDR3 de-reddened CMD of the six GCs, with the PARSEC isochrones (solid orange lines). The ZAHB is drawn with green lines (see text for details), the red dots are the selected extra-tidal candidates, the black dots are the stars inside 1$r_t$, and the blue dashed line is the boundary for selecting extra-tidal candidates.}
    \label{fig:CMD}%
\end{figure*}

\subsubsection{Color-magnitude diagram selection}

After obtaining preliminary samples by building the GMM in the PM space, we further eliminated outliers through their CMD. To construct the CMD we used Gaia EDR3 photometry. In this work we use isochrones to define outliers. We complement reddened PARSEC isochrones \citep{2012MNRAS.427..127B,2017ApJ...835...77M} with horizontal branch models from \citet{2012A&A...547A...5V}. The model requires [$\alpha$/Fe] in the range 0-0.3 and IMHB\footnote{Initial mass on HB} in the range 0.7-0.9. We sampled IMHB values at intervals of 0.05 between 0.7 and 0.9, and [$\alpha$/Fe] values at intervals of 0.05 between 0 and 0.3. Based on [Fe/H] and each set of [$\alpha$/Fe], we calculate the corresponding metallicity Z and input it into the model to generate a stellar evolutionary track. We then choose the track with the best-fitting results and record the parameters. We collected the basic parameters ([Fe/H], age, distance, [$\alpha$/Fe], IMHB) of six GCs from literature, and list  them in Table \ref{Tabel:GC iso}.

In order to  minimize the contamination of foreground stars, we imposed constraints on parallax. Since all six clusters in this work are located at   relatively distant positions ($>5$ kpc), where the parallax provided by Gaia has a relatively large uncertainty.  We ultimately limited the parallax to be less than 0.5$^{''}$ (D > 2 kpc) in order to exclude foreground stars with larger parallax values.

Figure \ref{fig:CMD} displays the cluster CMDs, PARSEC isochrones (orange lines), and zero age horizontal branch plots  (ZAHB, yellow lines). Cluster stars (i.e., within 1$r_t$) with PMs that match the cluster's PM (black points) are plotted to verify the robustness of isochrones and tracks. We defined regions of $\pm$ 0.1 mag around isochrones and tracks (blue lines) to select extra-tidal candidates (red dots). Due to observational errors, fainter stars with G $>$ 20 mag were excluded.

\begin{figure}[htbp]
    \centering
    \includegraphics[width=1\columnwidth]{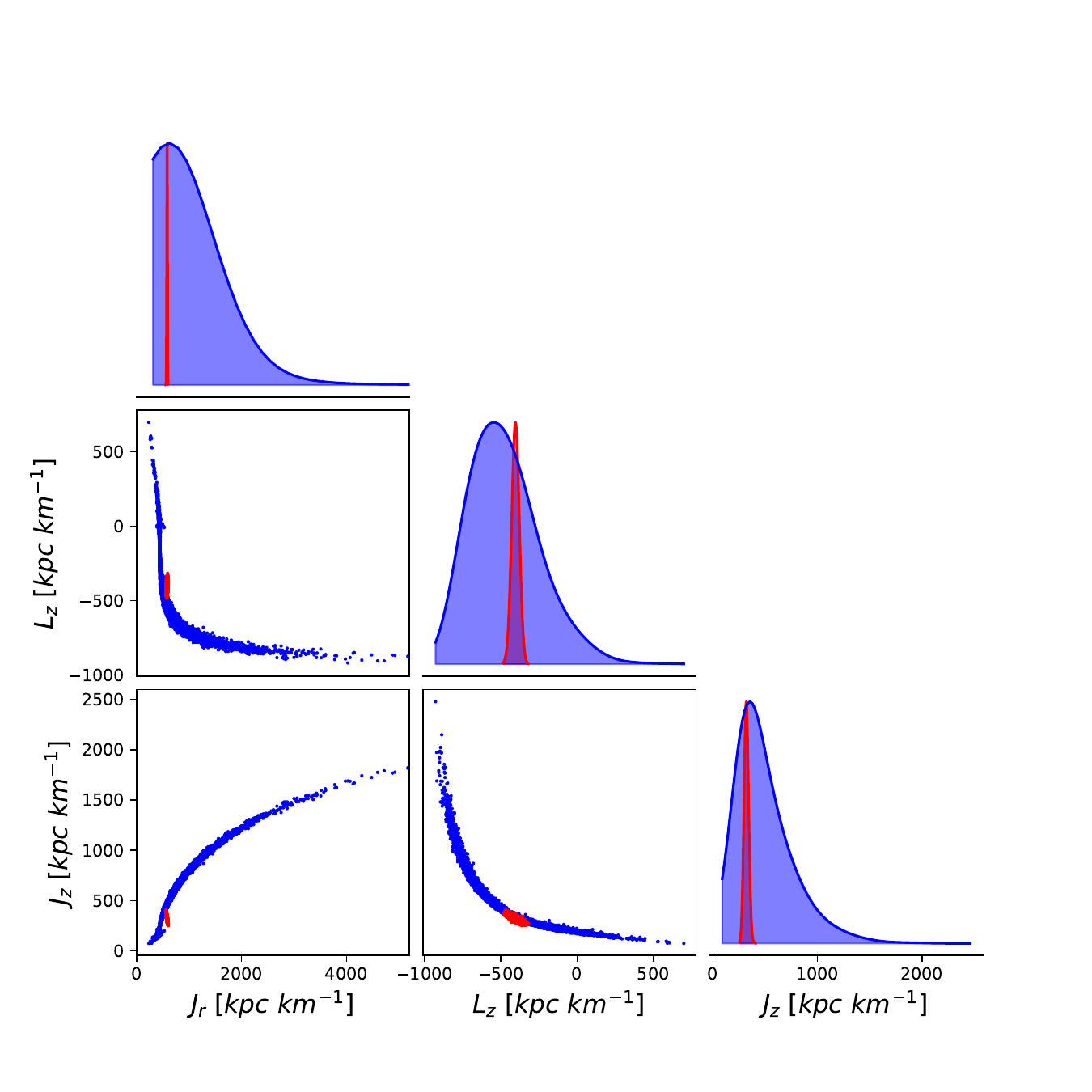}
    \caption{Monte Carlo sample distribution of the motion integral of NGC 6779 (red) and star Gaia EDR3 1189399628220805760. The metallicity difference and $p$ of this pair are 0.27 and 0.87, respectively. The marginal two-dimensional and one-dimensional kernel density estimates are represented by the frame and top histogram, respectively. The histograms are normalized to their maximum values for ease of presentation.}
    \label{fig:example_action}%
\end{figure}

\section{Results and discussion}
\label{sec:result}

\subsection{Associating N-rich field stars with globular clusters}
\label{sec:result-method1}

In this work we used IoMs to investigate the chemical and dynamical connections between N-rich field stars found in \citet{2020ApJ...891...28T} and existing GCs. Similar studies were recently conducted by \citet{2019A&A...624L...9S} and \citet{2020A&A...637A..98H}, where CN-strong field stars identified from SDSS were explored. We note that  the chemically peculiar stars in these three studies are all identified using the CN and CH molecular lines around 4000 \AA. These ``CN-strong stars'' here are called ``CN-strong, CH-normal stars'' to avoid confusion with other CN-strong stars caused by carbon enhancement. Furthermore, we   observed a substantial portion of stars in \citet{2020ApJ...891...28T} with high-resolution spectra, and they are confirmed to be N-rich \citep[][Tang et al. in prep.]{2021ApJ...913...23Y}.


We compared the action distributions of N-rich field stars and GCs to search for possible associations. Significant overlap may be found in individual action distribution, as shown in Fig. \ref{fig:example_action}, but the overlap may be trivial in multidimensional distribution. Therefore, we use a probability ($p$) that considers the full distribution of actions in this work. To calculate this probability, we established a four-dimensional \textbf{($J_r, J_z, L_z, [Fe/H]$)} probability density function, $\psi(\bm{v})$, of the different vector $\bm{v}$, for each pair of N-rich stars and GCs \citep{2020A&A...637A..98H}: 
\begin{equation}
    p = \int\limits_{\psi(\bm{v}^{ '}) < \psi(\bm{0})}\psi(\bm{v}^{ '})d\bm{v}^{ '}
.\end{equation}

\begin{figure}[htbp]
    \centering
    \includegraphics[width=1\columnwidth]{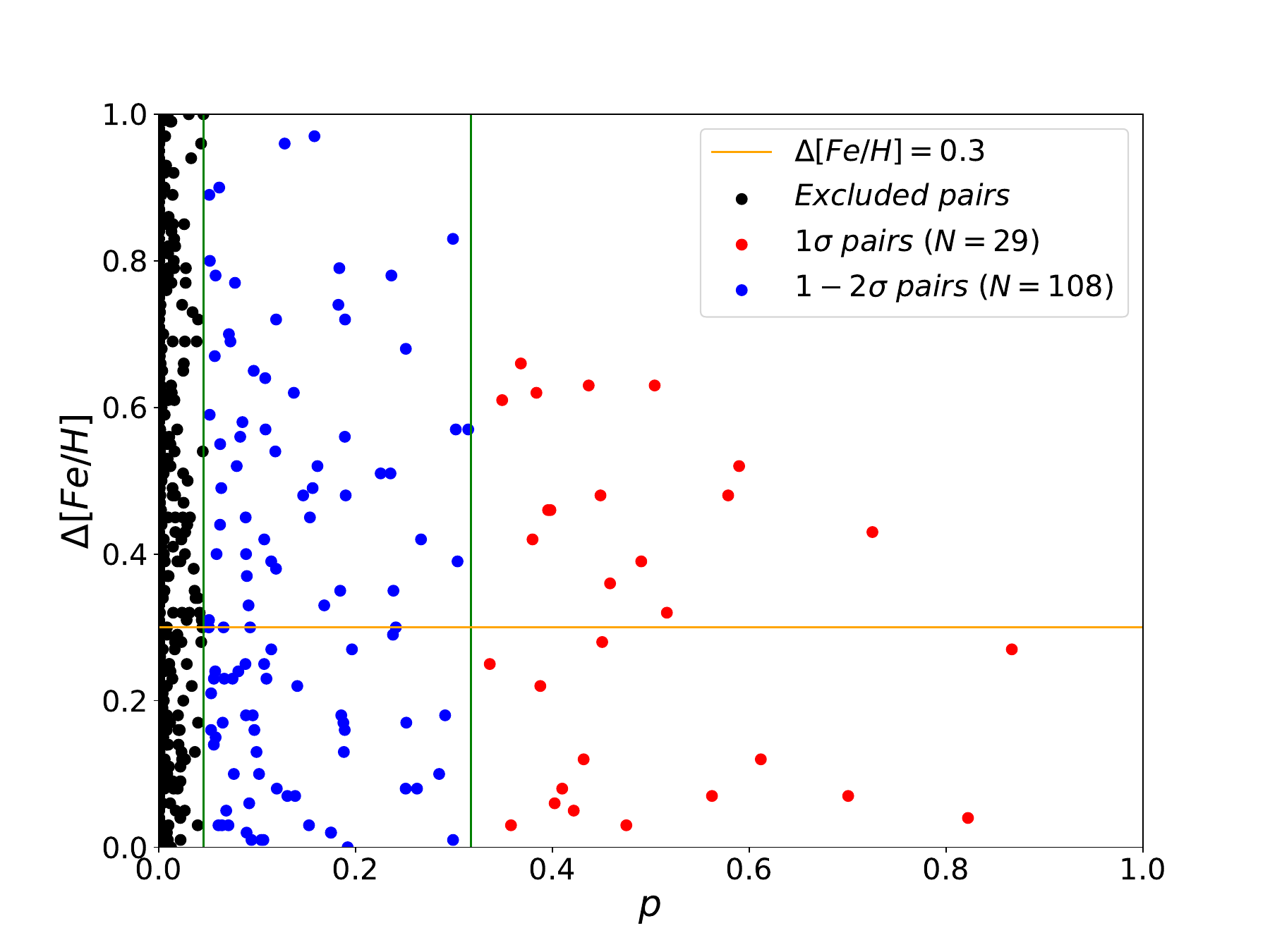}
    \caption{For each GC--star pair, confidence level p for the pair against absolute difference in metallicity. The green lines indicate the 0.05 (1$\sigma$) and 0.32 (2$\sigma$) confidence levels. The yellow line represents an absolute difference in metallicity of 0.3. The blue dots represent pairs with a confidence between 0.05 and 0.32, while the orange dots represent pairs with a confidence greater than 0.32. The most probable pair is represented by the red dots in the bottom right region.}
    \label{fig:confidence}%
\end{figure}

We obtained matching probability ($p$)
for 14,500 pairs of N-rich field stars an GCs. Figure \ref{fig:confidence} shows the distribution of $p$ values versus metallicity. We categorized the results into three groups, where 1$\sigma$ corresponds to p > 0.32, indicating that N-rich stars are closely associated with the four-dimensional distribution of the star cluster. The 1-2$\sigma$ range, with p values between 0.05 and 0.32, suggests a potential association that cannot be entirely ruled out. Ultimately, we exclusively considered pairs within the 1$\sigma$ range as the most likely associated pairs. We note that a $p$ value close to 1 does not necessarily mean that the N-rich field star escaped from the corresponding GC, but rather that we cannot simply rule out the possibility of their association. Since significant kinematic errors (such as PM, radial velocity, and especially distance) can greatly affect the reliability of the results in orbit integration. In such cases, the association between the star and the GC will be determined almost solely by the distribution of metallicity. Therefore, we adopted the full distribution of actions rather than marginalized distributions for each component.

We found 137 pairs with p values greater than 0.05, of which 29 pairs had p values greater than 0.32. Among these 29 pairs, only 16 pairs had metallicities consistent within 0.3 dex. We  show all results with p greater than 0.32 in Table \ref{Tabel: GC-star pair}. Interestingly, the majority of GCs in these pairs (17 out of 29 pairs) might be associated with accretion events (e.g., Gaia-Enceladus, Sagittarius dwarf, Helmi Streams, and Sequoia galaxy.) These pairs are labeled  by the last column of Table \ref{Tabel: GC-star pair}. This is consistent with the speculation that the MW accretion event may stimulate the escape: GC members in dissolved dwarf galaxies receive a large amount of  kinematic energy from the MW during the accretion \citep{2000gaun.book.....S}, leading to a higher possibility of leaving the cluster. However, precise chemical abundances are required to establish a firm connection between each star--GC pair.

A significant portion of the N-rich field stars do not seem to be chemodynamically associated with any known GC, which was also seen in the works of \cite{2019A&A...624L...9S} and 
 \cite{2020A&A...637A..98H}. We found that nearly 58\% of N-rich stars could not be connected with any existing GC, which is slightly higher than the ratio reported by (\citealt{2020A&A...637A..98H}, 38\%) and comparable to the ratio reported  by (\citealt{2019A&A...624L...9S}, 50\% ). This might be due to more accurate distance measurements in this work, where smaller uncertainties lead to a less allowable tolerance for errors, and thus lower matching probabilities. 
 Under the hypothesis that GCs are the only birth scenario for these chemically peculiar stars, these authors suggest the following reasons: 1) these stars may have escaped from GCs that have completely dissolved; 2) these stars may actually come from one of the known existing GCs, but do not retain the cluster's orbital signature. Possible explanations for the latter scenario include  high ejection velocities involving three-body interactions and early escape from the host GC, but the dramatic changes in Galactic potential energy render the steady-state potential assumption invalid.

\begin{table*}[h]
\caption{N-rich stars and potential GC origin from Method \uppercase\expandafter{\romannumeral1}. The columns list the Gaia EDR3 star ID and cluster name, the confidence measured from full action distribution combining the metallicity, and the merger events. Entries in bold are the five most probable associations}              
\label{Tabel: GC-star pair}      
\centering                                      
\begin{tabular}{ccccc}         
\hline\hline
\\[-5pt]

Gaia EDR3 ID & GC & Confidence ($p$) & m.e.$^1$ \\

\hline\\[-8pt]
4565255983145857920 & NGC 5139 & 0.41 & GE/Seq\\
\hline
4553384792324044928 & NGC 5897 & 0.44 & GE\\
4553384792324044928 & NGC 6355 & 0.42 & ...\\
4553384792324044928 & NGC 6652 & 0.35 & ...\\
4553384792324044928 & NGC 6809 & 0.58 & ...\\
\hline
4526616017814192768 & NGC 6544 & 0.38 & ...\\
\hline
4457311841406602752 & NGC 5946 & 0.36 & ...\\
4457311841406602752 & NGC 6284 & 0.40 & GE\\
\hline
4452741961844020992 & NGC 6712 & 0.56 & ...\\
\hline
3903971791408436352 & NGC 2419 & 0.37 & Sag\\
3903971791408436352 & NGC 5272 & 0.48 & H99\\
3903971791408436352 & Pal 5    & 0.43 & H99?\\
3903971791408436352 & NGC 6235 & 0.34 & GE\\
\textbf{3903971791408436352} & \textbf{NGC 6715} & \textbf{0.82} & \textbf{Sag}\\
3903971791408436352 & Ter 8    & 0.50 & Sag\\
\hline
3628877246314114176 & Pal 11   & 0.44 & ...\\
\hline
\textbf{3628603811516372864} & \textbf{NGC 6235} & \textbf{0.61} & \textbf{GE}\\
\hline
2485902851604869504 & NGC 4147 & 0.57 & GE\\
\hline
2105942793032574592 & NGC 6352 & 0.36 & ...\\
\hline
2095506851436919296 & NGC 6121 & 0.52 & ...\\
\hline
1542187382923726848 & NGC 5139 & 0.40 & GE/Seq\\
\hline
1314804426927128960 & NGC 6121 & 0.46 & ...\\
\hline
1189399628220805760 & NGC 288  & 0.48 & GE\\
1189399628220805760 & NGC 6325 & 0.38 & ...\\
\textbf{1189399628220805760} & \textbf{NGC 6779} & \textbf{0.87} & \textbf{G-E}\\
\hline
915304504835769216  & NGC 5904 & 0.45 & GE/H99\\
\textbf{915304504835769216}  & \textbf{NGC 6584} & \textbf{0.70} & \textbf{...}\\
\hline
912107709138952192  & NGC 2419 & 0.39 & Sag\\
\textbf{912107709138952192}  & \textbf{Ter 8}    & \textbf{0.74} & \textbf{Sag}\\

\hline                                             
\end{tabular}

\tablefoot{$1$: Merger events (m.e.) that might be associated with the host GCs \citep{2019A&A...630L...4M}. GE stands for Gaia-Enceladus \citep{2018Natur.563...85H}; Sag stands for Sagittarius dwarf \citep{1994Natur.370..194I}; H99 stands for Helmi Streams \citep{1999Natur.402...53H}; Seq stands for Sequoia galaxy \citep{2019MNRAS.488.1235M}.}
\end{table*}

\subsection{Extra-tidal candidates around clusters}
\label{sec:result-method2}

\begin{figure*}[htbp]
    \centering
    \includegraphics[width=0.95\textwidth, height=22.85cm]{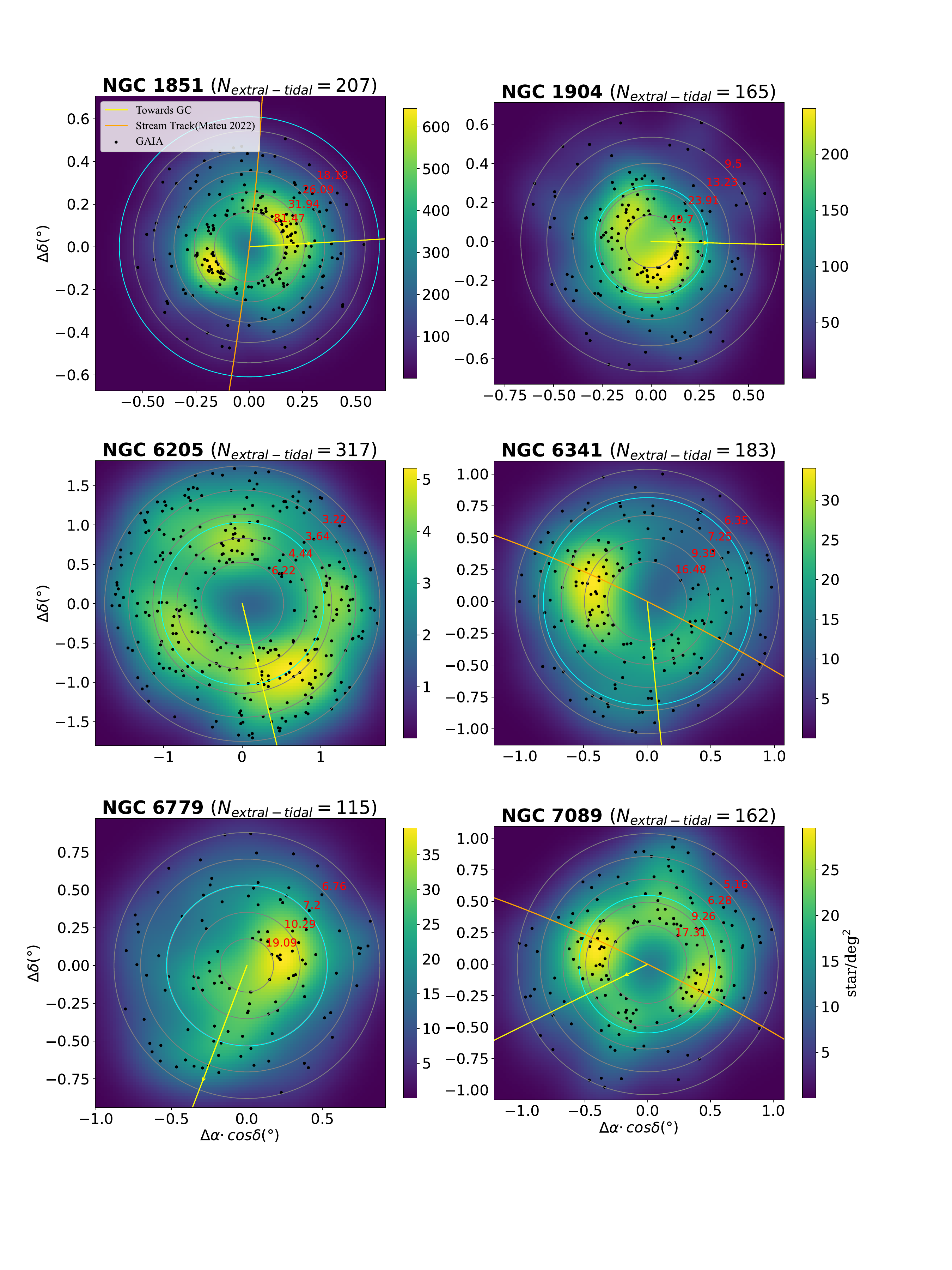}
    \caption{Density map of extra-tidal candidates for six GCs. The spatial density of extra-tidal candidates (black dots) are indicated by their colors. The yellow line represents the direction pointing toward the Galactic center, while the orange line represents the stream track provided by \citet{2023MNRAS.520.5225M} with a parallel shift to make sure that the track goes through the cluster center. The gray circles represent the selected range, where the innermost gray circle represents the tidal radius, the outermost gray circle represents five times the tidal radius, and the blue circle represents the Jacobi radius. The average radial Poisson noise is calculated in four regions, and the values are indicated by the red text.}
    \label{fig:density map}
\end{figure*}

Under the assumption of similar kinematic properties between the host GC and escaped stars in its vicinity, we explored extra-tidal candidates for six GE-related GCs. 
Following the selection method described in Section \ref{sec:method2},
we found 444, 165, 427, 286, 115, and  216 extra-tidal candidates in the vicinity of NGC 1851, NGC 1904, NGC 6205, NGC 6341, NGC 6779, and NGC 7089, respectively. The majority of these extra-tidal candidates are RGB stars or MS stars. HB extra-tidal candidates are rare: NGC 1851 (7 stars), NGC 1904 (1 star), NGC 6205 (4 stars), NGC 6341 (5 stars), NGC 6779 (5 stars), and NGC 7089 (7 stars). To investigate the possible extra-tidal features around GCs, we constructed a density map of the identified candidates (see  Fig. \ref{fig:density map}). Two density peaks (i.e., NGC 1851) located in the opposite directions of the inner boundary are a strong sign of stellar streams: stars were preferentially lost from these two points, and formed stellar streams due to differential Galactic rotation.

To further confirm if these extra-tidal candidates share similar RVs and metallicities with their host GCs, we cross-matched the candidates with large spectroscopic surveys. Only a small number of stars were observed by APOGEE DR17 \citep{2022ApJS..259...35A} and SEGUE DR12 \citep{2009AJ....137.4377Y,2011AJ....142...72E}, and they are located in the vicinity of four GCs (NGC 1851, NGC 6205, NGC 6341, and NGC 7089). We define an RV match if there is an overlap between the uncertainty range of the extra-tidal candidate and the cluster mean RV$\pm$dispersion. A metallicity match is defined similarly. In this section the cluster's RV and dispersion are provided by \citet{2018MNRAS.478.1520B};\footnote{https://people.smp.uq.edu.au/HolgerBaumgardt/globular/} the cluster's mean metallicity is listed in Table \ref{Tabel:GC iso}, with an uncertainty of 0.3 dex. 

\begin{table*}
\caption{Identified extra-tidal stars from Method \uppercase \expandafter{\romannumeral2}.}
\label{Table: extra-tidal stars}
\centering
\resizebox{\textwidth}{!}{%
\begin{tabular}{@{\extracolsep{18pt}}rccccrcccc}
\hline\hline
\\[-5pt]
Gaia EDR3 ID & GC & RV & [Fe/H] & Spectral source  & Gaia EDR3 ID & GC & RV & [Fe/H] & Spectral source\\[8pt]
& & [km$\cdot$s$^{-1}$] & dex & & & [km$\cdot$s$^{-1}$] & dex\\

\cline{1-5}\cline{6-10}\\[-4pt]

4819295189806128512 & NGC 1851 & 318.21 & -1.10 & APOGEE DR17 & 1360418074825605888 & NGC 6341 & -124.24 & NaN & SEGUE DR12\\
4819281785211829248 & NGC 1851 & 317.53 & -1.44 & APOGEE DR17 & 1360395706636272384 & NGC 6341 & -131.71 & NaN & SEGUE DR12\\
4819292681545239552 & NGC 1851 & 323.27 & -1.17 & APOGEE DR17 & 1360192198202792960 & NGC 6341 & -117.86 & -2.10 & SEGUE DR12\\
4819195546567853440 & NGC 1851 & 319.64 & -1.15 & APOGEE DR17 & 1360457352307630080 & NGC 6341 & -124.80 & -2.04 & SEGUE DR12\\
4819202040557645568 & NGC 1851 & 321.22 & -0.96 & APOGEE DR17 & 1360197970638757120 & NGC 6341 & -118.26 & -2.55 & SEGUE DR12\\
1328127243689705728 & NGC 6205 & -245.26 & -1.59 & SEGUE DR12 & 1360224668155504896 & NGC 6341 & -114.24 & -2.41 & SEGUE DR12\\
1331385646389205632 & NGC 6205 & -239.46 & -1.68 & SEGUE DR12 & 1360206968594956032 & NGC 6341 & -113.24 & -2.39 & SEGUE DR12\\
1328116244270330368 & NGC 6205 & -234.83 & -1.58 & SEGUE DR12 & 1360219720353705344 & NGC 6341 & -116.13 & -2.19 & SEGUE DR12\\
1328118718171299328 & NGC 6205 & -241.40 & -1.70 & SEGUE DR12 & 1360211508375698688 & NGC 6341 & -122.29 & -2.34 & SEGUE DR12\\
1328123391094186752 & NGC 6205 & -249.97 & -1.63 & SEGUE DR12 & 1360413023944914816 & NGC 6341 & -121.24 & -1.99 & SEGUE DR12\\
1328126002435827840 & NGC 6205 & -246.05 & -1.56 & SEGUE DR12 & 1360186009153430144 & NGC 6341 & -122.56 & -1.99 & SEGUE DR12\\
1328125006003407360 & NGC 6205 & -252.85 & -1.82 & SEGUE DR12 & 1354437216903606144 & NGC 6341 & -117.40 & -2.29 & SEGUE DR12\\
1327315426141900672 & NGC 6205 & -245.22 & -1.51 & SEGUE DR12 & 1360201505396486016 & NGC 6341 & -133.85 & -2.21 & SEGUE DR12\\
1328071954566404736 & NGC 6205 & -255.31 & -1.67 & SEGUE DR12 & 1360211577095721088 & NGC 6341 & -120.55 & -2.23 & SEGUE DR12\\
1331385646389205632 & NGC 6205 & -248.86 & -1.65 & SEGUE DR12 & 1360417215833016576 & NGC 6341 & -122.24 & -2.43 & SEGUE DR12\\
1328063437654350976 & NGC 6205 & -252.65 & -1.50 & SEGUE DR12 & 1360381898322080384 & NGC 6341 & -126.08 & -2.46 & SEGUE DR12\\
1328123601557429888 & NGC 6205 & -247.03 & -1.60 & SEGUE DR12 & 1360401208495084288 & NGC 6341 & -121.80 & -2.15 & SEGUE DR12\\
1328136688314520960 & NGC 6205 & -236.12 & -1.38 & SEGUE DR12 & 1354372758034296576 & NGC 6341 & -119.76 & -2.58 & SEGUE DR12\\
1360144021554095744 & NGC 6341 & -129.98 & NaN & SEGUE DR12 & 2686871720073157504 & NGC 7089 & -3.85 & NaN & SEGUE DR12\\
1360396222033324032 & NGC 6341 & -133.22 & NaN & SEGUE DR12 & 2686795819414993792 & NGC 7089 & -5.97 & NaN & SEGUE DR12\\
1360191781589539072 & NGC 6341 & -144.51 & NaN & SEGUE DR12 & 2686857254624358656 & NGC 7089 & 8.13 & NaN & SEGUE DR12\\
1360427420676038144 & NGC 6341 & -139.68 & NaN & SEGUE DR12 & 2686927211051932032 & NGC 7089 & 3.38 & -1.66 & SEGUE DR12\\
1360399898525361024 & NGC 6341 & -127.29 & NaN & SEGUE DR12 & 2686854437125995648 & NGC 7089 & 10.42 & -1.51 & SEGUE DR12\\
1360212603591009664 & NGC 6341 & -105.03 & NaN & SEGUE DR12 & 2686833237167039232 & NGC 7089 & 12.15 & -1.50 & SEGUE DR12\\
1360412061872257920 & NGC 6341 & -111.33 & NaN & SEGUE DR12 & 2686896806981288704 & NGC 7089 & 6.04 & -1.36 & SEGUE DR12\\
1360454118192779776 & NGC 6341 & -102.27 & NaN & SEGUE DR12 & 2686893710305935616 & NGC 7089 & -6.79 & -1.67 & SEGUE DR12\\
1360454358710919040 & NGC 6341 & -123.53 & NaN & SEGUE DR12 & 2686632065193630848 & NGC 7089 & -9.44 & -1.42 & SEGUE DR12\\
\hline

\end{tabular}}

\tablefoot{A list of all confirmed extra-tidal stars.  NaN  indicates no metallicity determination.}
\end{table*}

\subsubsection{\textbf{NGC 1851}}

NGC 1851 is an old \citep[9.98 Gyr,][]{2010MNRAS.404.1203F} mildly metal-poor (-1.18 dex, \citealt{2010arXiv1012.3224H}) GC located at a distance of about 11.95 kpc \citep{2021MNRAS.505.5957B}. NGC 1851 is thought to have an extended halo structure that is   continuously expanding \citep{2009AJ....138.1570O,2018MNRAS.473.2881K,2014MNRAS.445.2971C,2018MNRAS.474..683C}. With deep photometric data from the Dark Energy Survey (DES) and Gaia, \citet{2018ApJ...862..114S} and \citet{2021ApJ...914..123I} identified diffuse and long ($\sim$8$^\circ$) tidal tails around NGC 1851. 

\begin{figure}[htbp]
    \centering
    \includegraphics[width=1\columnwidth]{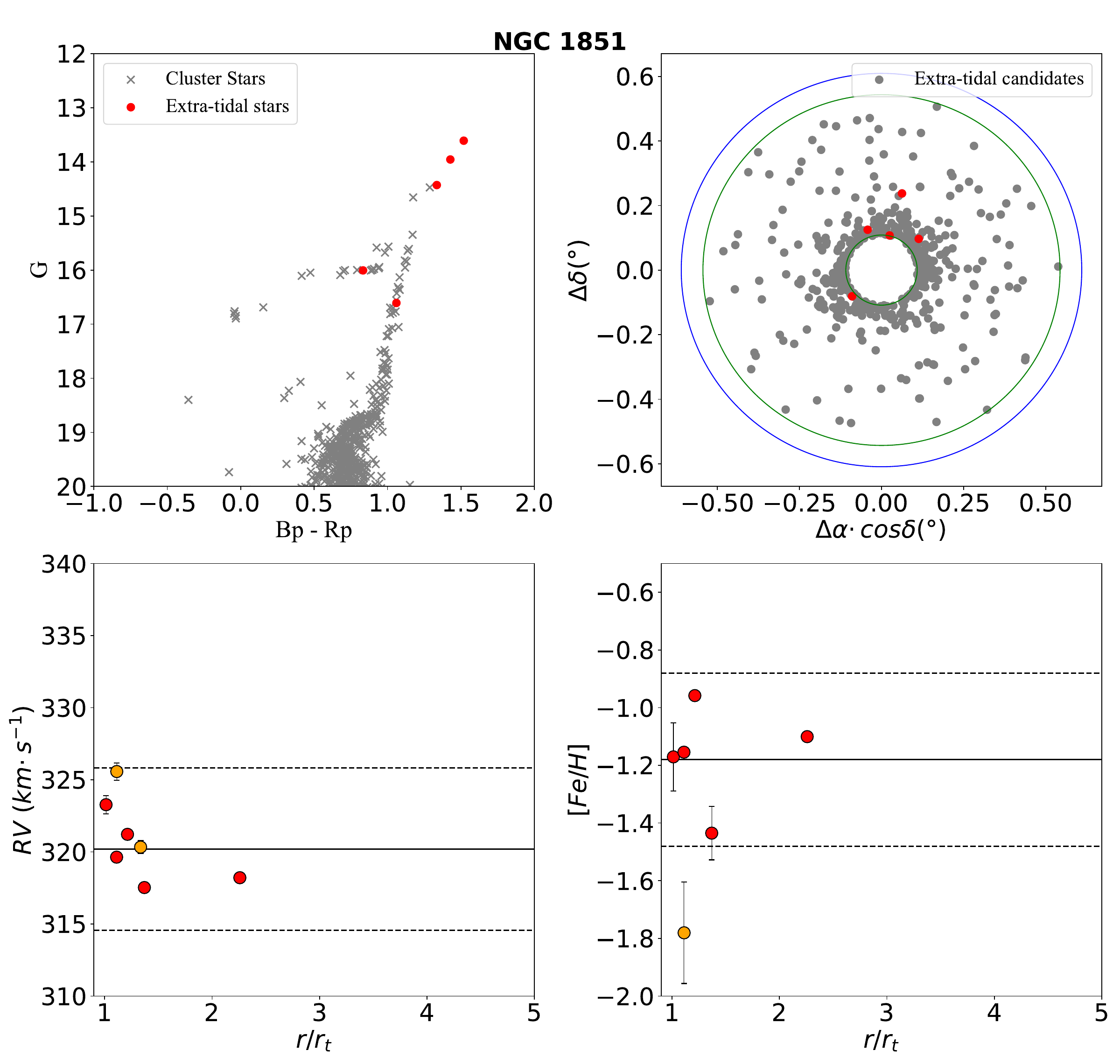}
    \caption{CMD, spatial distribution, radial velocity, and  metallicity distribution of the cluster and extra-tidal candidates based on Gaia EDR3 photometry \citep{2021A&A...649A...1G}. The gray crosses in the upper left panel represents cluster stars that are located within 1 $r_t$ and whose  star PM matches that of the GC. The gray points in the upper right panel represent the extra-tidal candidates   selected following the method described in Section \ref{sec:method2}. The red points represent extra-tidal candidates with  RV and [Fe/H] compatible with the cluster. The green circles represent 1 and 5 tidal radii, and the blue circle represents the Jacobi radius. In the bottom two panels  the orange points represent stars with radial velocity and metallicity observations that do not match the cluster. The solid horizontal line represents the mean value of the cluster, and the dashed horizontal line represents the corresponding dispersion or error range for cluster. The error bar represents the observation error of the individual stars.}
    \label{fig:NGC 1851}%
\end{figure}

In this work we found 444 extra-tidal candidate stars near NGC 1851, as shown in Fig. \ref{fig:NGC 1851}. Using high-precision PMs and CMDs from Gaia EDR3, we removed foreground stars and greatly expanded the sample of extra-tidal candidates. All stars are located within $r_J$, indicating that these stars are loosely bound to the cluster. 
Due to the high stellar density near the tidal radius of NGC 1851, we slightly expanded the inner boundary range to investigate whether NGC 1851 exhibits extended structures or stellar streams. The density distribution of the candidates (Fig. \ref{fig:density map}) appears to have two high-density peaks (> 3 Poisson noise) in the opposite direction (east--west direction), which indicates that mass loss preferentially happened in these two points. Due to the differential Galactic rotation, the GC escapees would form long stellar streams, similar to what was found in \citet{2018ApJ...862..114S} and \citet{2021ApJ...914..123I}.

To further verify their association with NGC 1851, we looked for the detection of their RVs and metallicities. Seven stars were observed by APOGEE, and their RVs all lie within the range of cluster mean RV$\pm$dispersion. Among them, five stars show metallicities consistent with that of the cluster, while one star does not have metallicity measurement, which does not exclude its membership. We note that extra-tidal candidates of NGC 1851 have been explored by several studies \citep[e.g.,][]{2014MNRAS.442.3044M,2021ApJ...914..123I}, but no common star is found between these studies and our five extra-tidal stars, mostly due to spatial difference. 

\subsubsection{\textbf{NGC 1904}}
NGC 1904 is an old (11.16 Gyr, \citealt{2010MNRAS.404.1203F}) intermediately metal-poor (-1.6 dex, \citealt{2010arXiv1012.3224H}) halo GC (D $\approx$ 13.08 Kpc, \citealt{2021MNRAS.505.5957B}). In the last few years, \citet{2019MNRAS.485.4906D} have shown that the radial density distribution of this cluster can be described by an extended model within the Jacobi radius. However, deep DES photometry presented by \citet{2018ApJ...862..114S} suggested the possible presence of a tidal tail around NGC 1904, oriented approximately north-south and extending to a radius of 1.5 $^\circ$. It was later confirmed by \citet{2020MNRAS.495.2222S} with Gaia DR2 data.

\begin{figure}[htbp]
    \centering
    \includegraphics[width=1\columnwidth]{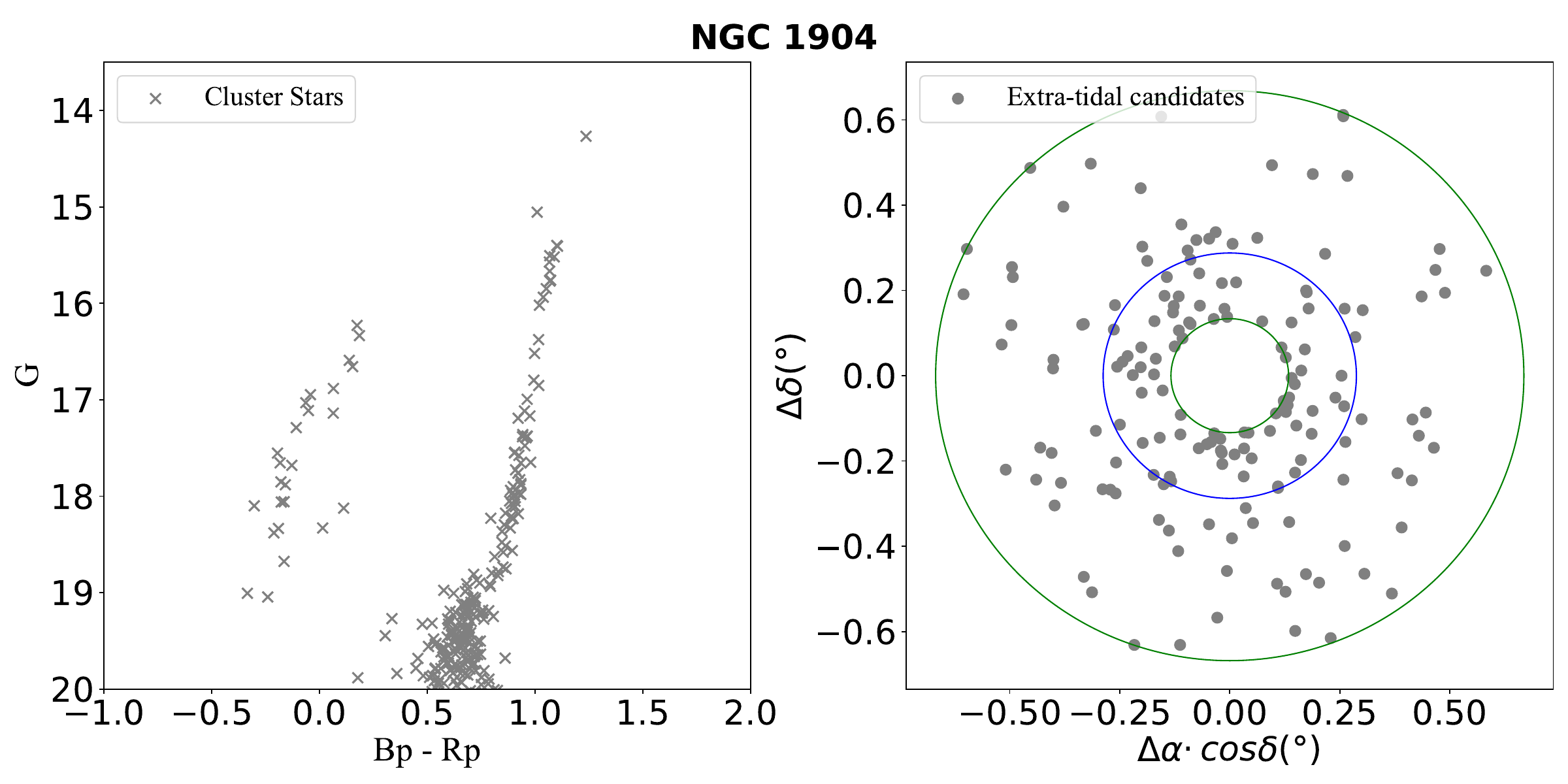}
    \caption{Same as Fig. \ref{fig:NGC 1851}, but only CMD and spatial distribution for NGC 1904.}
    \label{fig:NGC 1904}%
\end{figure}

In this study we found 165 extra-tidal candidates around NGC 1904. Among them, 85 stars are outside $r_J$, which means these stars are totally unbound from the NGC 1904 (Fig. \ref{fig:NGC 1904}). Unfortunately, we found no spectroscopic observations for these stars in APOGEE, SEGUE, and LAMOST. 

We then constructed density maps of extra-tidal candidates to check the extended structure of NGC 1904. We found clear density peaks along the N-S direction in the vicinity of NGC 1904 (Fig. \ref{fig:density map}), with a density exceeding three times the Poisson noise level, suggesting that NGC 1904 might have experienced significant stellar loss in these two directions (possibly L1 and L2 points). Similarly, this is a strong sign of stellar streams, which were reported by \citet{2018ApJ...862..114S} and \citet{2020MNRAS.495.2222S}. 

\subsubsection{\textbf{NGC 6205}}

NGC 6205 is an old (11.65 Gyr, \citealt{2010MNRAS.404.1203F}) intermediately metal-poor (-1.53 dex, \citealt{2010arXiv1012.3224H}) GC, with multiple population detections (e.g., \citealt{2004AJ....127.2162S,2012ApJ...754L..38J}).
In this work we identified a total of 427 extra-tidal candidates, of which 210 are located outside of $r_J$.  
 In the density map of extra-tidal candidates, we found multiple peaks ($>2$), but with density values comparable to the Poisson noise level, indicating that stars might have escaped this GC more homogeneously. The possible explanations include high contamination from field stars and  GC roation, but the lack of two significant peaks indicates that the tidal stream may not be easily formed, which is consistent with the conclusions drawn from the analysis of \citet{2010A&A...522A..71J} and \citet{2020MNRAS.495.2222S}. 

Among our extra-tidal candidates, 29 stars were observed by the SEGUE survey. We identified 13 stars with  RVs and metallicity consistent with those of  the cluster.\footnote{Stars with significantly different RVs and [Fe/H] are not shown here due to the limited dynamical range.} 

\citet{2020A&A...637A..98H} identified 32 extra-tidal candidates, with 13 located within 1.2$r_t$ and 19 beyond 1.6$r_t$. The latter   are classified as extra-tidal stars in their work. This may due to the high stellar density near the $r_t$. Among our 29 extra-tidal candidates, five stars overlap with Hanke's extra-tidal sample. Only three stars\footnote{GAIA EDR3 ID 1328125006003407360, 1328123601557429888, 1328116244270330368} show  RVs and [Fe/H] consistent with the cluster under our  stricter selection criteria.

\begin{figure}[htbp]
    \centering
    \includegraphics[width=1\columnwidth]{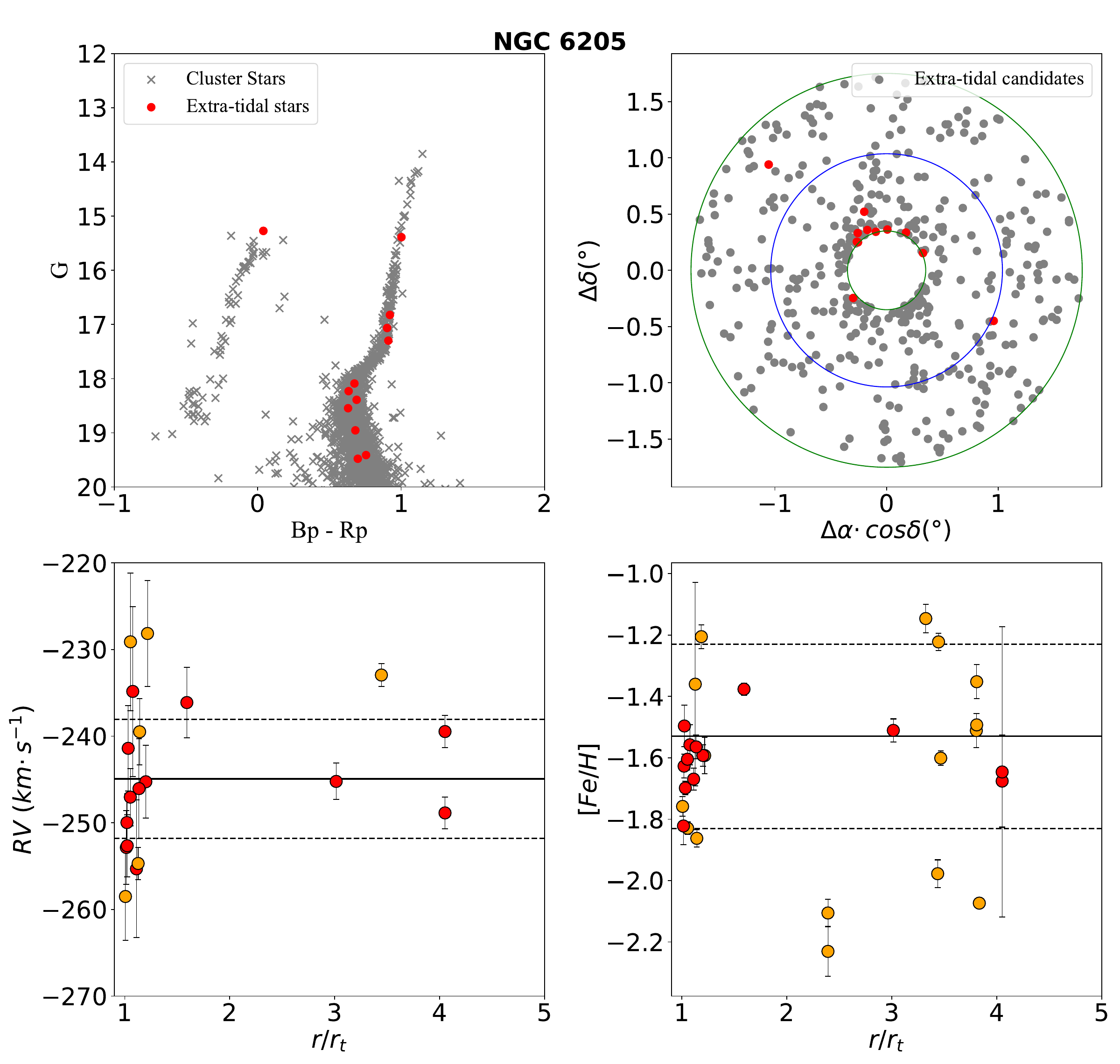}
    \caption{Same as Fig. \ref{fig:NGC 1851}, but for NGC 6205.}
    \label{fig:NGC 6205}%
\end{figure}

\subsubsection{\textbf{NGC 6341}}

NGC 6341, one of the most metal-poor GCs (-2.31 dex, \citealt{2010arXiv1012.3224H}) in the MW, was found to have an extended outer halo \citep{2000A&A...356..127T,2010A&A...522A..71J}. Recently, deep photometric observation revealed a long tidal tail (spanning $\sim$ 17$^\circ$ on the sky) around this cluster \citep{2021ApJ...914..123I, 2020MNRAS.495.2222S, 2020ApJ...902...89T}. \citet{2020A&A...637A..98H} identified 26 extra-tidal candidates, of which 12 were located between 12.5 and 75.2 $r_t$, greatly expanding the previously assumed range of tidal debris. These authors concluded that NGC 6341 might have experienced a strong tidal field, leaving highly dispersed extra-tidal stars after tidal disruption. In this work we found a total of 286 extra-tidal candidates, of which 53 are located beyond $r_J$. These stars are also highly dispersed in their spatial location. In the density map (Fig. \ref{fig:density map}) we found one significant density peak (> 2 Poisson noise) along the stellar stream track toward the northwest of NGC 6341. A much weaker density peak is also found in the opposite direction, indicating that a stellar stream could have formed \citep{2021ApJ...914..123I, 2020MNRAS.495.2222S, 2020ApJ...902...89T}.

\begin{figure}[htbp]
    \centering
    \includegraphics[width=1\columnwidth]{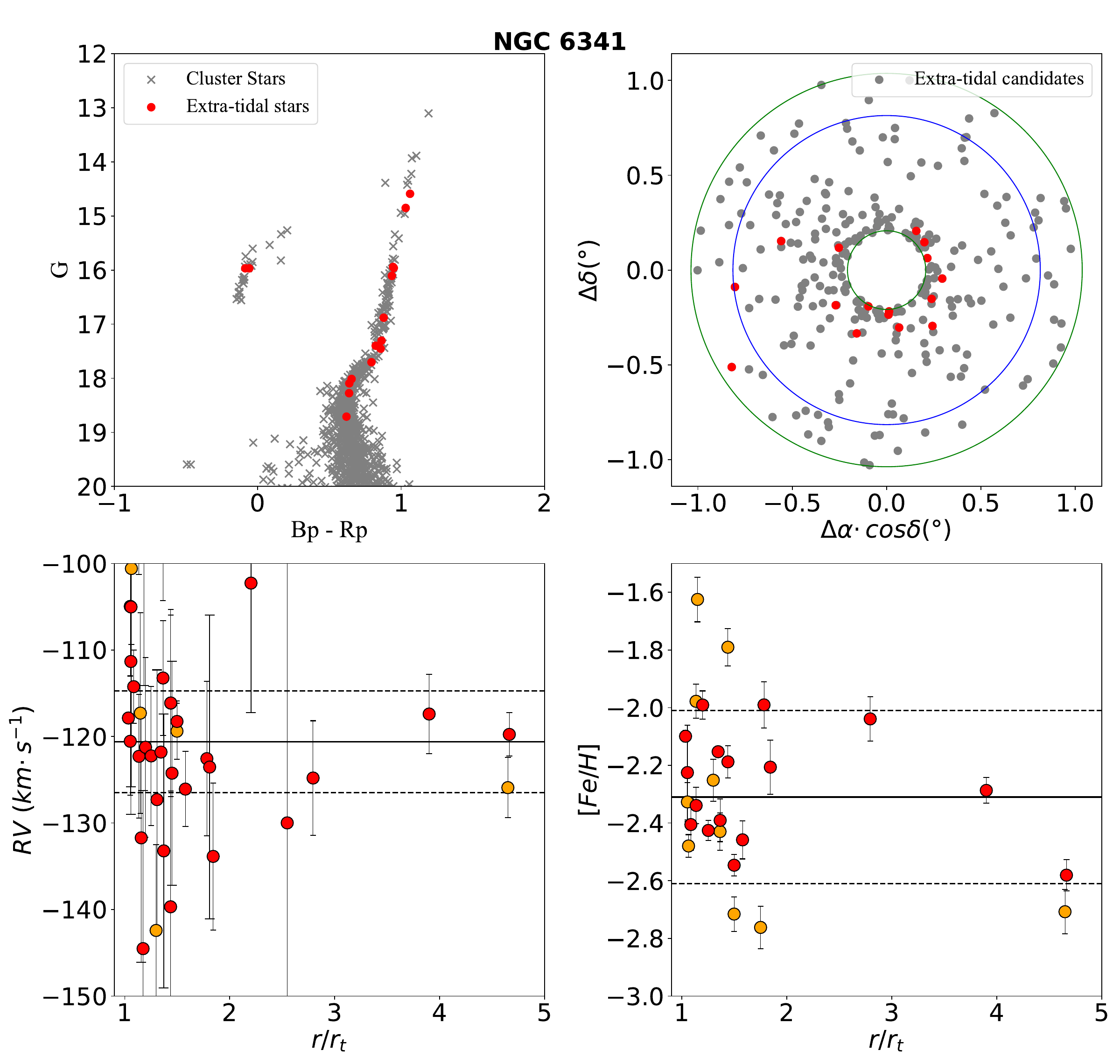}
    \caption{Same as Fig. \ref{fig:NGC 1851}, but for NGC 6341.}
    \label{fig:NGC 6341}%
\end{figure}

After cross-referencing with the spectral survey, we found a total of 48 stars with SEGUE observations among our candidates, of which 13 lacked metallicity information. Among the remaining 35 stars, we compared their RV distribution and [Fe/H], and ultimately identified 16 stars that are consistent with the cluster distribution, including 7 stars with two observations that both meet the cluster RV and [Fe/H] requirement. In addition, 7 of the 35 candidates overlapped with Hanke's results, of which 6 were retained as extra-tidal stars, and 1 star\footnote{Gaia EDR3 ID 1360223598707925376} was excluded due to a large difference in metallicity (-1.6 dex) compared to the cluster (-2.31 dex). For the 13 extra-tidal candidates with only RV observations, we also examined their RV distribution. We include them in left bottom panel of Fig. \ref{fig:NGC 6341} where 11 stars fall within the expected range of the cluster's RV. We will consider them to be extra-tidal stars until subsequent metallicity measurements are available.

\subsubsection{\textbf{NGC 6779}}

NGC 6779 is an extremely old (13.7 Gyr, \citealt{2010MNRAS.404.1203F}), metal-poor (-1.98 dex, \citealt{2010arXiv1012.3224H}) GC (D $\approx$ 10.43 Kpc, \citealt{2021MNRAS.505.5957B}). 
Recently, \citet{2019MNRAS.485.1029P} constructed the radial distribution map of the outer regions of the cluster using horizontal branch and main sequence stars and found that NGC 6779 may have an extended halo structure. However, due to the limited depth of the observations, the data did not cover the fainter main sequence stars, and there has been no clear observational evidence in recent years to confirm the extended structure. 

Our work identified 115 extra-tidal candidates, of which 58 are located beyond the $r_J$, as shown in Fig. \ref{fig:NGC 1904}. Due to the low number density of extra-tidal candidates, no reliable structure was found in its density map (Fig. \ref{fig:density map}). 

\begin{figure}[htbp]
    \centering
    \includegraphics[width=1\columnwidth]{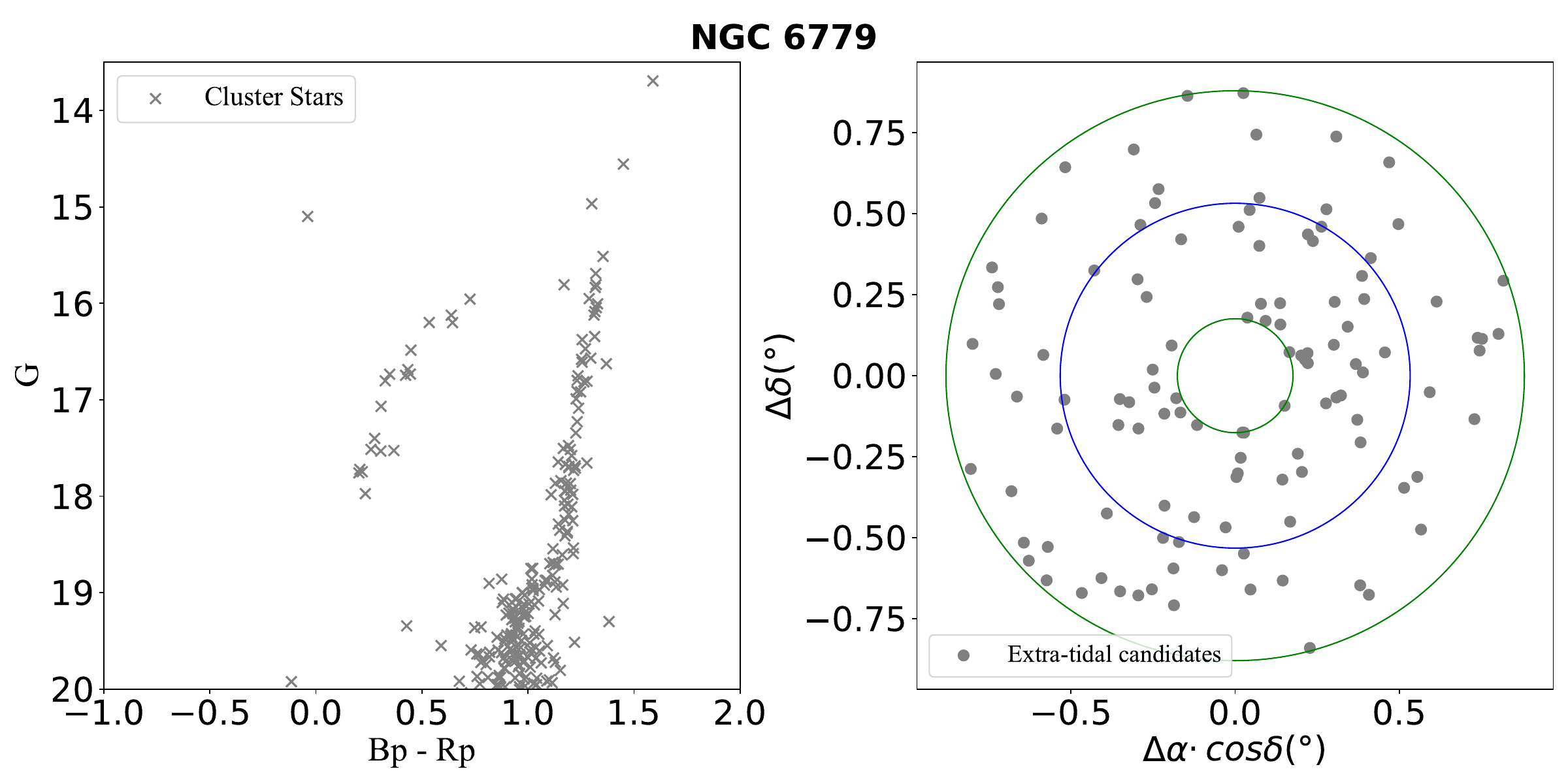}
    \caption{Same as Fig. \ref{fig:NGC 1851}, but only CMD and spatial distribution for NGC 6779.}
    \label{fig:NGC 6779}%
\end{figure}

\subsubsection{\textbf{NGC 7089}}

\begin{figure}[htbp]
    \centering
    \includegraphics[width=1\columnwidth]{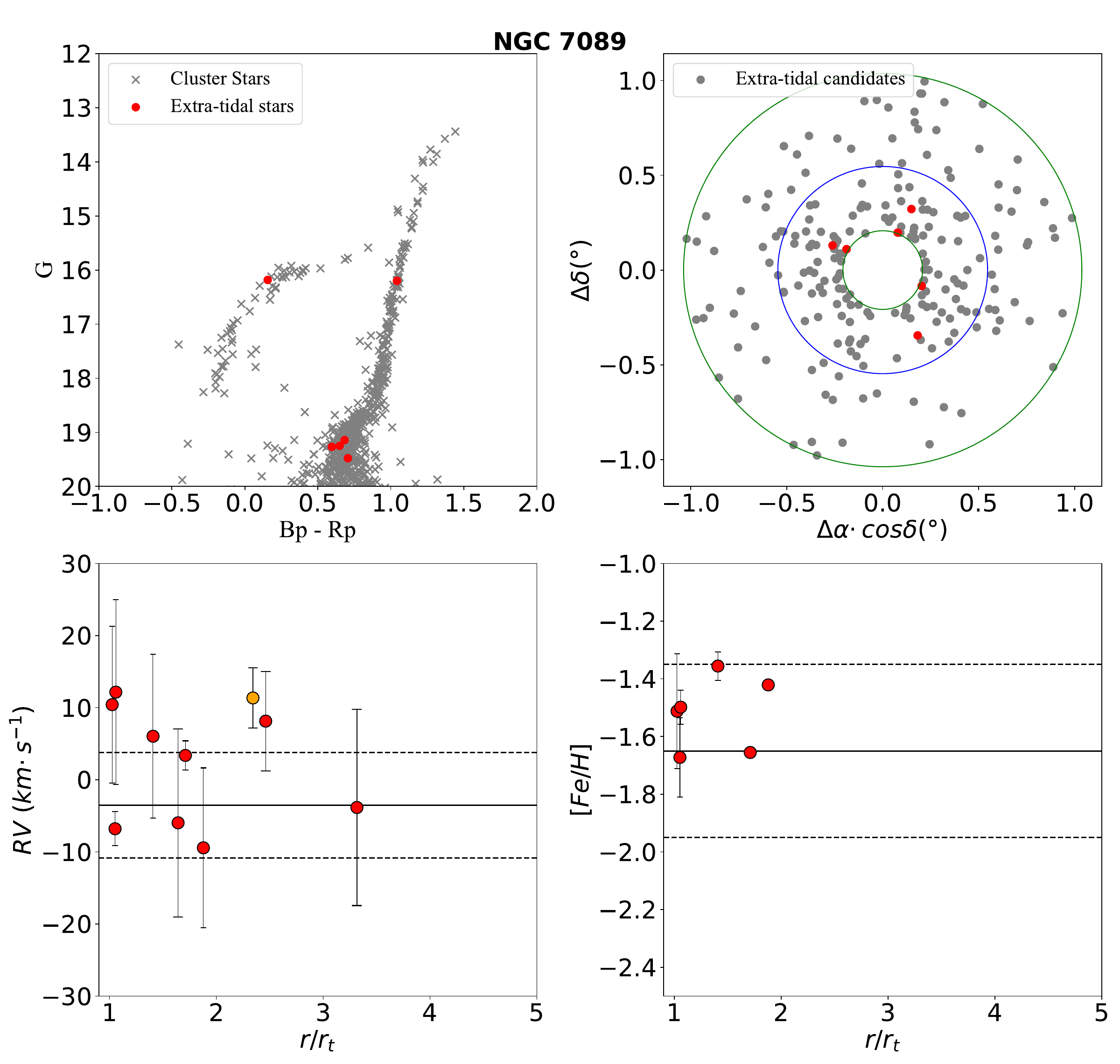}
    \caption{Same as Fig. \ref{fig:NGC 1851} but for NGC 7089.}
    \label{fig:NGC 7089}%
\end{figure}

NGC 7089 is an old (11.78 Gyr, \citealt{2010MNRAS.404.1203F}) intermediately metal-poor (-1.65 dex, \citealt{2010arXiv1012.3224H}) halo GC (D $\approx$ 11.65 Kpc, \citealt{2021MNRAS.505.5957B}), which shows multiple stellar populations (e.g., \citealt{2012ApJ...760...39P,2015MNRAS.447..927M,2014MNRAS.441.3396Y}). There are possible detections of extra-tidal structures associated with this GC: \citet{2016MNRAS.461.3639K} found evidence of a power-law envelope, \citet{2021ApJ...914..123I} and \citet{2022ApJ...929...89G} discovered a long tidal tail. We have identified a total of 216 extra-tidal candidates, of which 87 are located outside of $r_J$, as shown in Fig. \ref{fig:NGC 7089}. In the density distribution of extra-tidal candidates (Fig. \ref{fig:density map}), we discovered two density peaks along the east--west direction (with an upward trend toward the northwest by $\sim$10$^\circ$). Although this density peak is weaker ($\sim$ 2 Poisson noise) compared to NGC 1851 and NGC 1904, it still indicates the presence of a stellar stream structure. This alignment is in good agreement with the direction of the stellar stream structure described in \citet{2021ApJ...914..123I}. Although their analysis of the NGC 7089 stellar stream structure was based on a limited number of data points, and our investigated extra-tidal region of NGC 7089 was also limited, both results indicate the potential presence of a stellar stream near NGC 7089. To confirm the existence or determine the length of the stellar stream for NGC 7089, deep photometry and RV  will be required to identify additional stream member stars.

Out of the 216 candidates that we identified, 11  have SEGUE observation. By comparing the RVs and [Fe/H] of these stars with the individual RV distribution of NGC 7089, we identified six stars that are consistent with the cluster. Among these 11 candidates, 2 stars\footnote{GAIA EDR3 ID: 2686896806981288704, 2686893710305935616} were previously identified in \citet{2020A&A...637A..98H}, which were also identified by this work as extra-tidal stars. Similar to NGC 6341, for the five extra-tidal candidates with only RV observations, three stars are identified as extra-tidal stars, which are also plotted in the   bottom left  panel of Fig. \ref{fig:NGC 7089}.

\vspace{15pt}

In total, we confirmed 54 extra-tidal stars around six GE-related GCs out of 95 stars with spectroscopic information (Table \ref{Table: extra-tidal stars}). Except for \citet{2020A&A...637A..98H}, we do not find overlap with other literature studies to our best knowledge. 
According to our experiments of extra-tidal candidates with spectroscopic observations, the successful rate of selecting extra-tidal candidates that have RVs and metallicities consistent with clusters is high ($\sim 57 \%$). This is a necessary step for target selection in spectroscopic surveys (e.g., SDSS, LAMOST, 4MOST) in order to maximize the scientific outcome. With RVs from a large sample of extra-tidal candidates, it is promising to construct the dynamical environment of GCs, and even their rotation \citep{2021MNRAS.502.4513W}.

In these six GE-related GCs, we found that stellar streams may be related to their extra-tidal structures and orbits. The density map of extra-tidal candidates that we identified with PMs and CMDs, is an efficient way to discover extra-tidal structures. The two possible   density peaks at opposite directions of the inner boundary are a good indicator of a long stellar stream.
Furthermore, we calculated their orbits to investigate whether some GCs are more susceptible to mass loss,   thus forming stellar streams. Using \textsc{galpy}, we calculated the orbits of these six GCs, beginning from 5 Gyr ago to present (Appendix \ref{app:orb}). All six GCs have high eccentricity and have traversed the Galactic disk multiple times. Notably, the pericenters of NGC 1851, NGC 1904, NGC 6341, and NGC 7089 (GCs with stellar streams) are closer to the Galactic center than the other two GCs (without stellar streams). This demonstrates that these clusters have experienced stronger tidal effects, fostering the formation of long stellar streams.

\section{Summary}
\label{sec:conclusion}

As one of the most ancient objects in our Galaxy, GCs are the ideal candidates to trace Galactic evolution. During their internal dynamical evolution (e.g., two-body relaxation) and their interaction with the evolving MW, GCs continuously lose stars to the field. These GC escapees are the key to fully investigating the contribution of GCs to MW halo populations. In this work we inspected GC escaped candidates under the assumption of tidal evaporation, which may account for as much as 80\% of the escaped population. 

If a star   left its host GC   a long time ago (e.g., $\sim$1 Gyr), it is difficult to identify this star given its similar kinematic properties to other halo field stars. However, the distinctive chemical patterns between enriched GC stars and normal halo stars provide a possible solution. The recently found N-rich field stars are promising candidates of GC escapees. It has been suggested that the dense environment required for the production of peculiar chemical abundance, such as N enrichment, is unique to GCs. To investigate the chemodynamical links between N-rich field stars and existing GCs, we used low-resolution spectroscopic data from the LAMOST survey and astrometric data from Gaia. We created a dataset that allows   the characterization of 100 N-rich field stars and existing MW GCs through their full phase-space actions and metallicity. 

In this work we used dynamical action integrals to connect N-rich stars with GCs. We assumed that   stars that escaped from the cluster retained their original dynamical parameters, which allowed   distant stars to be considered  members without being spatially adjacent to the cluster. 
We reported 137 possible pairs of N-rich field stars and individual GCs, of which 29 pairs were highly probable, and 16 pairs had a difference in metallicity within 0.3. Among the 29 highly probable pairs, 17 were possibly related to accretion events, such as GE, Helmi Streams, or the  Sequoia galaxy. The fact that not every N-rich field star finds a GC host is probably related to the complex dynamical evolution of GCs under the ever-changing Galactic potential.

If a star left its host GC recently with tidal evaporation, it will maintain similar kinematic properties to those of its host GC. In order to identify escaped stars near clusters and provide observational samples for subsequent work (such as field stars contributed by dwarf galaxy GCs and the distribution of multiple populations), we analyzed the outer regions of six GCs that potentially relate to the GE accretion event. We identified extra-tidal candidates based on their spatial locations, PMs, and their CMD locations. We used a Gaussian mixture model (GMM) consisting of two components to estimate the mean and dispersion of the cluster PM. Then, we further confirmed the candidates by plotting the CMD of the cluster stars and corresponding isochrones based on the cluster's parameters. We found more than 1600 extra-tidal candidates in the vicinity of six clusters, including 54 extra-tidal stars, where most of the extra-tidal candidates are RGB, HB, and upper MS stars. Interestingly, we found that the density map of the extra-tidal candidates is an efficient way to discover extra-tidal structures. The two possible density peaks at opposite directions of the inner boundary is a good indicator of a  long stellar stream. 
Additionally, spectroscopic observations are available for some candidates in four clusters. Under the assumption that GC escaped stars have the same  radial velocities and metallicity as their host GCs, we confirmed that more than half ($\sim$ 57\%) of the extra-tidal candidates are escaped stars beyond the tidal radius of the clusters. The high success rate bodes well for follow-up observations of our extra-tidal candidates, where the dynamical environment of GCs can be investigated.

\begin{acknowledgements}

C.X., B.T., C.L., and L.W. gratefully acknowledge support from the National Natural Science Foundation of China through grants No. 12233013, No.12073090, the Natural Science Foundation of Guangdong Province under grant No. 2022A1515010732, and  the China Manned Space Project Nos. CMS-CSST-2021-B03, CMS-CSST-2021-A08, etc. J.G.F-T gratefully acknowledges the grant support provided by Proyecto Fondecyt Iniciaci\'on No. 11220340, and also from the Joint Committee ESO-Government of Chile 2021 (ORP 023/2021). L.W. thanks the support from the National Natural Science Foundation of China through grant 21BAA00619, and the one-hundred-talent project of Sun Yat-sen University, the Fundamental Research Funds for the Central Universities, Sun Yat-sen University  (22hytd09).
\end{acknowledgements}

%
%

\bibliographystyle{aa}
\bibliography{gc_escapees.bib}

\begin{thebibliography}{110}
\expandafter\ifx\csname natexlab\endcsname\relax\def\natexlab#1{#1}\fi

\bibitem[{{Abdurro'uf} {et~al.}(2022){Abdurro'uf}, {Accetta}, {Aerts}, {Silva
  Aguirre}, {Ahumada}, {Ajgaonkar}, {Filiz Ak}, {Alam}, {Allende Prieto},
  {Almeida}, {Anders}, {Anderson}, {Andrews}, {Anguiano}, {Aquino-Ort{\'\i}z},
  {Arag{\'o}n-Salamanca}, {Argudo-Fern{\'a}ndez}, {Ata}, {Aubert},
  {Avila-Reese}, {Badenes}, {Barb{\'a}}, {Barger}, {Barrera-Ballesteros},
  {Beaton}, {Beers}, {Belfiore}, {Bender}, {Bernardi}, {Bershady}, {Beutler},
  {Bidin}, {Bird}, {Bizyaev}, {Blanc}, {Blanton}, {Boardman}, {Bolton},
  {Boquien}, {Borissova}, {Bovy}, {Brandt}, {Brown}, {Brownstein}, {Brusa},
  {Buchner}, {Bundy}, {Burchett}, {Bureau}, {Burgasser}, {Cabang}, {Campbell},
  {Cappellari}, {Carlberg}, {Wanderley}, {Carrera}, {Cash}, {Chen}, {Chen},
  {Cherinka}, {Chiappini}, {Choi}, {Chojnowski}, {Chung}, {Clerc}, {Cohen},
  {Comerford}, {Comparat}, {da Costa}, {Covey}, {Crane}, {Cruz-Gonzalez},
  {Culhane}, {Cunha}, {Dai}, {Damke}, {Darling}, {Davidson}, {Davies},
  {Dawson}, {De Lee}, {Diamond-Stanic}, {Cano-D{\'\i}az}, {S{\'a}nchez},
  {Donor}, {Duckworth}, {Dwelly}, {Eisenstein}, {Elsworth}, {Emsellem},
  {Eracleous}, {Escoffier}, {Fan}, {Farr}, {Feng}, {Fern{\'a}ndez-Trincado},
  {Feuillet}, {Filipp}, {Fillingham}, {Frinchaboy}, {Fromenteau}, {Galbany},
  {Garc{\'\i}a}, {Garc{\'\i}a-Hern{\'a}ndez}, {Ge}, {Geisler}, {Gelfand},
  {G{\'e}ron}, {Gibson}, {Goddy}, {Godoy-Rivera}, {Grabowski}, {Green},
  {Greener}, {Grier}, {Griffith}, {Guo}, {Guy}, {Hadjara}, {Harding},
  {Hasselquist}, {Hayes}, {Hearty}, {Hern{\'a}ndez}, {Hill}, {Hogg},
  {Holtzman}, {Horta}, {Hsieh}, {Hsu}, {Hsu}, {Huber}, {Huertas-Company},
  {Hutchinson}, {Hwang}, {Ibarra-Medel}, {Chitham}, {Ilha}, {Imig}, {Jaekle},
  {Jayasinghe}, {Ji}, {Johnson}, {Jones}, {J{\"o}nsson}, {Katkov}, {Khalatyan},
  {Kinemuchi}, {Kisku}, {Knapen}, {Kneib}, {Kollmeier}, {Kong}, {Kounkel},
  {Kreckel}, {Krishnarao}, {Lacerna}, {Lane}, {Langgin}, {Lavender}, {Law},
  {Lazarz}, {Leung}, {Leung}, {Lewis}, {Li}, {Li}, {Lian}, {Liang}, {Lin},
  {Lin}, {Lin}, {Lintott}, {Long}, {Longa-Pe{\~n}a}, {L{\'o}pez-Cob{\'a}},
  {Lu}, {Lundgren}, {Luo}, {Mackereth}, {de la Macorra}, {Mahadevan},
  {Majewski}, {Manchado}, {Mandeville}, {Maraston}, {Margalef-Bentabol},
  {Masseron}, {Masters}, {Mathur}, {McDermid}, {Mckay}, {Merloni},
  {Merrifield}, {Meszaros}, {Miglio}, {Di Mille}, {Minniti}, {Minsley},
  {Monachesi}, {Moon}, {Mosser}, {Mulchaey}, {Muna}, {Mu{\~n}oz}, {Myers},
  {Myers}, {Nadathur}, {Nair}, {Nandra}, {Neumann}, {Newman}, {Nidever},
  {Nikakhtar}, {Nitschelm}, {O'Connell}, {Garma-Oehmichen}, {Luan Souza de
  Oliveira}, {Olney}, {Oravetz}, {Ortigoza-Urdaneta}, {Osorio}, {Otter},
  {Pace}, {Padilla}, {Pan}, {Pan}, {Parikh}, {Parker}, {Peirani}, {Pe{\~n}a
  Ram{\'\i}rez}, {Penny}, {Percival}, {Perez-Fournon}, {Pinsonneault},
  {Poidevin}, {Poovelil}, {Price-Whelan}, {B{\'a}rbara de Andrade Queiroz},
  {Raddick}, {Ray}, {Rembold}, {Riddle}, {Riffel}, {Riffel}, {Rix}, {Robin},
  {Rodr{\'\i}guez-Puebla}, {Roman-Lopes}, {Rom{\'a}n-Z{\'u}{\~n}iga}, {Rose},
  {Ross}, {Rossi}, {Rubin}, {Salvato}, {S{\'a}nchez}, {S{\'a}nchez-Gallego},
  {Sanderson}, {Santana Rojas}, {Sarceno}, {Sarmiento}, {Sayres}, {Sazonova},
  {Schaefer}, {Schiavon}, {Schlegel}, {Schneider}, {Schultheis}, {Schwope},
  {Serenelli}, {Serna}, {Shao}, {Shapiro}, {Sharma}, {Shen}, {Shetrone}, {Shu},
  {Simon}, {Skrutskie}, {Smethurst}, {Smith}, {Sobeck}, {Spoo}, {Sprague},
  {Stark}, {Stassun}, {Steinmetz}, {Stello}, {Stone-Martinez},
  {Storchi-Bergmann}, {Stringfellow}, {Stutz}, {Su}, {Taghizadeh-Popp},
  {Talbot}, {Tayar}, {Telles}, {Teske}, {Thakar}, {Theissen}, {Tkachenko},
  {Thomas}, {Tojeiro}, {Hernandez Toledo}, {Troup}, {Trump}, {Trussler},
  {Turner}, {Tuttle}, {Unda-Sanzana}, {V{\'a}zquez-Mata}, {Valentini},
  {Valenzuela}, {Vargas-Gonz{\'a}lez}, {Vargas-Maga{\~n}a}, {Alfaro},
  {Villanova}, {Vincenzo}, {Wake}, {Warfield}, {Washington}, {Weaver},
  {Weijmans}, {Weinberg}, {Weiss}, {Westfall}, {Wild}, {Wilde}, {Wilson},
  {Wilson}, {Wilson}, {Wolf}, {Wood-Vasey}, {Yan}, {Zamora}, {Zasowski},
  {Zhang}, {Zhao}, {Zheng}, {Zheng}, \& {Zhu}}]{2022ApJS..259...35A}
{Abdurro'uf}, {Accetta}, K., {Aerts}, C., {et~al.} 2022, \apjs, 259, 35

\bibitem[{{Abolfathi} {et~al.}(2018){Abolfathi}, {Aguado}, {Aguilar}, {Allende
  Prieto}, {Almeida}, {Ananna}, {Anders}, {Anderson}, {Andrews}, {Anguiano},
  {Arag{\'o}n-Salamanca}, {Argudo-Fern{\'a}ndez}, {Armengaud}, {Ata},
  {Aubourg}, {Avila-Reese}, {Badenes}, {Bailey}, {Balland}, {Barger},
  {Barrera-Ballesteros}, {Bartosz}, {Bastien}, {Bates}, {Baumgarten},
  {Bautista}, {Beaton}, {Beers}, {Belfiore}, {Bender}, {Bernardi}, {Bershady},
  {Beutler}, {Bird}, {Bizyaev}, {Blanc}, {Blanton}, {Blomqvist}, {Bolton},
  {Boquien}, {Borissova}, {Bovy}, {Bradna Diaz}, {Brandt}, {Brinkmann},
  {Brownstein}, {Bundy}, {Burgasser}, {Burtin}, {Busca}, {Ca{\~n}as},
  {Cano-D{\'\i}az}, {Cappellari}, {Carrera}, {Casey}, {Cervantes Sodi}, {Chen},
  {Cherinka}, {Chiappini}, {Choi}, {Chojnowski}, {Chuang}, {Chung}, {Clerc},
  {Cohen}, {Comerford}, {Comparat}, {Correa do Nascimento}, {da Costa},
  {Cousinou}, {Covey}, {Crane}, {Cruz-Gonzalez}, {Cunha}, {da Silva Ilha},
  {Damke}, {Darling}, {Davidson}, {Dawson}, {de Icaza Lizaola}, {de la
  Macorra}, {de la Torre}, {De Lee}, {de Sainte Agathe}, {Deconto Machado},
  {Dell'Agli}, {Delubac}, {Diamond-Stanic}, {Donor}, {Downes}, {Drory}, {du Mas
  des Bourboux}, {Duckworth}, {Dwelly}, {Dyer}, {Ebelke}, {Davis Eigenbrot},
  {Eisenstein}, {Elsworth}, {Emsellem}, {Eracleous}, {Erfanianfar},
  {Escoffier}, {Fan}, {Fern{\'a}ndez Alvar}, {Fernandez-Trincado}, {Fernando
  Cirolini}, {Feuillet}, {Finoguenov}, {Fleming}, {Font-Ribera}, {Freischlad},
  {Frinchaboy}, {Fu}, {G{\'o}mez Maqueo Chew}, {Galbany}, {Garc{\'\i}a
  P{\'e}rez}, {Garcia-Dias}, {Garc{\'\i}a-Hern{\'a}ndez}, {Garma Oehmichen},
  {Gaulme}, {Gelfand}, {Gil-Mar{\'\i}n}, {Gillespie}, {Goddard}, {Gonz{\'a}lez
  Hern{\'a}ndez}, {Gonzalez-Perez}, {Grabowski}, {Green}, {Grier}, {Gueguen},
  {Guo}, {Guy}, {Hagen}, {Hall}, {Harding}, {Hasselquist}, {Hawley}, {Hayes},
  {Hearty}, {Hekker}, {Hernandez}, {Hernandez Toledo}, {Hogg},
  {Holley-Bockelmann}, {Holtzman}, {Hou}, {Hsieh}, {Hunt}, {Hutchinson},
  {Hwang}, {Jimenez Angel}, {Johnson}, {Jones}, {J{\"o}nsson}, {Jullo}, {Khan},
  {Kinemuchi}, {Kirkby}, {Kirkpatrick}, {Kitaura}, {Knapp}, {Kneib},
  {Kollmeier}, {Lacerna}, {Lane}, {Lang}, {Law}, {Le Goff}, {Lee}, {Li}, {Li},
  {Lian}, {Liang}, {Lima}, {Lin}, {Long}, {Lucatello}, {Lundgren}, {Mackereth},
  {MacLeod}, {Mahadevan}, {Maia}, {Majewski}, {Manchado}, {Maraston},
  {Mariappan}, {Marques-Chaves}, {Masseron}, {Masters}, {McDermid}, {McGreer},
  {Melendez}, {Meneses-Goytia}, {Merloni}, {Merrifield}, {Meszaros}, {Meza},
  {Minchev}, {Minniti}, {Mueller}, {Muller-Sanchez}, {Muna}, {Mu{\~n}oz},
  {Myers}, {Nair}, {Nandra}, {Ness}, {Newman}, {Nichol}, {Nidever},
  {Nitschelm}, {Noterdaeme}, {O'Connell}, {Oelkers}, {Oravetz}, {Oravetz},
  {Ort{\'\i}z}, {Osorio}, {Pace}, {Padilla}, {Palanque-Delabrouille},
  {Palicio}, {Pan}, {Pan}, {Parikh}, {P{\^a}ris}, {Park}, {Peirani},
  {Pellejero-Ibanez}, {Penny}, {Percival}, {Perez-Fournon}, {Petitjean},
  {Pieri}, {Pinsonneault}, {Pisani}, {Prada}, {Prakash}, {Queiroz}, {Raddick},
  {Raichoor}, {Barboza Rembold}, {Richstein}, {Riffel}, {Riffel}, {Rix},
  {Robin}, {Rodr{\'\i}guez Torres}, {Rom{\'a}n-Z{\'u}{\~n}iga}, {Ross},
  {Rossi}, {Ruan}, {Ruggeri}, {Ruiz}, {Salvato}, {S{\'a}nchez}, {S{\'a}nchez},
  {Sanchez Almeida}, {S{\'a}nchez-Gallego}, {Santana Rojas}, {Santiago},
  {Schiavon}, {Schimoia}, {Schlafly}, {Schlegel}, {Schneider}, {Schuster},
  {Schwope}, {Seo}, {Serenelli}, {Shen}, {Shen}, {Shetrone}, {Shull}, {Silva
  Aguirre}, {Simon}, {Skrutskie}, {Slosar}, {Smethurst}, {Smith}, {Sobeck},
  {Somers}, {Souter}, {Souto}, {Spindler}, {Stark}, {Stassun}, {Steinmetz},
  {Stello}, {Storchi-Bergmann}, {Streblyanska}, {Stringfellow}, {Su{\'a}rez},
  {Sun}, {Szigeti}, {Taghizadeh-Popp}, {Talbot}, {Tang}, {Tao}, {Tayar},
  {Tembe}, {Teske}, {Thakar}, {Thomas}, {Tissera}, {Tojeiro}, {Tremonti},
  {Troup}, {Urry}, {Valenzuela}, {van den Bosch}, {Vargas-Gonz{\'a}lez},
  {Vargas-Maga{\~n}a}, {Vazquez}, {Villanova}, {Vogt}, {Wake}, {Wang},
  {Weaver}, {Weijmans}, {Weinberg}, {Westfall}, {Whelan}, {Wilcots}, {Wild},
  {Williams}, {Wilson}, {Wood-Vasey}, {Wylezalek}, {Xiao}, {Yan}, {Yang},
  {Ybarra}, {Y{\`e}che}, {Zakamska}, {Zamora}, {Zarrouk}, {Zasowski}, {Zhang},
  {Zhao}, {Zhao}, {Zheng}, {Zheng}, {Zhou}, {Zhu}, {Zinn}, \&
  {Zou}}]{2018ApJS..235...42A}
{Abolfathi}, B., {Aguado}, D.~S., {Aguilar}, G., {et~al.} 2018, \apjs, 235, 42

\bibitem[{{Anguiano} {et~al.}(2016){Anguiano}, {De Silva}, {Freeman}, {Da
  Costa}, {Zwitter}, {Quillen}, {Zucker}, {Navarro}, {Kunder}, {Siebert},
  {Wyse}, {Grebel}, {Kordopatis}, {Gibson}, {Seabroke}, {Sharma}, {Wojno},
  {Bland-Hawthorn}, {Parker}, {Steinmetz}, {Boeche}, {Gilmore}, {Bienaym{\'e}},
  {Reid}, \& {Watson}}]{2016MNRAS.457.2078A}
{Anguiano}, B., {De Silva}, G.~M., {Freeman}, K., {et~al.} 2016, \mnras, 457,
  2078

\bibitem[{{Anguiano} {et~al.}(2015){Anguiano}, {Zucker}, {Scholz}, {Grebel},
  {Seabroke}, {Kunder}, {Binney}, {McMillan}, {Zwitter}, {Wyse}, {Kordopatis},
  {Bienaym{\'e}}, {Bland-Hawthorn}, {Boeche}, {Freeman}, {Gibson}, {Gilmore},
  {Munari}, {Navarro}, {Parker}, {Reid}, {Siebert}, {Siviero}, {Steinmetz}, \&
  {Watson}}]{2015MNRAS.451.1229A}
{Anguiano}, B., {Zucker}, D.~B., {Scholz}, R.~D., {et~al.} 2015, \mnras, 451,
  1229

\bibitem[{{Bastian} \& {Lardo}(2018)}]{2018ARA&A..56...83B}
{Bastian}, N. \& {Lardo}, C. 2018, \araa, 56, 83

\bibitem[{{Baumgardt} \& {Hilker}(2018)}]{2018MNRAS.478.1520B}
{Baumgardt}, H. \& {Hilker}, M. 2018, \mnras, 478, 1520

\bibitem[{{Baumgardt} {et~al.}(2010){Baumgardt}, {Parmentier}, {Gieles}, \&
  {Vesperini}}]{2010MNRAS.401.1832B}
{Baumgardt}, H., {Parmentier}, G., {Gieles}, M., \& {Vesperini}, E. 2010,
  \mnras, 401, 1832

\bibitem[{{Baumgardt} \& {Vasiliev}(2021)}]{2021MNRAS.505.5957B}
{Baumgardt}, H. \& {Vasiliev}, E. 2021, \mnras, 505, 5957

\bibitem[{{Belokurov} {et~al.}(2018){Belokurov}, {Erkal}, {Evans}, {Koposov},
  \& {Deason}}]{2018MNRAS.478..611B}
{Belokurov}, V., {Erkal}, D., {Evans}, N.~W., {Koposov}, S.~E., \& {Deason},
  A.~J. 2018, \mnras, 478, 611

\bibitem[{{Belokurov} {et~al.}(2006){Belokurov}, {Zucker}, {Evans}, {Gilmore},
  {Vidrih}, {Bramich}, {Newberg}, {Wyse}, {Irwin}, {Fellhauer}, {Hewett},
  {Walton}, {Wilkinson}, {Cole}, {Yanny}, {Rockosi}, {Beers}, {Bell},
  {Brinkmann}, {Ivezi{\'c}}, \& {Lupton}}]{2006ApJ...642L.137B}
{Belokurov}, V., {Zucker}, D.~B., {Evans}, N.~W., {et~al.} 2006, \apjl, 642,
  L137

\bibitem[{{Binney}(2012)}]{2012MNRAS.426.1324B}
{Binney}, J. 2012, \mnras, 426, 1324

\bibitem[{{Binney} \& {Tremaine}(2008)}]{2008gady.book.....B}
{Binney}, J. \& {Tremaine}, S. 2008, {Galactic Dynamics: Second Edition}

\bibitem[{{Bovy}(2015)}]{2015ApJS..216...29B}
{Bovy}, J. 2015, \apjs, 216, 29

\bibitem[{{Bressan} {et~al.}(2012){Bressan}, {Marigo}, {Girardi}, {Salasnich},
  {Dal Cero}, {Rubele}, \& {Nanni}}]{2012MNRAS.427..127B}
{Bressan}, A., {Marigo}, P., {Girardi}, L., {et~al.} 2012, \mnras, 427, 127

\bibitem[{{Buder} {et~al.}(2021){Buder}, {Sharma}, {Kos}, {Amarsi},
  {Nordlander}, {Lind}, {Martell}, {Asplund}, {Bland-Hawthorn}, {Casey}, {de
  Silva}, {D'Orazi}, {Freeman}, {Hayden}, {Lewis}, {Lin}, {Schlesinger},
  {Simpson}, {Stello}, {Zucker}, {Zwitter}, {Beeson}, {Buck}, {Casagrande},
  {Clark}, {{\v{C}}otar}, {da Costa}, {de Grijs}, {Feuillet}, {Horner},
  {Kafle}, {Khanna}, {Kobayashi}, {Liu}, {Montet}, {Nandakumar}, {Nataf},
  {Ness}, {Spina}, {Tepper-Garc{\'\i}a}, {Ting}, {Traven},
  {Vogrin{\v{c}}i{\v{c}}}, {Wittenmyer}, {Wyse}, {{\v{Z}}erjal}, \& {Galah
  Collaboration}}]{2021MNRAS.506..150B}
{Buder}, S., {Sharma}, S., {Kos}, J., {et~al.} 2021, \mnras, 506, 150

\bibitem[{{Carballo-Bello} {et~al.}(2018){Carballo-Bello},
  {Mart{\'\i}nez-Delgado}, {Navarrete}, {Catelan}, {Mu{\~n}oz}, {Antoja}, \&
  {Sollima}}]{2018MNRAS.474..683C}
{Carballo-Bello}, J.~A., {Mart{\'\i}nez-Delgado}, D., {Navarrete}, C., {et~al.}
  2018, \mnras, 474, 683

\bibitem[{{Carballo-Bello} {et~al.}(2014){Carballo-Bello}, {Sollima},
  {Mart{\'\i}nez-Delgado}, {Pila-D{\'\i}ez}, {Leaman}, {Fliri}, {Mu{\~n}oz}, \&
  {Corral-Santana}}]{2014MNRAS.445.2971C}
{Carballo-Bello}, J.~A., {Sollima}, A., {Mart{\'\i}nez-Delgado}, D., {et~al.}
  2014, \mnras, 445, 2971

\bibitem[{{Carlin} {et~al.}(2015){Carlin}, {Liu}, {Newberg}, {Beers}, {Chen},
  {Deng}, {Guhathakurta}, {Hou}, {Hou}, {L{\'e}pine}, {Li}, {Luo}, {Smith},
  {Wu}, {Yang}, {Yanny}, {Zhang}, \& {Zheng}}]{2015AJ....150....4C}
{Carlin}, J.~L., {Liu}, C., {Newberg}, H.~J., {et~al.} 2015, \aj, 150, 4

\bibitem[{{Carretta} {et~al.}(2010){Carretta}, {Bragaglia}, {Gratton},
  {Recio-Blanco}, {Lucatello}, {D'Orazi}, \& {Cassisi}}]{2010A&A...516A..55C}
{Carretta}, E., {Bragaglia}, A., {Gratton}, R.~G., {et~al.} 2010, \aap, 516,
  A55

\bibitem[{{Cui} {et~al.}(2012){Cui}, {Zhao}, {Chu}, {Li}, {Li}, {Zhang}, {Su},
  {Yao}, {Wang}, {Xing}, {Li}, {Zhu}, {Wang}, {Gu}, {Luo}, {Xu}, {Zhang},
  {Liu}, {Zhang}, {Yang}, {Cao}, {Chen}, {Chen}, {Chen}, {Chen}, {Chu}, {Feng},
  {Gong}, {Hou}, {Hu}, {Hu}, {Hu}, {Jia}, {Jiang}, {Jiang}, {Jiang}, {Jin},
  {Li}, {Li}, {Li}, {Liu}, {Liu}, {Lu}, {Mao}, {Men}, {Qi}, {Qi}, {Shi},
  {Tang}, {Tao}, {Wang}, {Wang}, {Wang}, {Wang}, {Wang}, {Wang}, {Wang},
  {Wang}, {Wang}, {Wang}, {Wang}, {Wang}, {Xu}, {Xu}, {Yang}, {Yu}, {Yuan},
  {Yuan}, {Zhai}, {Zhang}, {Zhang}, {Zhang}, {Zhao}, {Zhou}, {Zhou}, {Zhu}, \&
  {Zou}}]{2012RAA....12.1197C}
{Cui}, X.-Q., {Zhao}, Y.-H., {Chu}, Y.-Q., {et~al.} 2012, Research in Astronomy
  and Astrophysics, 12, 1197

\bibitem[{{de Boer} {et~al.}(2019){de Boer}, {Gieles}, {Balbinot},
  {H{\'e}nault-Brunet}, {Sollima}, {Watkins}, \&
  {Claydon}}]{2019MNRAS.485.4906D}
{de Boer}, T.~J.~L., {Gieles}, M., {Balbinot}, E., {et~al.} 2019, \mnras, 485,
  4906

\bibitem[{{Deng} {et~al.}(2012){Deng}, {Newberg}, {Liu}, {Carlin}, {Beers},
  {Chen}, {Chen}, {Christlieb}, {Grillmair}, {Guhathakurta}, {Han}, {Hou},
  {Lee}, {L{\'e}pine}, {Li}, {Liu}, {Pan}, {Sellwood}, {Wang}, {Wang}, {Yang},
  {Yanny}, {Zhang}, {Zhang}, {Zheng}, \& {Zhu}}]{2012RAA....12..735D}
{Deng}, L.-C., {Newberg}, H.~J., {Liu}, C., {et~al.} 2012, Research in
  Astronomy and Astrophysics, 12, 735

\bibitem[{{Eisenstein} {et~al.}(2011){Eisenstein}, {Weinberg}, {Agol},
  {Aihara}, {Allende Prieto}, {Anderson}, {Arns}, {Aubourg}, {Bailey},
  {Balbinot}, {Barkhouser}, {Beers}, {Berlind}, {Bickerton}, {Bizyaev},
  {Blanton}, {Bochanski}, {Bolton}, {Bosman}, {Bovy}, {Brandt}, {Breslauer},
  {Brewington}, {Brinkmann}, {Brown}, {Brownstein}, {Burger}, {Busca},
  {Campbell}, {Cargile}, {Carithers}, {Carlberg}, {Carr}, {Chang}, {Chen},
  {Chiappini}, {Comparat}, {Connolly}, {Cortes}, {Croft}, {Cunha}, {da Costa},
  {Davenport}, {Dawson}, {De Lee}, {Porto de Mello}, {de Simoni}, {Dean},
  {Dhital}, {Ealet}, {Ebelke}, {Edmondson}, {Eiting}, {Escoffier}, {Esposito},
  {Evans}, {Fan}, {Femen{\'\i}a Castell{\'a}}, {Dutra Ferreira}, {Fitzgerald},
  {Fleming}, {Font-Ribera}, {Ford}, {Frinchaboy}, {Garc{\'\i}a P{\'e}rez},
  {Gaudi}, {Ge}, {Ghezzi}, {Gillespie}, {Gilmore}, {Girardi}, {Gott}, {Gould},
  {Grebel}, {Gunn}, {Hamilton}, {Harding}, {Harris}, {Hawley}, {Hearty},
  {Hennawi}, {Gonz{\'a}lez Hern{\'a}ndez}, {Ho}, {Hogg}, {Holtzman},
  {Honscheid}, {Inada}, {Ivans}, {Jiang}, {Jiang}, {Johnson}, {Jordan},
  {Jordan}, {Kauffmann}, {Kazin}, {Kirkby}, {Klaene}, {Knapp}, {Kneib},
  {Kochanek}, {Koesterke}, {Kollmeier}, {Kron}, {Lampeitl}, {Lang}, {Lawler},
  {Le Goff}, {Lee}, {Lee}, {Leisenring}, {Lin}, {Liu}, {Long}, {Loomis},
  {Lucatello}, {Lundgren}, {Lupton}, {Ma}, {Ma}, {MacDonald}, {Mack},
  {Mahadevan}, {Maia}, {Majewski}, {Makler}, {Malanushenko}, {Malanushenko},
  {Mandelbaum}, {Maraston}, {Margala}, {Maseman}, {Masters}, {McBride},
  {McDonald}, {McGreer}, {McMahon}, {Mena Requejo}, {M{\'e}nard},
  {Miralda-Escud{\'e}}, {Morrison}, {Mullally}, {Muna}, {Murayama}, {Myers},
  {Naugle}, {Neto}, {Nguyen}, {Nichol}, {Nidever}, {O'Connell}, {Ogando},
  {Olmstead}, {Oravetz}, {Padmanabhan}, {Paegert}, {Palanque-Delabrouille},
  {Pan}, {Pandey}, {Parejko}, {P{\^a}ris}, {Pellegrini}, {Pepper}, {Percival},
  {Petitjean}, {Pfaffenberger}, {Pforr}, {Phleps}, {Pichon}, {Pieri}, {Prada},
  {Price-Whelan}, {Raddick}, {Ramos}, {Reid}, {Reyle}, {Rich}, {Richards},
  {Rieke}, {Rieke}, {Rix}, {Robin}, {Rocha-Pinto}, {Rockosi}, {Roe},
  {Rollinde}, {Ross}, {Ross}, {Rossetto}, {S{\'a}nchez}, {Santiago}, {Sayres},
  {Schiavon}, {Schlegel}, {Schlesinger}, {Schmidt}, {Schneider}, {Sellgren},
  {Shelden}, {Sheldon}, {Shetrone}, {Shu}, {Silverman}, {Simmerer}, {Simmons},
  {Sivarani}, {Skrutskie}, {Slosar}, {Smee}, {Smith}, {Snedden}, {Stassun},
  {Steele}, {Steinmetz}, {Stockett}, {Stollberg}, {Strauss}, {Szalay},
  {Tanaka}, {Thakar}, {Thomas}, {Tinker}, {Tofflemire}, {Tojeiro}, {Tremonti},
  {Vargas Maga{\~n}a}, {Verde}, {Vogt}, {Wake}, {Wan}, {Wang}, {Weaver},
  {White}, {White}, {Wilson}, {Wisniewski}, {Wood-Vasey}, {Yanny}, {Yasuda},
  {Y{\`e}che}, {York}, {Young}, {Zasowski}, {Zehavi}, \&
  {Zhao}}]{2011AJ....142...72E}
{Eisenstein}, D.~J., {Weinberg}, D.~H., {Agol}, E., {et~al.} 2011, \aj, 142, 72

\bibitem[{{Fern{\'a}ndez-Trincado} {et~al.}(2022){Fern{\'a}ndez-Trincado},
  {Beers}, {Barbuy}, {Minniti}, {Chiappini}, {Garro}, {Tang}, {Alves-Brito},
  {Villanova}, {Geisler}, {Lane}, \& {Diaz}}]{2022A&A...663A.126F}
{Fern{\'a}ndez-Trincado}, J.~G., {Beers}, T.~C., {Barbuy}, B., {et~al.} 2022,
  \aap, 663, A126

\bibitem[{{Fern{\'a}ndez-Trincado}
  {et~al.}(2020{\natexlab{a}}){Fern{\'a}ndez-Trincado}, {Beers}, \&
  {Minniti}}]{2020A&A...644A..83F}
{Fern{\'a}ndez-Trincado}, J.~G., {Beers}, T.~C., \& {Minniti}, D.
  2020{\natexlab{a}}, \aap, 644, A83

\bibitem[{{Fern{\'a}ndez-Trincado}
  {et~al.}(2020{\natexlab{b}}){Fern{\'a}ndez-Trincado}, {Beers}, {Minniti},
  {Carigi}, {Barbuy}, {Placco}, {Moni Bidin}, {Villanova}, {Roman-Lopes}, \&
  {Nitschelm}}]{2020ApJ...903L..17F}
{Fern{\'a}ndez-Trincado}, J.~G., {Beers}, T.~C., {Minniti}, D., {et~al.}
  2020{\natexlab{b}}, \apjl, 903, L17

\bibitem[{{Fern{\'a}ndez-Trincado}
  {et~al.}(2021{\natexlab{a}}){Fern{\'a}ndez-Trincado}, {Beers}, {Minniti},
  {Carigi}, {Placco}, {Chun}, {Lane}, {Geisler}, {Villanova}, {Souza},
  {Barbuy}, {P{\'e}rez-Villegas}, {Chiappini}, {Queiroz}, {Tang},
  {Alonso-Garc{\'\i}a}, {Piatti}, {Palma}, {Alves-Brito}, {Moni Bidin},
  {Roman-Lopes}, {Mu{\~n}oz}, {Singh}, {Kundu}, {Chaves-Velasquez},
  {Romero-Colmenares}, {Longa-Pe{\~n}a}, {Soto}, \&
  {Vieira}}]{2021A&A...647A..64F}
{Fern{\'a}ndez-Trincado}, J.~G., {Beers}, T.~C., {Minniti}, D., {et~al.}
  2021{\natexlab{a}}, \aap, 647, A64

\bibitem[{{Fern{\'a}ndez-Trincado}
  {et~al.}(2020{\natexlab{c}}){Fern{\'a}ndez-Trincado}, {Beers}, {Minniti},
  {Tang}, {Villanova}, {Geisler}, {P{\'e}rez-Villegas}, \&
  {Vieira}}]{2020A&A...643L...4F}
{Fern{\'a}ndez-Trincado}, J.~G., {Beers}, T.~C., {Minniti}, D., {et~al.}
  2020{\natexlab{c}}, \aap, 643, L4

\bibitem[{{Fern{\'a}ndez-Trincado}
  {et~al.}(2019{\natexlab{a}}){Fern{\'a}ndez-Trincado}, {Beers}, {Placco},
  {Moreno}, {Alves-Brito}, {Minniti}, {Tang}, {P{\'e}rez-Villegas},
  {Reyl{\'e}}, {Robin}, \& {Villanova}}]{2019ApJ...886L...8F}
{Fern{\'a}ndez-Trincado}, J.~G., {Beers}, T.~C., {Placco}, V.~M., {et~al.}
  2019{\natexlab{a}}, \apjl, 886, L8

\bibitem[{{Fern{\'a}ndez-Trincado}
  {et~al.}(2021{\natexlab{b}}){Fern{\'a}ndez-Trincado}, {Beers}, {Queiroz},
  {Chiappini}, {Minniti}, {Barbuy}, {Majewski}, {Ortigoza-Urdaneta}, {Moni
  Bidin}, {Robin}, {Moreno}, {Chaves-Velasquez}, {Villanova}, {Lane}, {Pan}, \&
  {Bizyaev}}]{2021ApJ...918L..37F}
{Fern{\'a}ndez-Trincado}, J.~G., {Beers}, T.~C., {Queiroz}, A. B.~A., {et~al.}
  2021{\natexlab{b}}, \apjl, 918, L37

\bibitem[{{Fern{\'a}ndez-Trincado}
  {et~al.}(2019{\natexlab{b}}){Fern{\'a}ndez-Trincado}, {Beers}, {Tang},
  {Moreno}, {P{\'e}rez-Villegas}, \& {Ortigoza-Urdaneta}}]{2019MNRAS.488.2864F}
{Fern{\'a}ndez-Trincado}, J.~G., {Beers}, T.~C., {Tang}, B., {et~al.}
  2019{\natexlab{b}}, \mnras, 488, 2864

\bibitem[{{Fern{\'a}ndez-Trincado}
  {et~al.}(2020{\natexlab{d}}){Fern{\'a}ndez-Trincado}, {Chaves-Velasquez},
  {P{\'e}rez-Villegas}, {Vieira}, {Moreno}, {Ortigoza-Urdaneta}, \&
  {Vega-Neme}}]{2020MNRAS.495.4113F}
{Fern{\'a}ndez-Trincado}, J.~G., {Chaves-Velasquez}, L., {P{\'e}rez-Villegas},
  A., {et~al.} 2020{\natexlab{d}}, \mnras, 495, 4113

\bibitem[{{Fern{\'a}ndez-Trincado} {et~al.}(2016){Fern{\'a}ndez-Trincado},
  {Robin}, {Moreno}, {Schiavon}, {Garc{\'\i}a P{\'e}rez}, {Vieira}, {Cunha},
  {Zamora}, {Sneden}, {Souto}, {Carrera}, {Johnson}, {Shetrone}, {Zasowski},
  {Garc{\'\i}a-Hern{\'a}ndez}, {Majewski}, {Reyl{\'e}}, {Blanco-Cuaresma},
  {Martinez-Medina}, {P{\'e}rez-Villegas}, {Valenzuela}, {Pichardo}, {Meza},
  {M{\'e}sz{\'a}ros}, {Sobeck}, {Geisler}, {Anders}, {Schultheis}, {Tang},
  {Roman-Lopes}, {Mennickent}, {Pan}, {Nitschelm}, \&
  {Allard}}]{2016ApJ...833..132F}
{Fern{\'a}ndez-Trincado}, J.~G., {Robin}, A.~C., {Moreno}, E., {et~al.} 2016,
  \apj, 833, 132

\bibitem[{{Fern{\'a}ndez-Trincado} {et~al.}(2017){Fern{\'a}ndez-Trincado},
  {Zamora}, {Garc{\'\i}a-Hern{\'a}ndez}, {Souto}, {Dell'Agli}, {Schiavon},
  {Geisler}, {Tang}, {Villanova}, {Hasselquist}, {Mennickent}, {Cunha},
  {Shetrone}, {Allende Prieto}, {Vieira}, {Zasowski}, {Sobeck}, {Hayes},
  {Majewski}, {Placco}, {Beers}, {Schleicher}, {Robin}, {M{\'e}sz{\'a}ros},
  {Masseron}, {Garc{\'\i}a P{\'e}rez}, {Anders}, {Meza}, {Alves-Brito},
  {Carrera}, {Minniti}, {Lane}, {Fern{\'a}ndez-Alvar}, {Moreno}, {Pichardo},
  {P{\'e}rez-Villegas}, {Schultheis}, {Roman-Lopes}, {Fuentes}, {Nitschelm},
  {Harding}, {Bizyaev}, {Pan}, {Oravetz}, {Simmons}, {Ivans},
  {Blanco-Cuaresma}, {Hern{\'a}ndez}, {Alonso-Garc{\'\i}a}, {Valenzuela}, \&
  {Chanam{\'e}}}]{2017ApJ...846L...2F}
{Fern{\'a}ndez-Trincado}, J.~G., {Zamora}, O., {Garc{\'\i}a-Hern{\'a}ndez},
  D.~A., {et~al.} 2017, \apjl, 846, L2

\bibitem[{{Forbes} \& {Bridges}(2010)}]{2010MNRAS.404.1203F}
{Forbes}, D.~A. \& {Bridges}, T. 2010, \mnras, 404, 1203

\bibitem[{{Fukushige} \& {Heggie}(2000)}]{2000MNRAS.318..753F}
{Fukushige}, T. \& {Heggie}, D.~C. 2000, \mnras, 318, 753

\bibitem[{{Gaia Collaboration} {et~al.}(2018){Gaia Collaboration}, {Brown},
  {Vallenari}, {Prusti}, {de Bruijne}, {Babusiaux}, {Bailer-Jones}, {Biermann},
  {Evans}, {Eyer}, {Jansen}, {Jordi}, {Klioner}, {Lammers}, {Lindegren},
  {Luri}, {Mignard}, {Panem}, {Pourbaix}, {Randich}, {Sartoretti}, {Siddiqui},
  {Soubiran}, {van Leeuwen}, {Walton}, {Arenou}, {Bastian}, {Cropper},
  {Drimmel}, {Katz}, {Lattanzi}, {Bakker}, {Cacciari}, {Casta{\~n}eda},
  {Chaoul}, {Cheek}, {De Angeli}, {Fabricius}, {Guerra}, {Holl}, {Masana},
  {Messineo}, {Mowlavi}, {Nienartowicz}, {Panuzzo}, {Portell}, {Riello},
  {Seabroke}, {Tanga}, {Th{\'e}venin}, {Gracia-Abril}, {Comoretto},
  {Garcia-Reinaldos}, {Teyssier}, {Altmann}, {Andrae}, {Audard},
  {Bellas-Velidis}, {Benson}, {Berthier}, {Blomme}, {Burgess}, {Busso},
  {Carry}, {Cellino}, {Clementini}, {Clotet}, {Creevey}, {Davidson}, {De
  Ridder}, {Delchambre}, {Dell'Oro}, {Ducourant},
  {Fern{\'a}ndez-Hern{\'a}ndez}, {Fouesneau}, {Fr{\'e}mat}, {Galluccio},
  {Garc{\'\i}a-Torres}, {Gonz{\'a}lez-N{\'u}{\~n}ez}, {Gonz{\'a}lez-Vidal},
  {Gosset}, {Guy}, {Halbwachs}, {Hambly}, {Harrison}, {Hern{\'a}ndez},
  {Hestroffer}, {Hodgkin}, {Hutton}, {Jasniewicz}, {Jean-Antoine-Piccolo},
  {Jordan}, {Korn}, {Krone-Martins}, {Lanzafame}, {Lebzelter}, {L{\"o}ffler},
  {Manteiga}, {Marrese}, {Mart{\'\i}n-Fleitas}, {Moitinho}, {Mora}, {Muinonen},
  {Osinde}, {Pancino}, {Pauwels}, {Petit}, {Recio-Blanco}, {Richards},
  {Rimoldini}, {Robin}, {Sarro}, {Siopis}, {Smith}, {Sozzetti}, {S{\"u}veges},
  {Torra}, {van Reeven}, {Abbas}, {Abreu Aramburu}, {Accart}, {Aerts},
  {Altavilla}, {{\'A}lvarez}, {Alvarez}, {Alves}, {Anderson}, {Andrei},
  {Anglada Varela}, {Antiche}, {Antoja}, {Arcay}, {Astraatmadja}, {Bach},
  {Baker}, {Balaguer-N{\'u}{\~n}ez}, {Balm}, {Barache}, {Barata}, {Barbato},
  {Barblan}, {Barklem}, {Barrado}, {Barros}, {Barstow}, {Bartholom{\'e}
  Mu{\~n}oz}, {Bassilana}, {Becciani}, {Bellazzini}, {Berihuete}, {Bertone},
  {Bianchi}, {Bienaym{\'e}}, {Blanco-Cuaresma}, {Boch}, {Boeche}, {Bombrun},
  {Borrachero}, {Bossini}, {Bouquillon}, {Bourda}, {Bragaglia}, {Bramante},
  {Breddels}, {Bressan}, {Brouillet}, {Br{\"u}semeister}, {Brugaletta},
  {Bucciarelli}, {Burlacu}, {Busonero}, {Butkevich}, {Buzzi}, {Caffau},
  {Cancelliere}, {Cannizzaro}, {Cantat-Gaudin}, {Carballo}, {Carlucci},
  {Carrasco}, {Casamiquela}, {Castellani}, {Castro-Ginard}, {Charlot},
  {Chemin}, {Chiavassa}, {Cocozza}, {Costigan}, {Cowell}, {Crifo}, {Crosta},
  {Crowley}, {Cuypers}, {Dafonte}, {Damerdji}, {Dapergolas}, {David}, {David},
  {de Laverny}, {De Luise}, {De March}, {de Martino}, {de Souza}, {de Torres},
  {Debosscher}, {del Pozo}, {Delbo}, {Delgado}, {Delgado}, {Di Matteo},
  {Diakite}, {Diener}, {Distefano}, {Dolding}, {Drazinos}, {Dur{\'a}n},
  {Edvardsson}, {Enke}, {Eriksson}, {Esquej}, {Eynard Bontemps}, {Fabre},
  {Fabrizio}, {Faigler}, {Falc{\~a}o}, {Farr{\`a}s Casas}, {Federici},
  {Fedorets}, {Fernique}, {Figueras}, {Filippi}, {Findeisen}, {Fonti},
  {Fraile}, {Fraser}, {Fr{\'e}zouls}, {Gai}, {Galleti}, {Garabato},
  {Garc{\'\i}a-Sedano}, {Garofalo}, {Garralda}, {Gavel}, {Gavras}, {Gerssen},
  {Geyer}, {Giacobbe}, {Gilmore}, {Girona}, {Giuffrida}, {Glass}, {Gomes},
  {Granvik}, {Gueguen}, {Guerrier}, {Guiraud}, {Guti{\'e}rrez-S{\'a}nchez},
  {Haigron}, {Hatzidimitriou}, {Hauser}, {Haywood}, {Heiter}, {Helmi}, {Heu},
  {Hilger}, {Hobbs}, {Hofmann}, {Holland}, {Huckle}, {Hypki}, {Icardi},
  {Jan{\ss}en}, {Jevardat de Fombelle}, {Jonker}, {Juh{\'a}sz}, {Julbe},
  {Karampelas}, {Kewley}, {Klar}, {Kochoska}, {Kohley}, {Kolenberg},
  {Kontizas}, {Kontizas}, {Koposov}, {Kordopatis}, {Kostrzewa-Rutkowska},
  {Koubsky}, {Lambert}, {Lanza}, {Lasne}, {Lavigne}, {Le Fustec}, {Le
  Poncin-Lafitte}, {Lebreton}, {Leccia}, {Leclerc}, {Lecoeur-Taibi},
  {Lenhardt}, {Leroux}, {Liao}, {Licata}, {Lindstr{\o}m}, {Lister}, {Livanou},
  {Lobel}, {L{\'o}pez}, {Managau}, {Mann}, {Mantelet}, {Marchal}, {Marchant},
  {Marconi}, {Marinoni}, {Marschalk{\'o}}, {Marshall}, {Martino}, {Marton},
  {Mary}, {Massari}, {Matijevi{\v{c}}}, {Mazeh}, {McMillan}, {Messina},
  {Michalik}, {Millar}, {Molina}, {Molinaro}, {Moln{\'a}r}, {Montegriffo},
  {Mor}, {Morbidelli}, {Morel}, {Morris}, {Mulone}, {Muraveva}, {Musella},
  {Nelemans}, {Nicastro}, {Noval}, {O'Mullane}, {Ord{\'e}novic},
  {Ord{\'o}{\~n}ez-Blanco}, {Osborne}, {Pagani}, {Pagano}, {Pailler},
  {Palacin}, {Palaversa}, {Panahi}, {Pawlak}, {Piersimoni}, {Pineau}, {Plachy},
  {Plum}, {Poggio}, {Poujoulet}, {Pr{\v{s}}a}, {Pulone}, {Racero}, {Ragaini},
  {Rambaux}, {Ramos-Lerate}, {Regibo}, {Reyl{\'e}}, {Riclet}, {Ripepi}, {Riva},
  {Rivard}, {Rixon}, {Roegiers}, {Roelens}, {Romero-G{\'o}mez}, {Rowell},
  {Royer}, {Ruiz-Dern}, {Sadowski}, {Sagrist{\`a} Sell{\'e}s}, {Sahlmann},
  {Salgado}, {Salguero}, {Sanna}, {Santana-Ros}, {Sarasso}, {Savietto},
  {Schultheis}, {Sciacca}, {Segol}, {Segovia}, {S{\'e}gransan}, {Shih},
  {Siltala}, {Silva}, {Smart}, {Smith}, {Solano}, {Solitro}, {Sordo}, {Soria
  Nieto}, {Souchay}, {Spagna}, {Spoto}, {Stampa}, {Steele},
  {Steidelm{\"u}ller}, {Stephenson}, {Stoev}, {Suess}, {Surdej}, {Szabados},
  {Szegedi-Elek}, {Tapiador}, {Taris}, {Tauran}, {Taylor}, {Teixeira},
  {Terrett}, {Teyssandier}, {Thuillot}, {Titarenko}, {Torra Clotet}, {Turon},
  {Ulla}, {Utrilla}, {Uzzi}, {Vaillant}, {Valentini}, {Valette}, {van Elteren},
  {Van Hemelryck}, {van Leeuwen}, {Vaschetto}, {Vecchiato}, {Veljanoski},
  {Viala}, {Vicente}, {Vogt}, {von Essen}, {Voss}, {Votruba}, {Voutsinas},
  {Walmsley}, {Weiler}, {Wertz}, {Wevers}, {Wyrzykowski}, {Yoldas},
  {{\v{Z}}erjal}, {Ziaeepour}, {Zorec}, {Zschocke}, {Zucker}, {Zurbach}, \&
  {Zwitter}}]{2018A&A...616A...1G}
{Gaia Collaboration}, {Brown}, A.~G.~A., {Vallenari}, A., {et~al.} 2018, \aap,
  616, A1

\bibitem[{{Gaia Collaboration} {et~al.}(2021){Gaia Collaboration}, {Brown},
  {Vallenari}, {Prusti}, {de Bruijne}, {Babusiaux}, {Biermann}, {Creevey},
  {Evans}, {Eyer}, {Hutton}, {Jansen}, {Jordi}, {Klioner}, {Lammers},
  {Lindegren}, {Luri}, {Mignard}, {Panem}, {Pourbaix}, {Randich}, {Sartoretti},
  {Soubiran}, {Walton}, {Arenou}, {Bailer-Jones}, {Bastian}, {Cropper},
  {Drimmel}, {Katz}, {Lattanzi}, {van Leeuwen}, {Bakker}, {Cacciari},
  {Casta{\~n}eda}, {De Angeli}, {Ducourant}, {Fabricius}, {Fouesneau},
  {Fr{\'e}mat}, {Guerra}, {Guerrier}, {Guiraud}, {Jean-Antoine Piccolo},
  {Masana}, {Messineo}, {Mowlavi}, {Nicolas}, {Nienartowicz}, {Pailler},
  {Panuzzo}, {Riclet}, {Roux}, {Seabroke}, {Sordo}, {Tanga}, {Th{\'e}venin},
  {Gracia-Abril}, {Portell}, {Teyssier}, {Altmann}, {Andrae}, {Bellas-Velidis},
  {Benson}, {Berthier}, {Blomme}, {Brugaletta}, {Burgess}, {Busso}, {Carry},
  {Cellino}, {Cheek}, {Clementini}, {Damerdji}, {Davidson}, {Delchambre},
  {Dell'Oro}, {Fern{\'a}ndez-Hern{\'a}ndez}, {Galluccio}, {Garc{\'\i}a-Lario},
  {Garcia-Reinaldos}, {Gonz{\'a}lez-N{\'u}{\~n}ez}, {Gosset}, {Haigron},
  {Halbwachs}, {Hambly}, {Harrison}, {Hatzidimitriou}, {Heiter},
  {Hern{\'a}ndez}, {Hestroffer}, {Hodgkin}, {Holl}, {Jan{\ss}en}, {Jevardat de
  Fombelle}, {Jordan}, {Krone-Martins}, {Lanzafame}, {L{\"o}ffler}, {Lorca},
  {Manteiga}, {Marchal}, {Marrese}, {Moitinho}, {Mora}, {Muinonen}, {Osborne},
  {Pancino}, {Pauwels}, {Petit}, {Recio-Blanco}, {Richards}, {Riello},
  {Rimoldini}, {Robin}, {Roegiers}, {Rybizki}, {Sarro}, {Siopis}, {Smith},
  {Sozzetti}, {Ulla}, {Utrilla}, {van Leeuwen}, {van Reeven}, {Abbas}, {Abreu
  Aramburu}, {Accart}, {Aerts}, {Aguado}, {Ajaj}, {Altavilla}, {{\'A}lvarez},
  {{\'A}lvarez Cid-Fuentes}, {Alves}, {Anderson}, {Anglada Varela}, {Antoja},
  {Audard}, {Baines}, {Baker}, {Balaguer-N{\'u}{\~n}ez}, {Balbinot}, {Balog},
  {Barache}, {Barbato}, {Barros}, {Barstow}, {Bartolom{\'e}}, {Bassilana},
  {Bauchet}, {Baudesson-Stella}, {Becciani}, {Bellazzini}, {Bernet}, {Bertone},
  {Bianchi}, {Blanco-Cuaresma}, {Boch}, {Bombrun}, {Bossini}, {Bouquillon},
  {Bragaglia}, {Bramante}, {Breedt}, {Bressan}, {Brouillet}, {Bucciarelli},
  {Burlacu}, {Busonero}, {Butkevich}, {Buzzi}, {Caffau}, {Cancelliere},
  {C{\'a}novas}, {Cantat-Gaudin}, {Carballo}, {Carlucci}, {Carnerero},
  {Carrasco}, {Casamiquela}, {Castellani}, {Castro-Ginard}, {Castro Sampol},
  {Chaoul}, {Charlot}, {Chemin}, {Chiavassa}, {Cioni}, {Comoretto}, {Cooper},
  {Cornez}, {Cowell}, {Crifo}, {Crosta}, {Crowley}, {Dafonte}, {Dapergolas},
  {David}, {David}, {de Laverny}, {De Luise}, {De March}, {De Ridder}, {de
  Souza}, {de Teodoro}, {de Torres}, {del Peloso}, {del Pozo}, {Delbo},
  {Delgado}, {Delgado}, {Delisle}, {Di Matteo}, {Diakite}, {Diener},
  {Distefano}, {Dolding}, {Eappachen}, {Edvardsson}, {Enke}, {Esquej}, {Fabre},
  {Fabrizio}, {Faigler}, {Fedorets}, {Fernique}, {Fienga}, {Figueras},
  {Fouron}, {Fragkoudi}, {Fraile}, {Franke}, {Gai}, {Garabato},
  {Garcia-Gutierrez}, {Garc{\'\i}a-Torres}, {Garofalo}, {Gavras}, {Gerlach},
  {Geyer}, {Giacobbe}, {Gilmore}, {Girona}, {Giuffrida}, {Gomel}, {Gomez},
  {Gonzalez-Santamaria}, {Gonz{\'a}lez-Vidal}, {Granvik},
  {Guti{\'e}rrez-S{\'a}nchez}, {Guy}, {Hauser}, {Haywood}, {Helmi}, {Hidalgo},
  {Hilger}, {H{\l}adczuk}, {Hobbs}, {Holland}, {Huckle}, {Jasniewicz},
  {Jonker}, {Juaristi Campillo}, {Julbe}, {Karbevska}, {Kervella}, {Khanna},
  {Kochoska}, {Kontizas}, {Kordopatis}, {Korn}, {Kostrzewa-Rutkowska},
  {Kruszy{\'n}ska}, {Lambert}, {Lanza}, {Lasne}, {Le Campion}, {Le Fustec},
  {Lebreton}, {Lebzelter}, {Leccia}, {Leclerc}, {Lecoeur-Taibi}, {Liao},
  {Licata}, {Lindstr{\o}m}, {Lister}, {Livanou}, {Lobel}, {Madrero Pardo},
  {Managau}, {Mann}, {Marchant}, {Marconi}, {Marcos Santos}, {Marinoni},
  {Marocco}, {Marshall}, {Martin Polo}, {Mart{\'\i}n-Fleitas}, {Masip},
  {Massari}, {Mastrobuono-Battisti}, {Mazeh}, {McMillan}, {Messina},
  {Michalik}, {Millar}, {Mints}, {Molina}, {Molinaro}, {Moln{\'a}r},
  {Montegriffo}, {Mor}, {Morbidelli}, {Morel}, {Morris}, {Mulone}, {Munoz},
  {Muraveva}, {Murphy}, {Musella}, {Noval}, {Ord{\'e}novic}, {Orr{\`u}},
  {Osinde}, {Pagani}, {Pagano}, {Palaversa}, {Palicio}, {Panahi}, {Pawlak},
  {Pe{\~n}alosa Esteller}, {Penttil{\"a}}, {Piersimoni}, {Pineau}, {Plachy},
  {Plum}, {Poggio}, {Poretti}, {Poujoulet}, {Pr{\v{s}}a}, {Pulone}, {Racero},
  {Ragaini}, {Rainer}, {Raiteri}, {Rambaux}, {Ramos}, {Ramos-Lerate}, {Re
  Fiorentin}, {Regibo}, {Reyl{\'e}}, {Ripepi}, {Riva}, {Rixon}, {Robichon},
  {Robin}, {Roelens}, {Rohrbasser}, {Romero-G{\'o}mez}, {Rowell}, {Royer},
  {Rybicki}, {Sadowski}, {Sagrist{\`a} Sell{\'e}s}, {Sahlmann}, {Salgado},
  {Salguero}, {Samaras}, {Sanchez Gimenez}, {Sanna}, {Santove{\~n}a},
  {Sarasso}, {Schultheis}, {Sciacca}, {Segol}, {Segovia}, {S{\'e}gransan},
  {Semeux}, {Shahaf}, {Siddiqui}, {Siebert}, {Siltala}, {Slezak}, {Smart},
  {Solano}, {Solitro}, {Souami}, {Souchay}, {Spagna}, {Spoto}, {Steele},
  {Steidelm{\"u}ller}, {Stephenson}, {S{\"u}veges}, {Szabados}, {Szegedi-Elek},
  {Taris}, {Tauran}, {Taylor}, {Teixeira}, {Thuillot}, {Tonello}, {Torra},
  {Torra}, {Turon}, {Unger}, {Vaillant}, {van Dillen}, {Vanel}, {Vecchiato},
  {Viala}, {Vicente}, {Voutsinas}, {Weiler}, {Wevers}, {Wyrzykowski}, {Yoldas},
  {Yvard}, {Zhao}, {Zorec}, {Zucker}, {Zurbach}, \&
  {Zwitter}}]{2021A&A...649A...1G}
{Gaia Collaboration}, {Brown}, A.~G.~A., {Vallenari}, A., {et~al.} 2021, \aap,
  649, A1

\bibitem[{{Gaia Collaboration} {et~al.}(2016){Gaia Collaboration}, {Prusti},
  {de Bruijne}, {Brown}, {Vallenari}, {Babusiaux}, {Bailer-Jones}, {Bastian},
  {Biermann}, {Evans}, {Eyer}, {Jansen}, {Jordi}, {Klioner}, {Lammers},
  {Lindegren}, {Luri}, {Mignard}, {Milligan}, {Panem}, {Poinsignon},
  {Pourbaix}, {Randich}, {Sarri}, {Sartoretti}, {Siddiqui}, {Soubiran},
  {Valette}, {van Leeuwen}, {Walton}, {Aerts}, {Arenou}, {Cropper}, {Drimmel},
  {H{\o}g}, {Katz}, {Lattanzi}, {O'Mullane}, {Grebel}, {Holland}, {Huc},
  {Passot}, {Bramante}, {Cacciari}, {Casta{\~n}eda}, {Chaoul}, {Cheek}, {De
  Angeli}, {Fabricius}, {Guerra}, {Hern{\'a}ndez}, {Jean-Antoine-Piccolo},
  {Masana}, {Messineo}, {Mowlavi}, {Nienartowicz}, {Ord{\'o}{\~n}ez-Blanco},
  {Panuzzo}, {Portell}, {Richards}, {Riello}, {Seabroke}, {Tanga},
  {Th{\'e}venin}, {Torra}, {Els}, {Gracia-Abril}, {Comoretto},
  {Garcia-Reinaldos}, {Lock}, {Mercier}, {Altmann}, {Andrae}, {Astraatmadja},
  {Bellas-Velidis}, {Benson}, {Berthier}, {Blomme}, {Busso}, {Carry},
  {Cellino}, {Clementini}, {Cowell}, {Creevey}, {Cuypers}, {Davidson}, {De
  Ridder}, {de Torres}, {Delchambre}, {Dell'Oro}, {Ducourant}, {Fr{\'e}mat},
  {Garc{\'\i}a-Torres}, {Gosset}, {Halbwachs}, {Hambly}, {Harrison}, {Hauser},
  {Hestroffer}, {Hodgkin}, {Huckle}, {Hutton}, {Jasniewicz}, {Jordan},
  {Kontizas}, {Korn}, {Lanzafame}, {Manteiga}, {Moitinho}, {Muinonen},
  {Osinde}, {Pancino}, {Pauwels}, {Petit}, {Recio-Blanco}, {Robin}, {Sarro},
  {Siopis}, {Smith}, {Smith}, {Sozzetti}, {Thuillot}, {van Reeven}, {Viala},
  {Abbas}, {Abreu Aramburu}, {Accart}, {Aguado}, {Allan}, {Allasia},
  {Altavilla}, {{\'A}lvarez}, {Alves}, {Anderson}, {Andrei}, {Anglada Varela},
  {Antiche}, {Antoja}, {Ant{\'o}n}, {Arcay}, {Atzei}, {Ayache}, {Bach},
  {Baker}, {Balaguer-N{\'u}{\~n}ez}, {Barache}, {Barata}, {Barbier}, {Barblan},
  {Baroni}, {Barrado y Navascu{\'e}s}, {Barros}, {Barstow}, {Becciani},
  {Bellazzini}, {Bellei}, {Bello Garc{\'\i}a}, {Belokurov}, {Bendjoya},
  {Berihuete}, {Bianchi}, {Bienaym{\'e}}, {Billebaud}, {Blagorodnova},
  {Blanco-Cuaresma}, {Boch}, {Bombrun}, {Borrachero}, {Bouquillon}, {Bourda},
  {Bouy}, {Bragaglia}, {Breddels}, {Brouillet}, {Br{\"u}semeister},
  {Bucciarelli}, {Budnik}, {Burgess}, {Burgon}, {Burlacu}, {Busonero}, {Buzzi},
  {Caffau}, {Cambras}, {Campbell}, {Cancelliere}, {Cantat-Gaudin}, {Carlucci},
  {Carrasco}, {Castellani}, {Charlot}, {Charnas}, {Charvet}, {Chassat},
  {Chiavassa}, {Clotet}, {Cocozza}, {Collins}, {Collins}, {Costigan}, {Crifo},
  {Cross}, {Crosta}, {Crowley}, {Dafonte}, {Damerdji}, {Dapergolas}, {David},
  {David}, {De Cat}, {de Felice}, {de Laverny}, {De Luise}, {De March}, {de
  Martino}, {de Souza}, {Debosscher}, {del Pozo}, {Delbo}, {Delgado},
  {Delgado}, {di Marco}, {Di Matteo}, {Diakite}, {Distefano}, {Dolding}, {Dos
  Anjos}, {Drazinos}, {Dur{\'a}n}, {Dzigan}, {Ecale}, {Edvardsson}, {Enke},
  {Erdmann}, {Escolar}, {Espina}, {Evans}, {Eynard Bontemps}, {Fabre},
  {Fabrizio}, {Faigler}, {Falc{\~a}o}, {Farr{\`a}s Casas}, {Faye}, {Federici},
  {Fedorets}, {Fern{\'a}ndez-Hern{\'a}ndez}, {Fernique}, {Fienga}, {Figueras},
  {Filippi}, {Findeisen}, {Fonti}, {Fouesneau}, {Fraile}, {Fraser}, {Fuchs},
  {Furnell}, {Gai}, {Galleti}, {Galluccio}, {Garabato}, {Garc{\'\i}a-Sedano},
  {Gar{\'e}}, {Garofalo}, {Garralda}, {Gavras}, {Gerssen}, {Geyer}, {Gilmore},
  {Girona}, {Giuffrida}, {Gomes}, {Gonz{\'a}lez-Marcos},
  {Gonz{\'a}lez-N{\'u}{\~n}ez}, {Gonz{\'a}lez-Vidal}, {Granvik}, {Guerrier},
  {Guillout}, {Guiraud}, {G{\'u}rpide}, {Guti{\'e}rrez-S{\'a}nchez}, {Guy},
  {Haigron}, {Hatzidimitriou}, {Haywood}, {Heiter}, {Helmi}, {Hobbs},
  {Hofmann}, {Holl}, {Holland}, {Hunt}, {Hypki}, {Icardi}, {Irwin}, {Jevardat
  de Fombelle}, {Jofr{\'e}}, {Jonker}, {Jorissen}, {Julbe}, {Karampelas},
  {Kochoska}, {Kohley}, {Kolenberg}, {Kontizas}, {Koposov}, {Kordopatis},
  {Koubsky}, {Kowalczyk}, {Krone-Martins}, {Kudryashova}, {Kull}, {Bachchan},
  {Lacoste-Seris}, {Lanza}, {Lavigne}, {Le Poncin-Lafitte}, {Lebreton},
  {Lebzelter}, {Leccia}, {Leclerc}, {Lecoeur-Taibi}, {Lemaitre}, {Lenhardt},
  {Leroux}, {Liao}, {Licata}, {Lindstr{\o}m}, {Lister}, {Livanou}, {Lobel},
  {L{\"o}ffler}, {L{\'o}pez}, {Lopez-Lozano}, {Lorenz}, {Loureiro},
  {MacDonald}, {Magalh{\~a}es Fernandes}, {Managau}, {Mann}, {Mantelet},
  {Marchal}, {Marchant}, {Marconi}, {Marie}, {Marinoni}, {Marrese},
  {Marschalk{\'o}}, {Marshall}, {Mart{\'\i}n-Fleitas}, {Martino}, {Mary},
  {Matijevi{\v{c}}}, {Mazeh}, {McMillan}, {Messina}, {Mestre}, {Michalik},
  {Millar}, {Miranda}, {Molina}, {Molinaro}, {Molinaro}, {Moln{\'a}r},
  {Moniez}, {Montegriffo}, {Monteiro}, {Mor}, {Mora}, {Morbidelli}, {Morel},
  {Morgenthaler}, {Morley}, {Morris}, {Mulone}, {Muraveva}, {Musella},
  {Narbonne}, {Nelemans}, {Nicastro}, {Noval}, {Ord{\'e}novic},
  {Ordieres-Mer{\'e}}, {Osborne}, {Pagani}, {Pagano}, {Pailler}, {Palacin},
  {Palaversa}, {Parsons}, {Paulsen}, {Pecoraro}, {Pedrosa}, {Pentik{\"a}inen},
  {Pereira}, {Pichon}, {Piersimoni}, {Pineau}, {Plachy}, {Plum}, {Poujoulet},
  {Pr{\v{s}}a}, {Pulone}, {Ragaini}, {Rago}, {Rambaux}, {Ramos-Lerate},
  {Ranalli}, {Rauw}, {Read}, {Regibo}, {Renk}, {Reyl{\'e}}, {Ribeiro},
  {Rimoldini}, {Ripepi}, {Riva}, {Rixon}, {Roelens}, {Romero-G{\'o}mez},
  {Rowell}, {Royer}, {Rudolph}, {Ruiz-Dern}, {Sadowski}, {Sagrist{\`a}
  Sell{\'e}s}, {Sahlmann}, {Salgado}, {Salguero}, {Sarasso}, {Savietto},
  {Schnorhk}, {Schultheis}, {Sciacca}, {Segol}, {Segovia}, {Segransan},
  {Serpell}, {Shih}, {Smareglia}, {Smart}, {Smith}, {Solano}, {Solitro},
  {Sordo}, {Soria Nieto}, {Souchay}, {Spagna}, {Spoto}, {Stampa}, {Steele},
  {Steidelm{\"u}ller}, {Stephenson}, {Stoev}, {Suess}, {S{\"u}veges}, {Surdej},
  {Szabados}, {Szegedi-Elek}, {Tapiador}, {Taris}, {Tauran}, {Taylor},
  {Teixeira}, {Terrett}, {Tingley}, {Trager}, {Turon}, {Ulla}, {Utrilla},
  {Valentini}, {van Elteren}, {Van Hemelryck}, {van Leeuwen}, {Varadi},
  {Vecchiato}, {Veljanoski}, {Via}, {Vicente}, {Vogt}, {Voss}, {Votruba},
  {Voutsinas}, {Walmsley}, {Weiler}, {Weingrill}, {Werner}, {Wevers},
  {Whitehead}, {Wyrzykowski}, {Yoldas}, {{\v{Z}}erjal}, {Zucker}, {Zurbach},
  {Zwitter}, {Alecu}, {Allen}, {Allende Prieto}, {Amorim},
  {Anglada-Escud{\'e}}, {Arsenijevic}, {Azaz}, {Balm}, {Beck}, {Bernstein},
  {Bigot}, {Bijaoui}, {Blasco}, {Bonfigli}, {Bono}, {Boudreault}, {Bressan},
  {Brown}, {Brunet}, {Bunclark}, {Buonanno}, {Butkevich}, {Carret}, {Carrion},
  {Chemin}, {Ch{\'e}reau}, {Corcione}, {Darmigny}, {de Boer}, {de Teodoro}, {de
  Zeeuw}, {Delle Luche}, {Domingues}, {Dubath}, {Fodor}, {Fr{\'e}zouls},
  {Fries}, {Fustes}, {Fyfe}, {Gallardo}, {Gallegos}, {Gardiol}, {Gebran},
  {Gomboc}, {G{\'o}mez}, {Grux}, {Gueguen}, {Heyrovsky}, {Hoar}, {Iannicola},
  {Isasi Parache}, {Janotto}, {Joliet}, {Jonckheere}, {Keil}, {Kim},
  {Klagyivik}, {Klar}, {Knude}, {Kochukhov}, {Kolka}, {Kos}, {Kutka}, {Lainey},
  {LeBouquin}, {Liu}, {Loreggia}, {Makarov}, {Marseille}, {Martayan},
  {Martinez-Rubi}, {Massart}, {Meynadier}, {Mignot}, {Munari}, {Nguyen},
  {Nordlander}, {Ocvirk}, {O'Flaherty}, {Olias Sanz}, {Ortiz}, {Osorio},
  {Oszkiewicz}, {Ouzounis}, {Palmer}, {Park}, {Pasquato}, {Peltzer}, {Peralta},
  {P{\'e}turaud}, {Pieniluoma}, {Pigozzi}, {Poels}, {Prat}, {Prod'homme},
  {Raison}, {Rebordao}, {Risquez}, {Rocca-Volmerange}, {Rosen}, {Ruiz-Fuertes},
  {Russo}, {Sembay}, {Serraller Vizcaino}, {Short}, {Siebert}, {Silva},
  {Sinachopoulos}, {Slezak}, {Soffel}, {Sosnowska}, {Strai{\v{z}}ys}, {ter
  Linden}, {Terrell}, {Theil}, {Tiede}, {Troisi}, {Tsalmantza}, {Tur},
  {Vaccari}, {Vachier}, {Valles}, {Van Hamme}, {Veltz}, {Virtanen}, {Wallut},
  {Wichmann}, {Wilkinson}, {Ziaeepour}, \& {Zschocke}}]{2016A&A...595A...1G}
{Gaia Collaboration}, {Prusti}, T., {de Bruijne}, J.~H.~J., {et~al.} 2016,
  \aap, 595, A1

\bibitem[{{Gratton} {et~al.}(2012){Gratton}, {Carretta}, \&
  {Bragaglia}}]{2012A&ARv..20...50G}
{Gratton}, R.~G., {Carretta}, E., \& {Bragaglia}, A. 2012, \aapr, 20, 50

\bibitem[{{Grillmair}(2022)}]{2022ApJ...929...89G}
{Grillmair}, C.~J. 2022, \apj, 929, 89

\bibitem[{{Grillmair} \& {Dionatos}(2006)}]{2006ApJ...643L..17G}
{Grillmair}, C.~J. \& {Dionatos}, O. 2006, \apjl, 643, L17

\bibitem[{{Hanke} {et~al.}(2020){Hanke}, {Koch}, {Prudil}, {Grebel}, \&
  {Bastian}}]{2020A&A...637A..98H}
{Hanke}, M., {Koch}, A., {Prudil}, Z., {Grebel}, E.~K., \& {Bastian}, U. 2020,
  \aap, 637, A98

\bibitem[{{Harris}(1996)}]{1996AJ....112.1487H}
{Harris}, W.~E. 1996, \aj, 112, 1487

\bibitem[{{Harris}(2010)}]{2010arXiv1012.3224H}
{Harris}, W.~E. 2010, arXiv e-prints, arXiv:1012.3224

\bibitem[{{Helmi} {et~al.}(2018){Helmi}, {Babusiaux}, {Koppelman}, {Massari},
  {Veljanoski}, \& {Brown}}]{2018Natur.563...85H}
{Helmi}, A., {Babusiaux}, C., {Koppelman}, H.~H., {et~al.} 2018, \nat, 563, 85

\bibitem[{{Helmi} {et~al.}(1999){Helmi}, {White}, {de Zeeuw}, \&
  {Zhao}}]{1999Natur.402...53H}
{Helmi}, A., {White}, S. D.~M., {de Zeeuw}, P.~T., \& {Zhao}, H. 1999, \nat,
  402, 53

\bibitem[{{Huang} {et~al.}(2015){Huang}, {Liu}, {Yuan}, {Xiang}, {Huo}, {Chen},
  {Zhang}, \& {Hou}}]{2015MNRAS.449..162H}
{Huang}, Y., {Liu}, X.~W., {Yuan}, H.~B., {et~al.} 2015, \mnras, 449, 162

\bibitem[{{Huang} {et~al.}(2016){Huang}, {Liu}, {Yuan}, {Xiang}, {Zhang},
  {Chen}, {Ren}, {Wang}, {Zhang}, {Hou}, {Wang}, \&
  {Cao}}]{2016MNRAS.463.2623H}
{Huang}, Y., {Liu}, X.~W., {Yuan}, H.~B., {et~al.} 2016, \mnras, 463, 2623

\bibitem[{{Ibata} {et~al.}(2021){Ibata}, {Malhan}, {Martin}, {Aubert},
  {Famaey}, {Bianchini}, {Monari}, {Siebert}, {Thomas}, {Bellazzini},
  {Bonifacio}, {Caffau}, \& {Renaud}}]{2021ApJ...914..123I}
{Ibata}, R., {Malhan}, K., {Martin}, N., {et~al.} 2021, \apj, 914, 123

\bibitem[{{Ibata} {et~al.}(2019){Ibata}, {Bellazzini}, {Malhan}, {Martin}, \&
  {Bianchini}}]{2019NatAs...3..667I}
{Ibata}, R.~A., {Bellazzini}, M., {Malhan}, K., {Martin}, N., \& {Bianchini},
  P. 2019, Nature Astronomy, 3, 667

\bibitem[{{Ibata} {et~al.}(1994){Ibata}, {Gilmore}, \&
  {Irwin}}]{1994Natur.370..194I}
{Ibata}, R.~A., {Gilmore}, G., \& {Irwin}, M.~J. 1994, \nat, 370, 194

\bibitem[{{Johnson} \& {Pilachowski}(2012)}]{2012ApJ...754L..38J}
{Johnson}, C.~I. \& {Pilachowski}, C.~A. 2012, \apjl, 754, L38

\bibitem[{{Jordi} \& {Grebel}(2010)}]{2010A&A...522A..71J}
{Jordi}, K. \& {Grebel}, E.~K. 2010, \aap, 522, A71

\bibitem[{{Koch} {et~al.}(2019){Koch}, {Grebel}, \&
  {Martell}}]{2019A&A...625A..75K}
{Koch}, A., {Grebel}, E.~K., \& {Martell}, S.~L. 2019, \aap, 625, A75

\bibitem[{{Kundu} {et~al.}(2019){Kundu}, {Fern{\'a}ndez-Trincado}, {Minniti},
  {Singh}, {Moreno}, {Reyl{\'e}}, {Robin}, \& {Soto}}]{2019MNRAS.489.4565K}
{Kundu}, R., {Fern{\'a}ndez-Trincado}, J.~G., {Minniti}, D., {et~al.} 2019,
  \mnras, 489, 4565

\bibitem[{{Kundu} {et~al.}(2021){Kundu}, {Navarrete}, {Fern{\'a}ndez-Trincado},
  {Minniti}, {Singh}, {Sbordone}, {Piatti}, \&
  {Reyl{\'e}}}]{2021A&A...645A.116K}
{Kundu}, R., {Navarrete}, C., {Fern{\'a}ndez-Trincado}, J.~G., {et~al.} 2021,
  \aap, 645, A116

\bibitem[{{Kundu} {et~al.}(2022){Kundu}, {Navarrete}, {Sbordone},
  {Carballo-Bello}, {Fern{\'a}ndez-Trincado}, {Minniti}, \&
  {Singh}}]{2022A&A...665A...8K}
{Kundu}, R., {Navarrete}, C., {Sbordone}, L., {et~al.} 2022, \aap, 665, A8

\bibitem[{{K{\"u}pper} {et~al.}(2010){K{\"u}pper}, {Kroupa}, {Baumgardt}, \&
  {Heggie}}]{2010MNRAS.407.2241K}
{K{\"u}pper}, A. H.~W., {Kroupa}, P., {Baumgardt}, H., \& {Heggie}, D.~C. 2010,
  \mnras, 407, 2241

\bibitem[{{Kuzma} {et~al.}(2018){Kuzma}, {Da Costa}, \&
  {Mackey}}]{2018MNRAS.473.2881K}
{Kuzma}, P.~B., {Da Costa}, G.~S., \& {Mackey}, A.~D. 2018, \mnras, 473, 2881

\bibitem[{{Kuzma} {et~al.}(2016){Kuzma}, {Da Costa}, {Mackey}, \&
  {Roderick}}]{2016MNRAS.461.3639K}
{Kuzma}, P.~B., {Da Costa}, G.~S., {Mackey}, A.~D., \& {Roderick}, T.~A. 2016,
  \mnras, 461, 3639

\bibitem[{{Li} {et~al.}(2021){Li}, {Tang}, {Milone}, {de Grijs}, {Hong},
  {Yang}, \& {Wang}}]{2021ApJ...906..133L}
{Li}, C., {Tang}, B., {Milone}, A.~P., {et~al.} 2021, \apj, 906, 133

\bibitem[{{Limberg} {et~al.}(2022){Limberg}, {Souza}, {P{\'e}rez-Villegas},
  {Rossi}, {Perottoni}, \& {Santucci}}]{2022ApJ...935..109L}
{Limberg}, G., {Souza}, S.~O., {P{\'e}rez-Villegas}, A., {et~al.} 2022, \apj,
  935, 109

\bibitem[{{Lind} {et~al.}(2015){Lind}, {Koposov}, {Battistini}, {Marino},
  {Ruchti}, {Serenelli}, {Worley}, {Alves-Brito}, {Asplund}, {Barklem},
  {Bensby}, {Bergemann}, {Blanco-Cuaresma}, {Bragaglia}, {Edvardsson},
  {Feltzing}, {Gruyters}, {Heiter}, {Jofre}, {Korn}, {Nordlander}, {Ryde},
  {Soubiran}, {Gilmore}, {Randich}, {Ferguson}, {Jeffries}, {Vallenari},
  {Allende Prieto}, {Pancino}, {Recio-Blanco}, {Romano}, {Smiljanic},
  {Bellazzini}, {Damiani}, {Hill}, {de Laverny}, {Jackson}, {Lardo}, \&
  {Zaggia}}]{2015A&A...575L..12L}
{Lind}, K., {Koposov}, S.~E., {Battistini}, C., {et~al.} 2015, \aap, 575, L12

\bibitem[{{Majewski} {et~al.}(2017){Majewski}, {Schiavon}, {Frinchaboy},
  {Allende Prieto}, {Barkhouser}, {Bizyaev}, {Blank}, {Brunner}, {Burton},
  {Carrera}, {Chojnowski}, {Cunha}, {Epstein}, {Fitzgerald}, {Garc{\'\i}a
  P{\'e}rez}, {Hearty}, {Henderson}, {Holtzman}, {Johnson}, {Lam}, {Lawler},
  {Maseman}, {M{\'e}sz{\'a}ros}, {Nelson}, {Nguyen}, {Nidever}, {Pinsonneault},
  {Shetrone}, {Smee}, {Smith}, {Stolberg}, {Skrutskie}, {Walker}, {Wilson},
  {Zasowski}, {Anders}, {Basu}, {Beland}, {Blanton}, {Bovy}, {Brownstein},
  {Carlberg}, {Chaplin}, {Chiappini}, {Eisenstein}, {Elsworth}, {Feuillet},
  {Fleming}, {Galbraith-Frew}, {Garc{\'\i}a}, {Garc{\'\i}a-Hern{\'a}ndez},
  {Gillespie}, {Girardi}, {Gunn}, {Hasselquist}, {Hayden}, {Hekker}, {Ivans},
  {Kinemuchi}, {Klaene}, {Mahadevan}, {Mathur}, {Mosser}, {Muna}, {Munn},
  {Nichol}, {O'Connell}, {Parejko}, {Robin}, {Rocha-Pinto}, {Schultheis},
  {Serenelli}, {Shane}, {Silva Aguirre}, {Sobeck}, {Thompson}, {Troup},
  {Weinberg}, \& {Zamora}}]{2017AJ....154...94M}
{Majewski}, S.~R., {Schiavon}, R.~P., {Frinchaboy}, P.~M., {et~al.} 2017, \aj,
  154, 94

\bibitem[{{Malhan} {et~al.}(2018){Malhan}, {Ibata}, \&
  {Martin}}]{2018MNRAS.481.3442M}
{Malhan}, K., {Ibata}, R.~A., \& {Martin}, N.~F. 2018, \mnras, 481, 3442

\bibitem[{{Malhan} {et~al.}(2022){Malhan}, {Ibata}, {Sharma}, {Famaey},
  {Bellazzini}, {Carlberg}, {D'Souza}, {Yuan}, {Martin}, \&
  {Thomas}}]{2022ApJ...926..107M}
{Malhan}, K., {Ibata}, R.~A., {Sharma}, S., {et~al.} 2022, \apj, 926, 107

\bibitem[{{Marigo} {et~al.}(2017){Marigo}, {Girardi}, {Bressan}, {Rosenfield},
  {Aringer}, {Chen}, {Dussin}, {Nanni}, {Pastorelli}, {Rodrigues}, {Trabucchi},
  {Bladh}, {Dalcanton}, {Groenewegen}, {Montalb{\'a}n}, \&
  {Wood}}]{2017ApJ...835...77M}
{Marigo}, P., {Girardi}, L., {Bressan}, A., {et~al.} 2017, \apj, 835, 77

\bibitem[{{Marino} {et~al.}(2014){Marino}, {Milone}, {Yong}, {Dotter}, {Da
  Costa}, {Asplund}, {Jerjen}, {Mackey}, {Norris}, {Cassisi}, {Sbordone},
  {Stetson}, {Weiss}, {Aparicio}, {Bedin}, {Lind}, {Monelli}, {Piotto},
  {Angeloni}, \& {Buonanno}}]{2014MNRAS.442.3044M}
{Marino}, A.~F., {Milone}, A.~P., {Yong}, D., {et~al.} 2014, \mnras, 442, 3044

\bibitem[{{Martell} \& {Grebel}(2010)}]{2010A&A...519A..14M}
{Martell}, S.~L. \& {Grebel}, E.~K. 2010, \aap, 519, A14

\bibitem[{{Martell} {et~al.}(2016){Martell}, {Shetrone}, {Lucatello},
  {Schiavon}, {M{\'e}sz{\'a}ros}, {Allende Prieto},
  {Garc{\'\i}a-Hern{\'a}ndez}, {Beers}, \& {Nidever}}]{2016ApJ...825..146M}
{Martell}, S.~L., {Shetrone}, M.~D., {Lucatello}, S., {et~al.} 2016, \apj, 825,
  146

\bibitem[{{Martell} {et~al.}(2011){Martell}, {Smolinski}, {Beers}, \&
  {Grebel}}]{2011A&A...534A.136M}
{Martell}, S.~L., {Smolinski}, J.~P., {Beers}, T.~C., \& {Grebel}, E.~K. 2011,
  \aap, 534, A136

\bibitem[{{Massari} {et~al.}(2019){Massari}, {Koppelman}, \&
  {Helmi}}]{2019A&A...630L...4M}
{Massari}, D., {Koppelman}, H.~H., \& {Helmi}, A. 2019, \aap, 630, L4

\bibitem[{{Mateu}(2023)}]{2023MNRAS.520.5225M}
{Mateu}, C. 2023, \mnras, 520, 5225

\bibitem[{{M{\'e}sz{\'a}ros} {et~al.}(2015){M{\'e}sz{\'a}ros}, {Martell},
  {Shetrone}, {Lucatello}, {Troup}, {Bovy}, {Cunha},
  {Garc{\'\i}a-Hern{\'a}ndez}, {Overbeek}, {Allende Prieto}, {Beers},
  {Frinchaboy}, {Garc{\'\i}a P{\'e}rez}, {Hearty}, {Holtzman}, {Majewski},
  {Nidever}, {Schiavon}, {Schneider}, {Sobeck}, {Smith}, {Zamora}, \&
  {Zasowski}}]{2015AJ....149..153M}
{M{\'e}sz{\'a}ros}, S., {Martell}, S.~L., {Shetrone}, M., {et~al.} 2015, \aj,
  149, 153

\bibitem[{{Milone} {et~al.}(2015{\natexlab{a}}){Milone}, {Marino}, {Piotto},
  {Bedin}, {Anderson}, {Renzini}, {King}, {Bellini}, {Brown}, {Cassisi},
  {D'Antona}, {Jerjen}, {Nardiello}, {Salaris}, {Marel}, {Vesperini}, {Yong},
  {Aparicio}, {Sarajedini}, \& {Zoccali}}]{2015MNRAS.447..927M}
{Milone}, A.~P., {Marino}, A.~F., {Piotto}, G., {et~al.} 2015{\natexlab{a}},
  \mnras, 447, 927

\bibitem[{{Milone} {et~al.}(2015{\natexlab{b}}){Milone}, {Marino}, {Piotto},
  {Renzini}, {Bedin}, {Anderson}, {Cassisi}, {D'Antona}, {Bellini}, {Jerjen},
  {Pietrinferni}, \& {Ventura}}]{2015ApJ...808...51M}
{Milone}, A.~P., {Marino}, A.~F., {Piotto}, G., {et~al.} 2015{\natexlab{b}},
  \apj, 808, 51

\bibitem[{{Myeong} {et~al.}(2019){Myeong}, {Vasiliev}, {Iorio}, {Evans}, \&
  {Belokurov}}]{2019MNRAS.488.1235M}
{Myeong}, G.~C., {Vasiliev}, E., {Iorio}, G., {Evans}, N.~W., \& {Belokurov},
  V. 2019, \mnras, 488, 1235

\bibitem[{{Odenkirchen} {et~al.}(2001){Odenkirchen}, {Grebel}, {Rockosi},
  {Dehnen}, {Ibata}, {Rix}, {Stolte}, {Wolf}, {Anderson}, {Bahcall},
  {Brinkmann}, {Csabai}, {Hennessy}, {Hindsley}, {Ivezi{\'c}}, {Lupton},
  {Munn}, {Pier}, {Stoughton}, \& {York}}]{2001ApJ...548L.165O}
{Odenkirchen}, M., {Grebel}, E.~K., {Rockosi}, C.~M., {et~al.} 2001, \apjl,
  548, L165

\bibitem[{{Olszewski} {et~al.}(2009){Olszewski}, {Saha}, {Knezek},
  {Subramaniam}, {de Boer}, \& {Seitzer}}]{2009AJ....138.1570O}
{Olszewski}, E.~W., {Saha}, A., {Knezek}, P., {et~al.} 2009, \aj, 138, 1570

\bibitem[{{Pedregosa} {et~al.}(2011){Pedregosa}, {Varoquaux}, {Gramfort},
  {Michel}, {Thirion}, {Grisel}, {Blondel}, {M{\"u}ller}, {Nothman}, {Louppe},
  {Prettenhofer}, {Weiss}, {Dubourg}, {Vanderplas}, {Passos}, {Cournapeau},
  {Brucher}, {Perrot}, \& {Duchesnay}}]{2011JMLR...12.2825P}
{Pedregosa}, F., {Varoquaux}, G., {Gramfort}, A., {et~al.} 2011, Journal of
  Machine Learning Research, 12, 2825

\bibitem[{{Piatti} \& {Carballo-Bello}(2019)}]{2019MNRAS.485.1029P}
{Piatti}, A.~E. \& {Carballo-Bello}, J.~A. 2019, \mnras, 485, 1029

\bibitem[{{Piotto} {et~al.}(2012){Piotto}, {Milone}, {Anderson}, {Bedin},
  {Bellini}, {Cassisi}, {Marino}, {Aparicio}, \&
  {Nascimbeni}}]{2012ApJ...760...39P}
{Piotto}, G., {Milone}, A.~P., {Anderson}, J., {et~al.} 2012, \apj, 760, 39

\bibitem[{{Randich} {et~al.}(2022){Randich}, {Gilmore}, {Magrini}, {Sacco},
  {Jackson}, {Jeffries}, {Worley}, {Hourihane}, {Gonneau}, {Viscasillas
  Vazquez}, {Franciosini}, {Lewis}, {Alfaro}, {Allende Prieto}, {Bensby},
  {Blomme}, {Bragaglia}, {Flaccomio}, {Fran{\c{c}}ois}, {Irwin}, {Koposov},
  {Korn}, {Lanzafame}, {Pancino}, {Recio-Blanco}, {Smiljanic}, {Van Eck},
  {Zwitter}, {Asplund}, {Bonifacio}, {Feltzing}, {Binney}, {Drew}, {Ferguson},
  {Micela}, {Negueruela}, {Prusti}, {Rix}, {Vallenari}, {Bayo}, {Bergemann},
  {Biazzo}, {Carraro}, {Casey}, {Damiani}, {Frasca}, {Heiter}, {Hill},
  {Jofr{\'e}}, {de Laverny}, {Lind}, {Marconi}, {Martayan}, {Masseron},
  {Monaco}, {Morbidelli}, {Prisinzano}, {Sbordone}, {Sousa}, {Zaggia},
  {Adibekyan}, {Bonito}, {Caffau}, {Daflon}, {Feuillet}, {Gebran}, {Gonzalez
  Hernandez}, {Guiglion}, {Herrero}, {Lobel}, {Maiz Apellaniz}, {Merle},
  {Mikolaitis}, {Montes}, {Morel}, {Soubiran}, {Spina}, {Tabernero},
  {Tautvai{\v{s}}iene}, {Traven}, {Valentini}, {Van der Swaelmen}, {Villanova},
  {Wright}, {Abbas}, {Aguirre B{\o}rsen-Koch}, {Alves}, {Balaguer-Nunez},
  {Barklem}, {Barrado}, {Berlanas}, {Binks}, {Bressan}, {Capuzzo-Dolcetta},
  {Casagrande}, {Casamiquela}, {Collins}, {D'Orazi}, {Dantas}, {Debattista},
  {Delgado-Mena}, {Di Marcantonio}, {Drazdauskas}, {Evans}, {Famaey},
  {Franchini}, {Fr{\'e}mat}, {Friel}, {Fu}, {Geisler}, {Gerhard}, {Gonzalez
  Solares}, {Grebel}, {Gutierrez Albarran}, {Hatzidimitriou}, {Held},
  {Jim{\'e}nez-Esteban}, {J{\"o}nsson}, {Jordi}, {Khachaturyants},
  {Kordopatis}, {Kos}, {Lagarde}, {Mahy}, {Mapelli}, {Marfil}, {Martell},
  {Messina}, {Miglio}, {Minchev}, {Moitinho}, {Montalban}, {Monteiro},
  {Morossi}, {Mowlavi}, {Mucciarelli}, {Murphy}, {Nardetto}, {Ortolani},
  {Paletou}, {Palou{\v{s}}}, {Paunzen}, {Pickering}, {Quirrenbach}, {Re
  Fiorentin}, {Read}, {Romano}, {Ryde}, {Sanna}, {Santos}, {Seabroke},
  {Spagna}, {Steinmetz}, {Stonkut{\'e}}, {Sutorius}, {Th{\'e}venin}, {Tosi},
  {Tsantaki}, {Vink}, {Wright}, {Wyse}, {Zoccali}, {Zorec}, {Zucker}, \&
  {Walton}}]{2022A&A...666A.121R}
{Randich}, S., {Gilmore}, G., {Magrini}, L., {et~al.} 2022, \aap, 666, A121

\bibitem[{{Sariya} \& {Yadav}(2015)}]{2015A&A...584A..59S}
{Sariya}, D.~P. \& {Yadav}, R.~K.~S. 2015, \aap, 584, A59

\bibitem[{{Savino} \& {Posti}(2019)}]{2019A&A...624L...9S}
{Savino}, A. \& {Posti}, L. 2019, \aap, 624, L9

\bibitem[{{Schiavon} {et~al.}(2017){Schiavon}, {Zamora}, {Carrera},
  {Lucatello}, {Robin}, {Ness}, {Martell}, {Smith},
  {Garc{\'\i}a-Hern{\'a}ndez}, {Manchado}, {Sch{\"o}nrich}, {Bastian},
  {Chiappini}, {Shetrone}, {Mackereth}, {Williams}, {M{\'e}sz{\'a}ros},
  {Allende Prieto}, {Anders}, {Bizyaev}, {Beers}, {Chojnowski}, {Cunha},
  {Epstein}, {Frinchaboy}, {Garc{\'\i}a P{\'e}rez}, {Hearty}, {Holtzman},
  {Johnson}, {Kinemuchi}, {Majewski}, {Muna}, {Nidever}, {Nguyen}, {O'Connell},
  {Oravetz}, {Pan}, {Pinsonneault}, {Schneider}, {Schultheis}, {Simmons},
  {Skrutskie}, {Sobeck}, {Wilson}, \& {Zasowski}}]{2017MNRAS.465..501S}
{Schiavon}, R.~P., {Zamora}, O., {Carrera}, R., {et~al.} 2017, \mnras, 465, 501

\bibitem[{{Sch{\"o}nrich} {et~al.}(2010){Sch{\"o}nrich}, {Binney}, \&
  {Dehnen}}]{2010MNRAS.403.1829S}
{Sch{\"o}nrich}, R., {Binney}, J., \& {Dehnen}, W. 2010, \mnras, 403, 1829

\bibitem[{{Shipp} {et~al.}(2018){Shipp}, {Drlica-Wagner}, {Balbinot},
  {Ferguson}, {Erkal}, {Li}, {Bechtol}, {Belokurov}, {Buncher}, {Carollo},
  {Carrasco Kind}, {Kuehn}, {Marshall}, {Pace}, {Rykoff}, {Sevilla-Noarbe},
  {Sheldon}, {Strigari}, {Vivas}, {Yanny}, {Zenteno}, {Abbott}, {Abdalla},
  {Allam}, {Avila}, {Bertin}, {Brooks}, {Burke}, {Carretero}, {Castander},
  {Cawthon}, {Crocce}, {Cunha}, {D'Andrea}, {da Costa}, {Davis}, {De Vicente},
  {Desai}, {Diehl}, {Doel}, {Evrard}, {Flaugher}, {Fosalba}, {Frieman},
  {Garc{\'\i}a-Bellido}, {Gaztanaga}, {Gerdes}, {Gruen}, {Gruendl}, {Gschwend},
  {Gutierrez}, {Hartley}, {Honscheid}, {Hoyle}, {James}, {Johnson}, {Krause},
  {Kuropatkin}, {Lahav}, {Lin}, {Maia}, {March}, {Martini}, {Menanteau},
  {Miller}, {Miquel}, {Nichol}, {Plazas}, {Romer}, {Sako}, {Sanchez},
  {Santiago}, {Scarpine}, {Schindler}, {Schubnell}, {Smith}, {Smith},
  {Sobreira}, {Suchyta}, {Swanson}, {Tarle}, {Thomas}, {Tucker}, {Walker},
  {Wechsler}, \& {DES Collaboration}}]{2018ApJ...862..114S}
{Shipp}, N., {Drlica-Wagner}, A., {Balbinot}, E., {et~al.} 2018, \apj, 862, 114

\bibitem[{{Sneden} {et~al.}(2004){Sneden}, {Kraft}, {Guhathakurta}, {Peterson},
  \& {Fulbright}}]{2004AJ....127.2162S}
{Sneden}, C., {Kraft}, R.~P., {Guhathakurta}, P., {Peterson}, R.~C., \&
  {Fulbright}, J.~P. 2004, \aj, 127, 2162

\bibitem[{{Sollima}(2020)}]{2020MNRAS.495.2222S}
{Sollima}, A. 2020, \mnras, 495, 2222

\bibitem[{{Sparke} \& {Gallagher}(2000)}]{2000gaun.book.....S}
{Sparke}, L.~S. \& {Gallagher}, John~S., I. 2000, {Galaxies in the universe :
  an introduction}

\bibitem[{{Steinmetz} {et~al.}(2020){Steinmetz}, {Matijevic}, {Enke},
  {Zwitter}, {Guiglion}, {McMillan}, {Kordopatis}, {Valentini}, {Chiappini},
  {Casagrande}, {Wojno}, {Anguiano}, {Bienayme}, {Bijaoui}, {Binney}, {Burton},
  {Cass}, {de Laverny}, {Fiegert}, {Freeman}, {Fulbright}, {Gibson}, {Gilmore},
  {Grebel}, {Helmi}, {Kunder}, {Munari}, {Navarro}, {Parker}, {Ruchti},
  {Recio-Blanco}, {Reid}, {Seabroke}, {Siviero}, {Siebert}, {Stupar}, {Watson},
  {Williams}, {Wyse}, {Anders}, {Antoja}, {Birko}, {Bland-Hawthorn}, {Bossini},
  {Garcia}, {Carrillo}, {Chaplin}, {Elsworth}, {Famaey}, {Gerhard}, {Jofre},
  {Just}, {Mathur}, {Miglio}, {Minchev}, {Monari}, {Mosser}, {Ritter},
  {Rodrigues}, {Scholz}, {Sharma}, \& {Sysoliatina}}]{2020yCat.3283....0S}
{Steinmetz}, M., {Matijevic}, G., {Enke}, H., {et~al.} 2020, VizieR Online Data
  Catalog, III/283

\bibitem[{{Tang} {et~al.}(2017){Tang}, {Cohen}, {Geisler}, {Schiavon},
  {Majewski}, {Villanova}, {Carrera}, {Zamora}, {Garcia-Hernandez}, {Shetrone},
  {Frinchaboy}, {Meza}, {Fern{\'a}ndez-Trincado}, {Mu{\~n}oz}, {Lin}, {Lane},
  {Nitschelm}, {Pan}, {Bizyaev}, {Oravetz}, \& {Simmons}}]{2017MNRAS.465...19T}
{Tang}, B., {Cohen}, R.~E., {Geisler}, D., {et~al.} 2017, \mnras, 465, 19

\bibitem[{{Tang} {et~al.}(2018){Tang}, {Fern{\'a}ndez-Trincado}, {Geisler},
  {Zamora}, {M{\'e}sz{\'a}ros}, {Masseron}, {Cohen},
  {Garc{\'\i}a-Hern{\'a}ndez}, {Dell'Agli}, {Beers}, {Schiavon}, {Sohn},
  {Hasselquist}, {Robin}, {Shetrone}, {Majewski}, {Villanova}, {Schiappacasse
  Ulloa}, {Lane}, {Minnti}, {Roman-Lopes}, {Almeida}, \&
  {Moreno}}]{2018ApJ...855...38T}
{Tang}, B., {Fern{\'a}ndez-Trincado}, J.~G., {Geisler}, D., {et~al.} 2018,
  \apj, 855, 38

\bibitem[{{Tang} {et~al.}(2020){Tang}, {Fern{\'a}ndez-Trincado}, {Liu}, {Yu},
  {Yan}, {Gao}, {Shi}, \& {Geisler}}]{2020ApJ...891...28T}
{Tang}, B., {Fern{\'a}ndez-Trincado}, J.~G., {Liu}, C., {et~al.} 2020, \apj,
  891, 28

\bibitem[{{Tang} {et~al.}(2019){Tang}, {Liu}, {Fern{\'a}ndez-Trincado},
  {Geisler}, {Shi}, {Zamora}, {Worthey}, \& {Moreno}}]{2019ApJ...871...58T}
{Tang}, B., {Liu}, C., {Fern{\'a}ndez-Trincado}, J.~G., {et~al.} 2019, \apj,
  871, 58

\bibitem[{{Tang} {et~al.}(2021){Tang}, {Wang}, {Huang}, {Li}, {Yu}, {Geisler},
  {Dias}, {Fern{\'a}ndez-Trincado}, {Carballo-Bello}, \&
  {Cabrera-Lavers}}]{2021ApJ...908..220T}
{Tang}, B., {Wang}, Y., {Huang}, R., {et~al.} 2021, \apj, 908, 220

\bibitem[{{Testa} {et~al.}(2000){Testa}, {Zaggia}, {Andreon}, {Longo},
  {Scaramella}, {Djorgovski}, \& {de Carvalho}}]{2000A&A...356..127T}
{Testa}, V., {Zaggia}, S.~R., {Andreon}, S., {et~al.} 2000, \aap, 356, 127

\bibitem[{{Thomas} {et~al.}(2020){Thomas}, {Jensen}, {McConnachie},
  {C{\^o}t{\'e}}, {Venn}, {Longeard}, {Carlberg}, {Chapman}, {Cuillandre},
  {Famaey}, {Ferrarese}, {Gwyn}, {Hammer}, {Ibata}, {Malhan}, {Martin}, {Mei},
  {Navarro}, {Reyl{\'e}}, \& {Starkenburg}}]{2020ApJ...902...89T}
{Thomas}, G.~F., {Jensen}, J., {McConnachie}, A., {et~al.} 2020, \apj, 902, 89

\bibitem[{{Valcarce} {et~al.}(2012){Valcarce}, {Catelan}, \&
  {Sweigart}}]{2012A&A...547A...5V}
{Valcarce}, A.~A.~R., {Catelan}, M., \& {Sweigart}, A.~V. 2012, \aap, 547, A5

\bibitem[{{Vasiliev} \& {Baumgardt}(2021)}]{2021MNRAS.505.5978V}
{Vasiliev}, E. \& {Baumgardt}, H. 2021, \mnras, 505, 5978

\bibitem[{{Wan} {et~al.}(2020){Wan}, {Lewis}, {Li}, {Simpson}, {Martell},
  {Zucker}, {Mould}, {Erkal}, {Pace}, {Mackey}, {Ji}, {Koposov}, {Kuehn},
  {Shipp}, {Balbinot}, {Bland-Hawthorn}, {Casey}, {Da Costa}, {Kafle},
  {Sharma}, \& {De Silva}}]{2020Natur.583..768W}
{Wan}, Z., {Lewis}, G.~F., {Li}, T.~S., {et~al.} 2020, \nat, 583, 768

\bibitem[{{Wan} {et~al.}(2021){Wan}, {Oliver}, {Baumgardt}, {Lewis}, {Gieles},
  {H{\'e}nault-Brunet}, {de Boer}, {Balbinot}, {Da Costa}, \&
  {Mackey}}]{2021MNRAS.502.4513W}
{Wan}, Z., {Oliver}, W.~H., {Baumgardt}, H., {et~al.} 2021, \mnras, 502, 4513

\bibitem[{{Weatherford} {et~al.}(2023){Weatherford}, {K{\i}ro{\u{g}}lu},
  {Fragione}, {Chatterjee}, {Kremer}, \& {Rasio}}]{2023ApJ...946..104W}
{Weatherford}, N.~C., {K{\i}ro{\u{g}}lu}, F., {Fragione}, G., {et~al.} 2023,
  \apj, 946, 104

\bibitem[{{Yanny} {et~al.}(2009){Yanny}, {Rockosi}, {Newberg}, {Knapp},
  {Adelman-McCarthy}, {Alcorn}, {Allam}, {Allende Prieto}, {An}, {Anderson},
  {Anderson}, {Bailer-Jones}, {Bastian}, {Beers}, {Bell}, {Belokurov},
  {Bizyaev}, {Blythe}, {Bochanski}, {Boroski}, {Brinchmann}, {Brinkmann},
  {Brewington}, {Carey}, {Cudworth}, {Evans}, {Evans}, {Gates}, {G{\"a}nsicke},
  {Gillespie}, {Gilmore}, {Nebot Gomez-Moran}, {Grebel}, {Greenwell}, {Gunn},
  {Jordan}, {Jordan}, {Harding}, {Harris}, {Hendry}, {Holder}, {Ivans},
  {Ivezi{\v{c}}}, {Jester}, {Johnson}, {Kent}, {Kleinman}, {Kniazev},
  {Krzesinski}, {Kron}, {Kuropatkin}, {Lebedeva}, {Lee}, {French Leger},
  {L{\'e}pine}, {Levine}, {Lin}, {Long}, {Loomis}, {Lupton}, {Malanushenko},
  {Malanushenko}, {Margon}, {Martinez-Delgado}, {McGehee}, {Monet}, {Morrison},
  {Munn}, {Neilsen}, {Nitta}, {Norris}, {Oravetz}, {Owen}, {Padmanabhan},
  {Pan}, {Peterson}, {Pier}, {Platson}, {Re Fiorentin}, {Richards}, {Rix},
  {Schlegel}, {Schneider}, {Schreiber}, {Schwope}, {Sibley}, {Simmons},
  {Snedden}, {Allyn Smith}, {Stark}, {Stauffer}, {Steinmetz}, {Stoughton},
  {SubbaRao}, {Szalay}, {Szkody}, {Thakar}, {Sivarani}, {Tucker}, {Uomoto},
  {Vanden Berk}, {Vidrih}, {Wadadekar}, {Watters}, {Wilhelm}, {Wyse}, {Yarger},
  \& {Zucker}}]{2009AJ....137.4377Y}
{Yanny}, B., {Rockosi}, C., {Newberg}, H.~J., {et~al.} 2009, \aj, 137, 4377

\bibitem[{{Yong} {et~al.}(2014){Yong}, {Roederer}, {Grundahl}, {Da Costa},
  {Karakas}, {Norris}, {Aoki}, {Fishlock}, {Marino}, {Milone}, \&
  {Shingles}}]{2014MNRAS.441.3396Y}
{Yong}, D., {Roederer}, I.~U., {Grundahl}, F., {et~al.} 2014, \mnras, 441, 3396

\bibitem[{{Yu} {et~al.}(2021){Yu}, {Tang}, {Fern{\'a}ndez-Trincado}, {Geisler},
  {Yan}, \& {Soto}}]{2021ApJ...913...23Y}
{Yu}, J., {Tang}, B., {Fern{\'a}ndez-Trincado}, J.~G., {et~al.} 2021, \apj,
  913, 23

\bibitem[{{Zhao} {et~al.}(2012){Zhao}, {Zhao}, {Chu}, {Jing}, \&
  {Deng}}]{2012RAA....12..723Z}
{Zhao}, G., {Zhao}, Y.-H., {Chu}, Y.-Q., {Jing}, Y.-P., \& {Deng}, L.-C. 2012,
  Research in Astronomy and Astrophysics, 12, 723

\bibitem[{{Zhou} {et~al.}(2023){Zhou}, {Li}, {Huang}, \&
  {Zhang}}]{2023ApJ...946...73Z}
{Zhou}, Y., {Li}, X., {Huang}, Y., \& {Zhang}, H. 2023, \apj, 946, 73

\end{thebibliography}

\begin{appendix} 
\appendix
\onecolumn
\centering
\section{Orbits of 6 GCs in Sec \ref{sec:result-method2}}
\label{app:orb}

\begin{figure}[htbp]
        \centering
        \begin{minipage}{0.49\linewidth}
                \centering
                \includegraphics[width=0.9\linewidth]{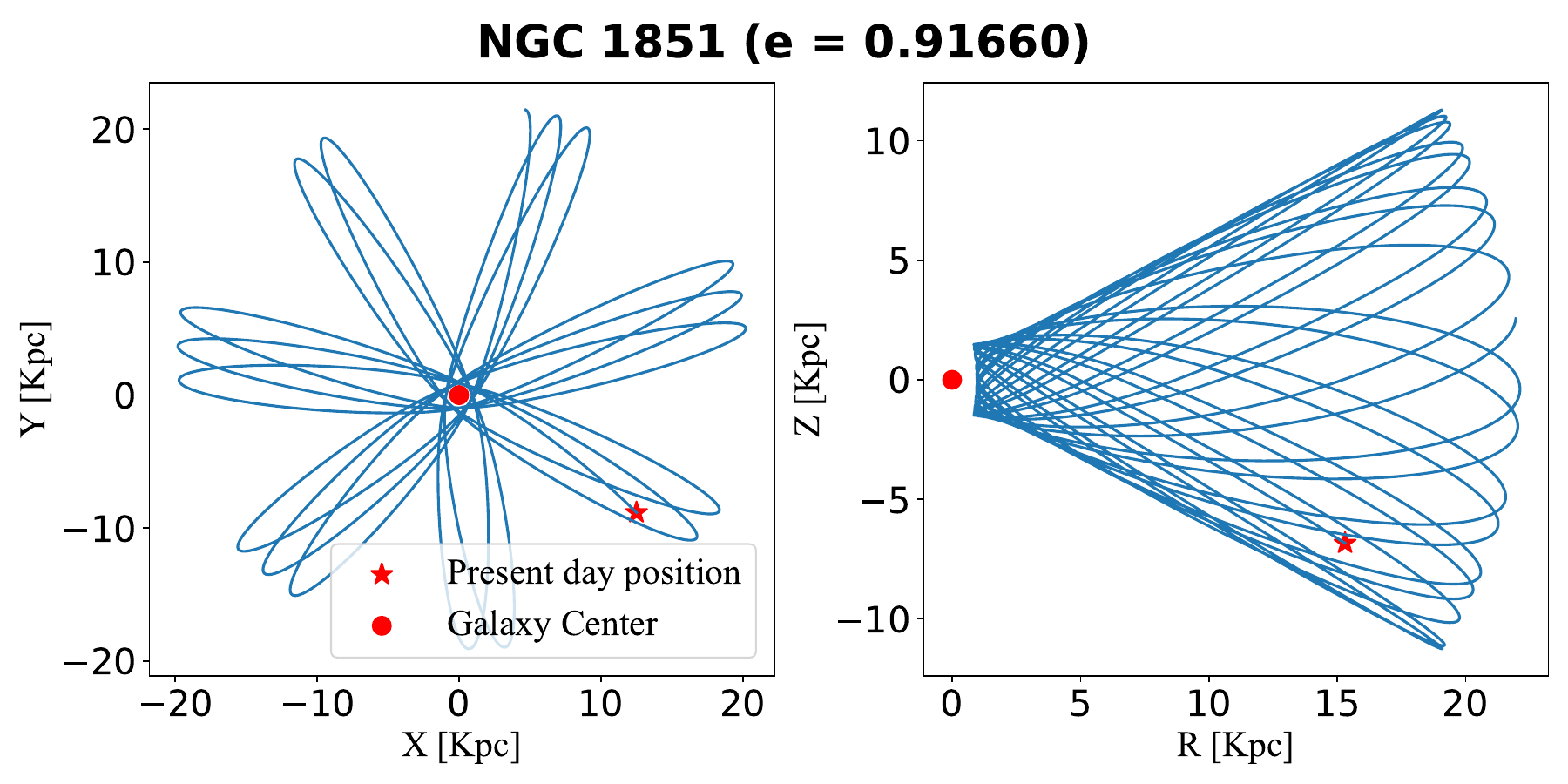}
        \end{minipage}
        \begin{minipage}{0.49\linewidth}
                \centering
                \includegraphics[width=0.9\linewidth]{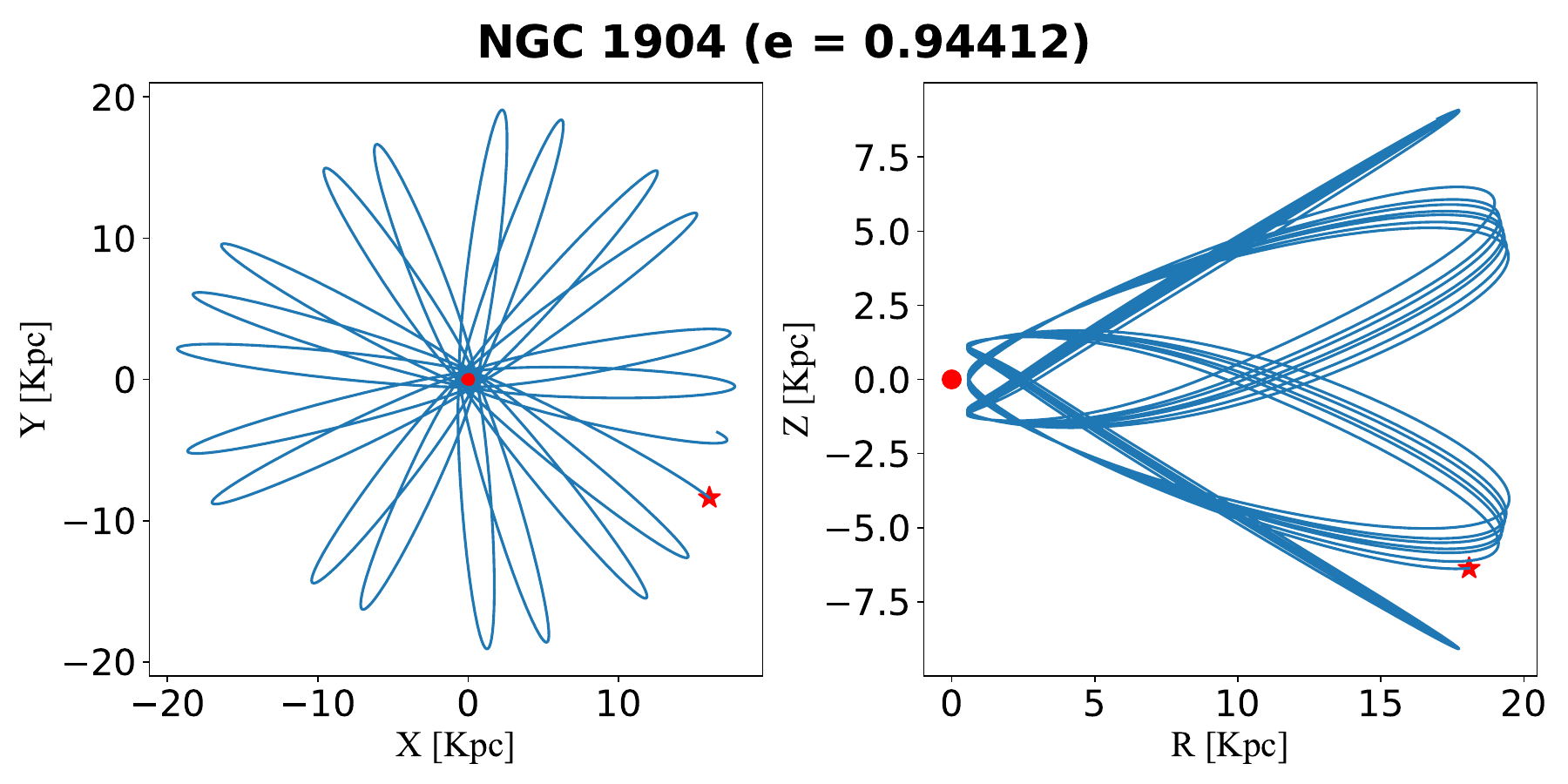}
        \end{minipage}
        
        \begin{minipage}{0.49\linewidth}
                \centering
                \includegraphics[width=0.9\linewidth]{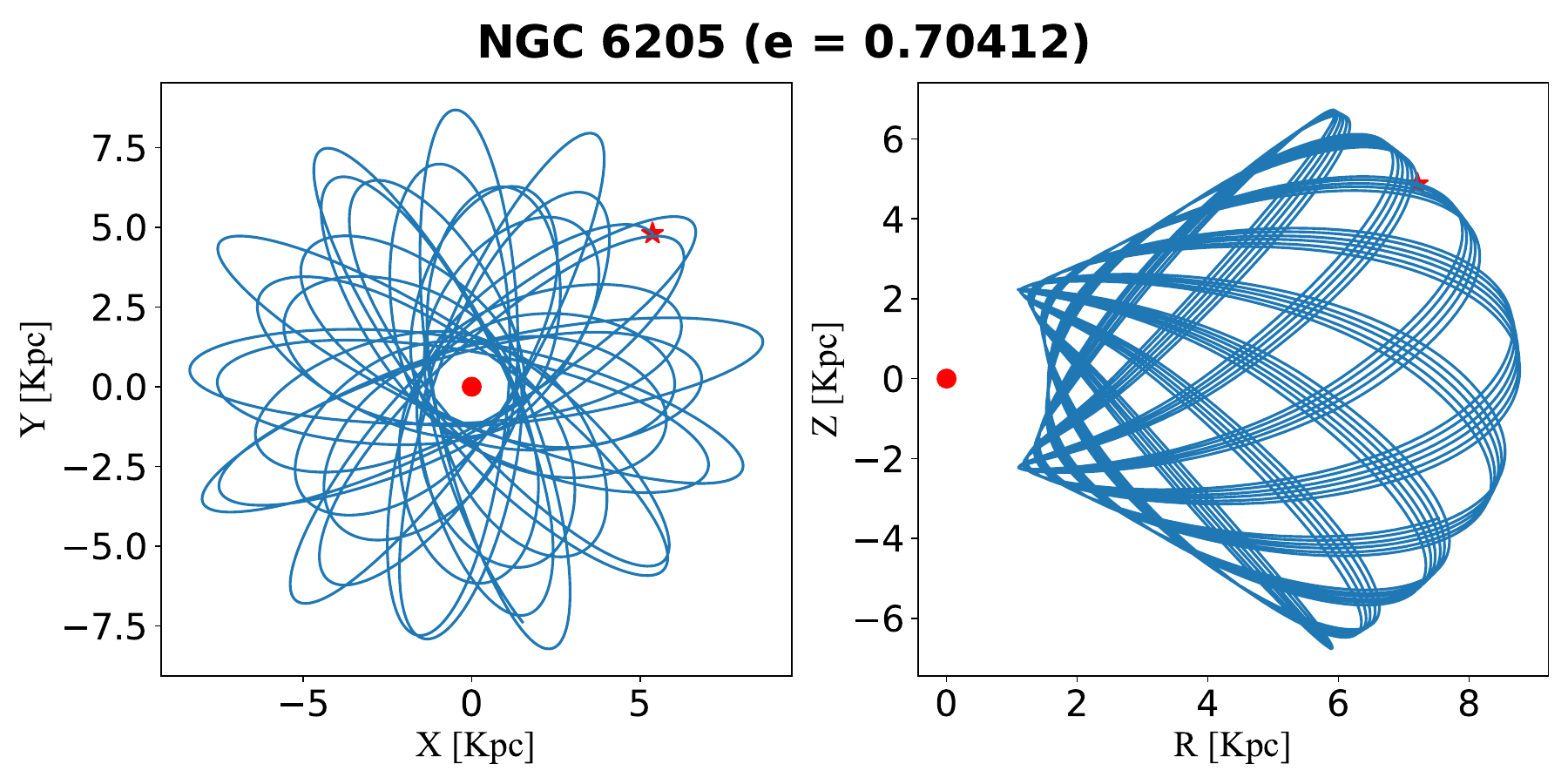}
                
        \end{minipage}
        \begin{minipage}{0.49\linewidth}
                \centering
                \includegraphics[width=0.9\linewidth]{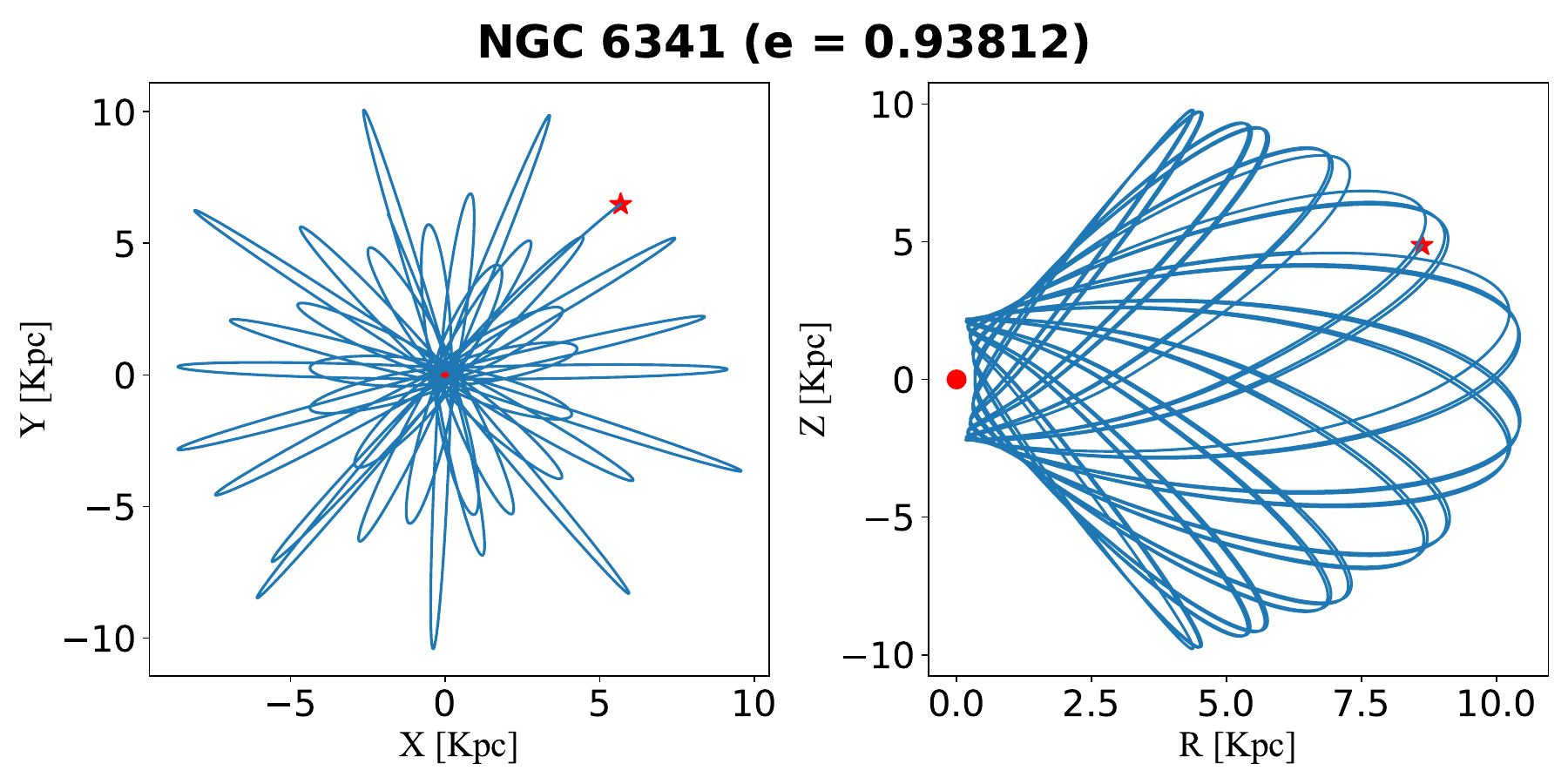}
        \end{minipage}

    \begin{minipage}{0.49\linewidth}
                \centering
                \includegraphics[width=0.9\linewidth]{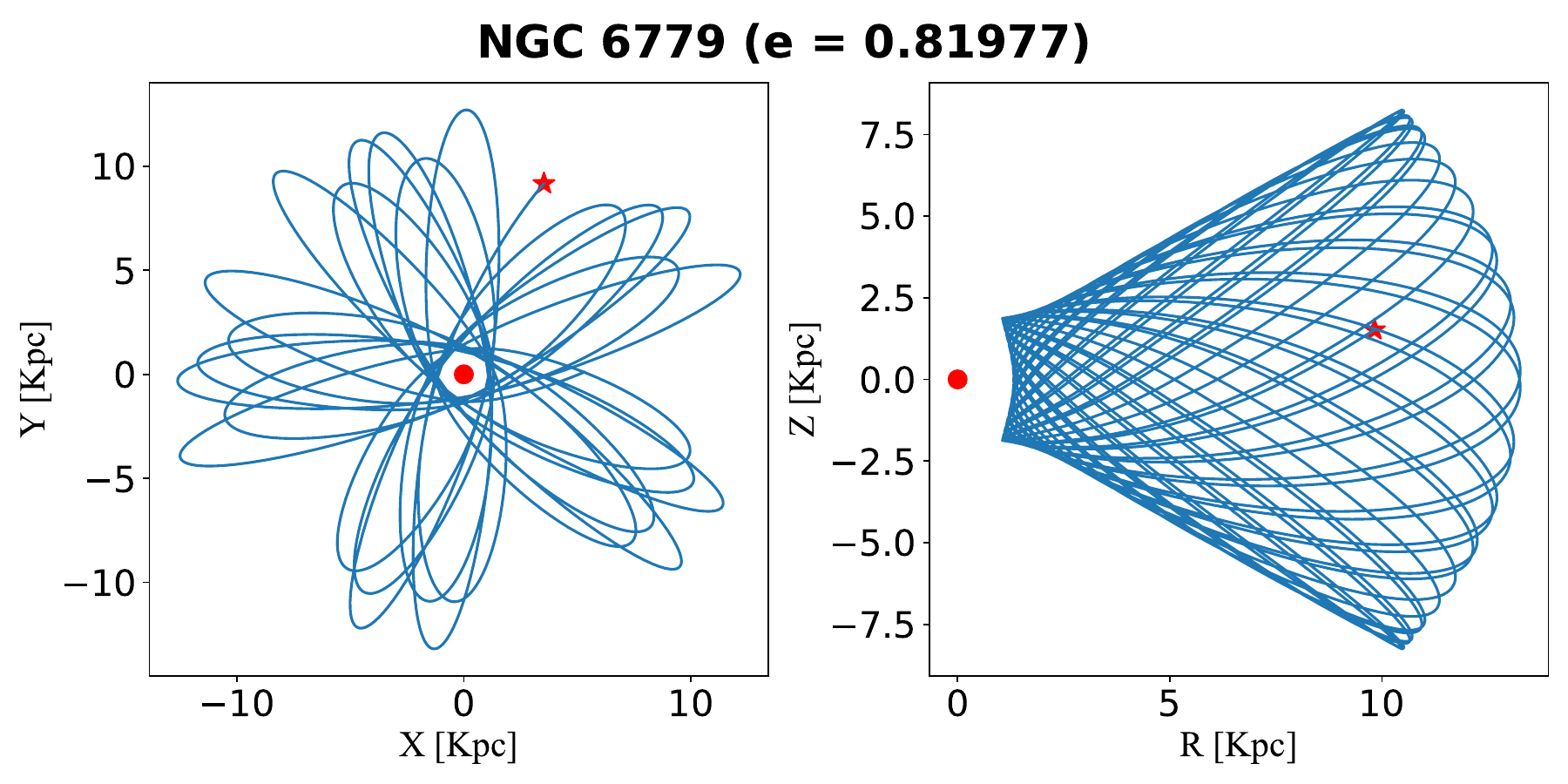}
        \end{minipage}
        \begin{minipage}{0.49\linewidth}
                \centering
                \includegraphics[width=0.9\linewidth]{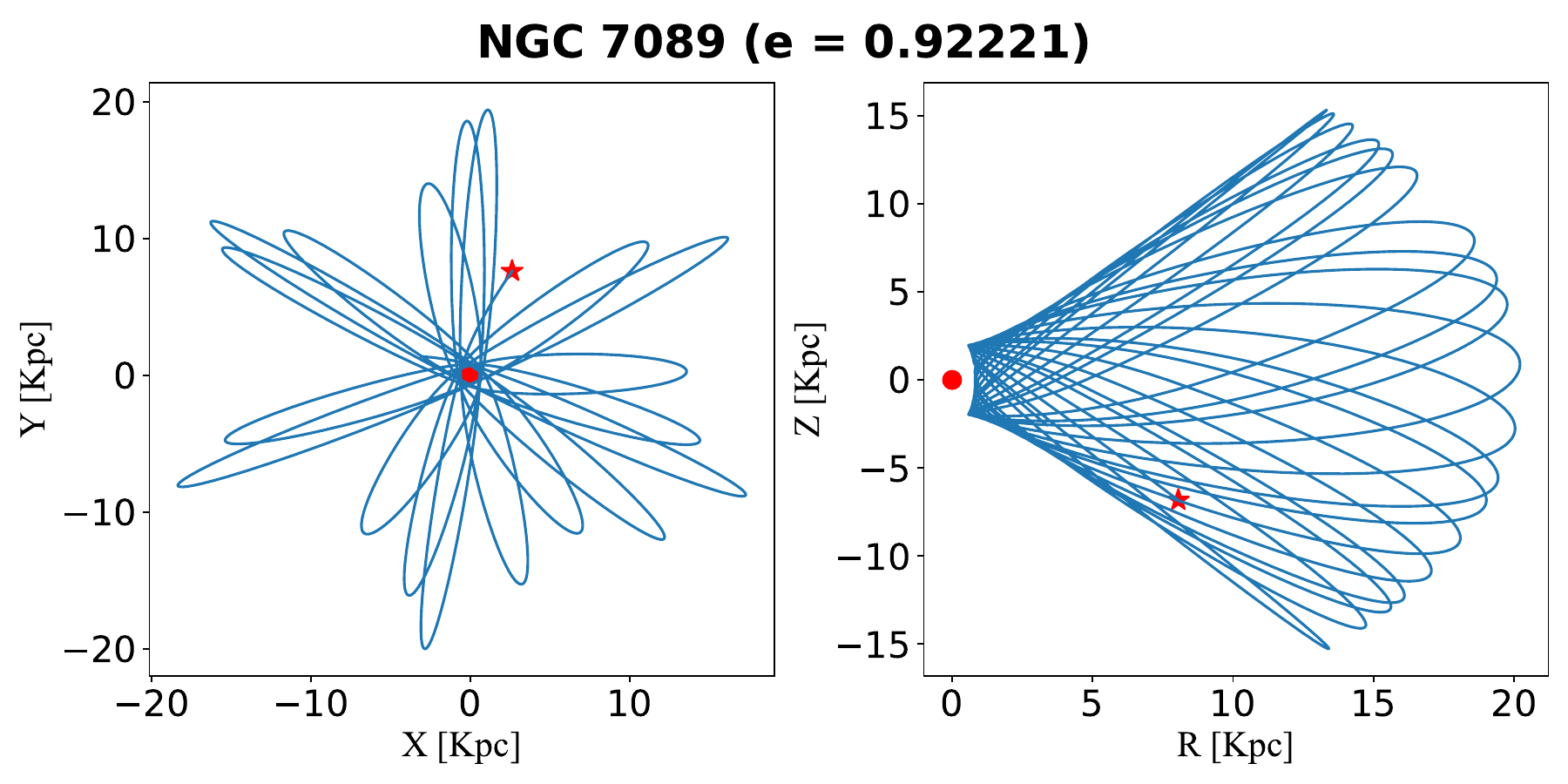}
        \end{minipage}
    \caption{Orbits of NGC 1851, NGC 1904, NGC 6205, NGC 6341, NGC 6779 and NGC 7089.}
\end{figure}

\end{appendix}

\end{document}